\documentclass[onecolumn,12pt]{IEEEtran}

\usepackage{amssymb,amsfonts,amsmath,amstext,mathrsfs}
\usepackage{latexsym}
\usepackage{eufrak}
\usepackage{epsfig,psfrag,color,graphics,epsf}
\usepackage{enumerate,cite}
\usepackage{bbm}
\usepackage{stmaryrd}
\usepackage{soul}
\usepackage{tcolorbox}

\usepackage{amsmath}
\usepackage{amssymb}
\usepackage{amsfonts}
\usepackage{amsthm}
\usepackage{graphicx}
\usepackage{float}
\usepackage{graphics}
\usepackage{graphicx}
\usepackage{euscript}
\usepackage{psfrag}

\usepackage{standalone}
\usepackage{tikz,pgfplots}

\DeclareMathOperator*{\argmax}{arg\,max}

\newtheorem{theorem}{Theorem}
\newtheorem{lemma}{Lemma}
\newtheorem{claim}{Claim}
\newtheorem{proposition}{Proposition}

\newtheorem{corollary}{Corollary}
\newtheorem{definition}{Definition}

\newcommand{\dotleq}{%
\DOTSB\mathrel{\mathop{\kern0pt \leq}\limits^{\textstyle.}}}
\newcommand{\dotgeq}{%
\DOTSB\mathrel{\mathop{\kern0pt \geq}\limits^{\textstyle.}}}

\newcommand{\reals} {\mathbb{R}}

\newcommand{\beq} {\begin{equation}}
\newcommand{\eeq} {\end{equation}}
\newcommand{\beqa} {\begin{align}}
\newcommand{\eeqa} {\end{align}}

\newcommand{\indicator}{\mathbbm{1}}

\newcommand {\bx} {\boldsymbol{x}}
\newcommand {\by} {\boldsymbol{y}}
\newcommand {\bz} {\boldsymbol{z}}

\newcommand {\bX} {\boldsymbol{X}}
\newcommand {\bY} {\boldsymbol{Y}}
\newcommand {\bZ} {\boldsymbol{Z}}

\newcommand{\calA}{{\mathcal A}}
\newcommand{\calB}{{\mathcal B}}
\newcommand{\calC}{{\mathcal C}}
\newcommand{\calD}{{\mathcal D}}
\newcommand{\calE}{{\mathcal E}}
\newcommand{\calF}{{\mathcal F}}

\newcommand{\calK}{{\mathcal K}}
\newcommand{\calL}{{\mathcal L}}
\newcommand{\calM}{{\mathcal M}}

\newcommand{\calP}{{\mathcal P}}
\newcommand{\calQ}{{\mathcal Q}}

\newcommand{\calS}{{\mathcal S}}
\newcommand{\calT}{{\mathcal T}}
\newcommand{\calU}{{\mathcal U}}

\newcommand{\calW}{{\mathcal W}}
\newcommand{\calX}{{\mathcal X}}
\newcommand{\calY}{{\mathcal Y}}
\newcommand{\calZ}{{\mathcal Z}}

\newcommand{\EE}{{\mathbb E}}

\allowdisplaybreaks
\IEEEoverridecommandlockouts

\begin{document}



\title{Upper Bounds on the Mismatched Reliability Function and Capacity Using a Genie Receiver
}

\author{\IEEEauthorblockN{Anelia Somekh-Baruch\\}
\IEEEauthorblockA{\tt{Faculty of Engineering\\ Bar-Ilan University \\
Ramat-Gan, Israel}
}
\thanks{
This work was supported by the Israel Science Foundation (ISF) grant no.\ 631/17.
}}

\maketitle
\begin{abstract}

We develop a novel framework for proving converse theorems for channel coding, which is based on the analysis technique of multicast transmission with an additional auxiliary receiver, which serves as a genie to the original receiver. The genie 
provides the original receiver a certain narrowed list of codewords to choose from that includes the transmitted one. This technique is used to derive upper bounds on the mismatch capacity of discrete memoryless channels as well as the reliability function with a mismatched decoding metric. Unlike previous works, our bounding technique exploits also the inherent symmetric requirement from the codewords, leading to these new upper bounds. Since the computations of most of the known bounds on the mismatch capacity are rather complicated, we further present a method to obtain relaxed bounds that are easier to compute. As an example, we analyze the obtained bounds in the binary-input channels case. We conclude by presenting simpler bounds on the reliability function, and provide sufficient conditions for their tightness in certain ranges of rates.

\end{abstract}

\section{Introduction}
This paper addresses the problem of determining the fundamental limits of reliable communication over a discrete memoryless channel $W$ with a given decoding metric $q$. This setup is usually referred to as mismatched decoding, since the decoding metric differs from the optimal maximum likelihood (ML) metric that is matched to the communication channel. 
While the ML decoder minimizes the error probability, in certain cases it is not applicable for various reasons such as channel estimation errors or for practical decoder implementation considerations. 
The highest achievable rate with decoding metric $q$ is referred to as the mismatch capacity. 
The problem of determining the fundamental bounds on channels with mismatched decoding is also related to other information-theoretic setups such as zero-error transmission over communication channels.

There have been quite a few works on achievable rates for channels with mismatched decoding from the information theoretic view point. A partial list of works is \cite{Hui83,ShamaiKaplan1993information,CsiszarKorner81graph,Lapidoth96,SomekhBaruch_mismatchachievableIT2014,SomekhBaruchISIT_2013,ScarlettMartinezGuilleniFabregasISIT_2013,MerhavKaplanLapidothShamai94,Lapidoth96b,GantiLapidothTelatar2000,ShamaiSason2002,ScarlettMartinezGuilleniFabregas_mismatch_2014_IT,ScalettMartinezGuilleniFabregas_IT_2016,ScarlettPengMerhavMartinezGuilleniFabregas_mismatch_2014_IT,ScarlettSomehkBaruchMartinezGuilleniFabregas2015}, and for a survey on the subject see \cite{ScarlettGuilleniFabregasSomekhBaruchMartinez2020}, and references therein.

Early results with a converse flavor were derived in \cite{CsiszarNarayan95}, where a necessary and sufficient condition for the positivity of the mismatch capacity was determined, as well as a single-letter expression for the mismatch capacity in the case of the binary input binary output channel. 

Later works \cite{SomekhBaruch_general_formula_IT2015,SomekhBaruchConverses_IT2018} presented among other results, a tight multi-letter soft converse result, and initiated multi-letter bounds, which involve a max-min form, where the maximum is taken over input distributions, and a minimum is taken over a set of auxiliary channels that satisfy certain constraints (see \cite[Theorem 4]{SomekhBaruchConverses_IT2018} and \cite[Theorems 4-5]{SomekhBaruch_general_formula_IT2015}).

The work \cite{Kangarshahi_GuilleniFabregas2020_IT} proved a max-min upper bound having a single-letter form. The bound of \cite{Kangarshahi_GuilleniFabregas2020_IT} was obtained by constructing a graph in the output space of the channel and using graph-theoretic and large deviations tools.

Tighter max-min single-letter upper bounds on the mismatch capacity were derived in \cite{SomekhBaruch2020singleletter_part1} (see also \cite{SomekhBaruch_ITW_2020}), using a proof technique based on multicast transmission of only one common message over a broadcast channel. This technique relies on extending the single-user channel $W$ with output
$Y$ to a channel that has an additional output $Z$,
with the property that the intersection event of correct $q$-decoding of the $Y$-receiver and erroneous 
decoding of the auxiliary $Z$-receiver has zero probability for any codebook of a certain composition $P$. This approach led to a strictly tighter bound with a significantly simpler proof, which holds also for continuous alphabet channels. 
A tighter bound of a more involved form was presented in \cite{SomekhBaruch2020singleletter_part1}), that relied on considering a $Z$-receiver which is genie-aided and is informed of the actual joint empirical distribution of the transmitted codeword and the $Z$-output sequence. 
Equivalence classes of isomorphic channel-metric pairs $(W,q)$ were further introduced in \cite{SomekhBaruch2020singleletter_part1}, that enabled to derive a sufficient condition for the tightness of the bound.

In a later work, \cite{SomekhBaruch_ISIT2021}, 
the class of broadcast channels was enlarged 
to include channels satisfying that if the $Z$-receiver makes an error, then with high probability (approaching $1$) so does the $Y$-receiver. 
An improved bound was derived in 
 \cite{KangarshahiGuilleniFabregas2021spherepackingArXiv,KangarshahiGuilleniFabregasISIT2021spherepacking}, which further enlarged the class of channels. 
The above mentioned bounds are described in detail in this paper, but most of them are quite complicated to compute in the sense that it is required to solve an optimization problem in order to determine whether a certain channel belongs to the set or not. 
The tightest bound 
known to date which is easily computable in this sense
is the basic bound of \cite{SomekhBaruch2020singleletter_part1}.

The study of the reliability function (error exponents) of channels with ML decoding has been quite extensive (see, e.g., \cite{Gallager68,CsiszarKorner81,SHANNONGallagerBerlekamp1967522,shannonGallagerBerlekamp1967lower}). 
Clearly, the known upper bounds are applicable also to mismatched decoding. 
Lower bounds on the exponents with 
mismatched decoding were derived in several works such as \cite{CsiszarKorner81graph,SomekhBaruch_mismatchachievableIT2014,ScarlettPengMerhavMartinezGuilleniFabregas_mismatch_2014_IT} (see also \cite{ScarlettGuilleniFabregasSomekhBaruchMartinez2020}, and references therein). 
Recently, an upper bound on the reliability function with mismatched decoding for zero-rate codes was derived in \cite{BondaschiGuilleniFabregasDalai-IT2021}. 
For a wide class of channel-metric pairs, this bound was shown to be tight at $R=0^+$. 
In \cite{KangarshahiGuilleniFabregas2021spherepackingArXiv}, an upper bound was derived for all rates up to the aforementioned bound of \cite{KangarshahiGuilleniFabregas2021spherepackingArXiv} on the mismatch capacity. 
 
In this paper, we refine our multicast approach to allow the genie-aided auxiliary $Z$-receiver of the channel to serve as a genie for the original $Y$-receiver. We call this approach ``transmission with a genie-aided-genie". 
The idea is very simple: the genie-aided auxiliary $Z$-receiver informs the original $Y$-receiver of the list of all the codewords that share the same empirical statistics (joint type-class) with the channel $Z$-output as that of the actual transmitted codeword. Doing so, it narrows down the list of competing hypothesized codewords that the original mismatched decoder needs to choose from.
We consider channels satisfying the condition that if the list  and yields a lower bound on the probability of error.
This leads to a basic upper bound on the mismatch capacity and the reliability function. 
We further present possibly looser bounds which are easily computable.

This paper is organized as follows: 
In Section \ref{sc: Notation} we present notation conventions. 
A formal statement of the mismatched decoding setup appears in Section \ref{sc: A Formal Statement of the Problem}. 
In Section \ref{sc: GAG setup}, we present the transmission with a Genie-Aided-Genie proof technique. 
Section \ref{sc: An Overview of the Main Results} summarizes our main results. 
In Section \ref{cs: Binary-Input-Any-Output Channels}, we study the case of binary-input DMCs. 
In Section \ref{sc: Comparison to Previous Results} we compare our results to former results. 
Sections \ref{sc: Proof of Exponent Theorem } and \ref{sc: Proof of Theorem Capacity} are dedicated to the proofs of the main theorems regarding reliability function and mismatch capacity theorems, respectively. 
In Section \ref{sc: aifugviudfg}, we establish some simpler bounds on the reliability function and sufficient conditions for tightness. 
Section \ref{sc: Concluding Remarks} discusses some concluding remarks.
Proofs of additional results and lemmas appear in the appendix.

\section{Notation}\label{sc: Notation}

Throughout this paper, scalar random variables (RVs) are denoted by capital letters, their sample values are denoted by their respective lower case letters, and their alphabets are denoted by their respective calligraphic letters; e.g.\ $X$, $x$, and $\calX$, respectively. A similar convention applies to random vectors of dimension $n$ and their sample values, which are 
denoted in boldface; e.g., $\bx$. The set of all $n$-vectors with components taking values in a certain finite alphabet are denoted by the same alphabet superscripted by $n$, e.g., $\calX^n$. 
Logarithms are taken to the natural base $e$, unless stated otherwise. 

For a given sequence $\bx \in \calX^n$, where $\calX$ is a finite alphabet,  $\widehat{P}_{\bx}$ denotes the empirical distribution on $\calX$ extracted from $\bx$; in other words, $\widehat{P}_{\bx}$ is the vector $\{ \widehat{P}_{\bx} (x), x\in\calX\}$, where $ \widehat{P}_{\bx} (x)$ is the relative frequency of the symbol $x$ in the vector $\bx$. The type-class of $\bx$ is the set of $\bx'\in\calX^n$ such that $\widehat{P}_{\bx'}=\widehat{P}_{\bx}$, which is denoted $\calT_n(\widehat{P}_{\bx})$. 
Similarly, the joint empirical distribution of two sequences $\bx,\by$, denoted $\widehat{P}_{\bx\by}$, 
is the vector $\{ \widehat{P}_{\bx\by} (x,y), (x,y)\in\calX\times \calY\}$, where $ \widehat{P}_{\bx\by} (x,y)$ is the relative frequency of the pair of symbols $(x,y)$ in the vector $(\bx,\by)$; i.e. the number of indices $i$ such that $(x_i,y_i)=(x,y)$ normalized by $n$. 
The conditional type-class of $\by$ given $\bx$ is the set of $\tilde{\by}$'s such that $\widehat{P}_{\bx,\tilde{\by}}=\widehat{P}_{\bx,\by}$, which is denoted $\calT_n(\widehat{P}_{\by|\bx} |\bx)$. The set of all probability distributions on $\calX$ is denoted by $\calP(\calX)$, the set of conditional distributions from $\calX$ to $\calY$ is denoted $\calP(\calY|\calX)$, and the set of empirical distributions of order $n$ on alphabet $\calX$ is denoted $\calP_n(\calX)$.

Information theoretic quantities, such as entropy, conditional entropy, and mutual information are denoted following the usual conventions in the information theory literature, e.g., $H (X )$, $H (X |Y )$, $I(X;Y)$ and so on. To emphasize the dependence of a quantity on a certain underlying probability distribution, say $\mu$, we at times use notations such as $H(\mu )$, $H(\mu_{X|Y})$, $I(\mu_{XY})$, or $H_{\mu}(X)$, $H_{\mu}(Y|X)$, etc. 
For $P\in\calP(\calX)$, and $V,Q\in\calP(\calY|\calX)$ we denote the conditional divergence as
$
D(V\|Q|P)= \sum_{x,y}P(x)V(y|x)\log\frac{V(y|x)}{Q(y|x)}$. 

The expectation operator is denoted by $\mathbb{E} (\cdot)$, and to make the dependence on the underlying distribution $\mu$ explicit, it is denoted by $\mathbb{E}_\mu(\cdot)$. The cardinality of a finite set $\calA$ is denoted by $|\calA|$. The indicator function of an event $\calE$ is denoted by $1_{\{\calE \}}$.

For two measures $P,Q$ defined on the same measurable space $(\Omega,\calF)$ the measure $P$ is said to be absolutely continuous w.r.t.\ $Q$ if for every $\calE\in \calF$ such that $Q(\calE)=0$ it also holds that $P(\calE)=0$: this is denoted $P\ll Q$.

\section{A Formal Statement of the Problem}\label{sc: A Formal Statement of the Problem}

Consider transmission over a memoryless channel described by a conditional probability $W(y|x)$, with input $x\in\calX$ and output $y\in\calY$ finite alphabets $\calX$ and $\calY$; in particular, $W(y|x)$ is a conditional probability mass function. We define 
$
W^n(\by|\bx) = \prod_{k=1}^n W(y_k|x_k)
$
for input/output sequences $\bx = (x_1,\dotsc,x_n)\in\calX^n$ and $\by= (y_1,\dotsc,y_n)\in\calY^n$. The corresponding RVs are denoted by $\bX$ and $\bY$.

An encoder maps a message $m\in \{1,\dotsc,\mathbb{M}_n\}$ to a channel input sequence $\bx_m\in\calX$, 
where the number of messages is denoted by $\mathbb{M}_n$.
The message, represented by the random variable $M$, is assumed to take values in $\{1,\dotsc,\mathbb{M}_n\}$ equi-probably. This mapping induces an $(n,\mathbb{M}_n)$-codebook $\calC_n=\{\bx_1,\dotsc,\bx_{\mathbb{M}_n}\}$ with rate $R_n=\frac{1}{n}\log \mathbb{M}_n$.

Upon observing the channel output $\by$, the decoder produces an estimate of the transmitted message $\widehat m \in \{1,\dotsc,\mathbb{M}_n\}$.
We consider the decoding rule 
\beq
\widehat m = \argmax_{i\in\{1,\dotsc,\mathbb{M}_n\}} \,q(\bx_i,\by), \label{eq:decoder}
\eeq
where $q(\bx_i,\by)$ is a certain additive decoding metric 
\cite[Ch.~2]{CsiszarKorner81} defined by a {\em single-letter} mapping $q:\calX\times\calY\rightarrow \reals$ such that 
\begin{equation}
   q(\bx,\by)=  \frac{1}{n} \sum_{i=1}^n q(x_i,y_i) = \EE_{\widehat{P}_{\bx\by}}[ q(X,Y) ]\triangleq q(\widehat{P}_{\bx\by}) ,
    \label{eq:q_additive}
\end{equation}
where for convenience we slightly abuse notation using $q$ for both the per-letter metric $q(x,y)$ and the $n$-letter metric $q(\widehat{P}_{\bx\by})$. 
Throughout the paper it is assumed that ties are broken uniformly between the maximizers. 

Denoting the RV corresponding to the decoded message by $\widehat M_q(\bY)$, we denote the average error probability as
$
P_e(W,\calC_n,q) = \Pr \bigl[\widehat M_q(\bY) \neq M\bigr]
$.

A rate $R$ is said to be achievable with decoding metric $q$ if there exists a sequence of codebooks $\calC_n$, $n=1,2,...$ such that $\frac{1}{n}\log|\calC_n|\geq R$ and $\lim_{n\rightarrow \infty}P_e(W,\calC_n,q)=0$. 
The channel capacity w.r.t.\ metric $q$, denoted $C_q(W)$, is defined as the supremum of achievable rates, and is referred to as the mismatch capacity.

Since the optimal decoding rule w.r.t.\ average error probability (of equiprobable messages) is ML, which is additive for DMCs, 
Shannon's channel capacity $C(W)$ can be viewed in fact as the channel capacity w.r.t.\ the metric $q(x,y)=\log W(y|x)$; that is, 
\begin{flalign}
C(W)=C_q(W)|_{q(x,y)=\log W(y|x)}.
\end{flalign} 
A rate-exponent pair $(R,E)$ is said to be achievable 
    for channel $W$ with decoding metric $q$ if there exists a sequence of codebooks $\calC_n$, $n=1,2,...$ such that 
    for all $n$, $\frac{1}{n}\log|\calC_n|\geq R$ and 
    \begin{flalign}
        \liminf_{n\to\infty} \, -\,\frac{1}{n}\log P_e(W,\calC_n,q)\geq E. 
    \end{flalign}
    Equivalently, we say that $E$ is an achievable error exponent at rate $R$ if $(R,E)$ is an achievable rate-exponent pair.

The reliability function of the channel with decoding metric $q$ is the supremum of achievable error exponents as a function of the code rate, and is denoted by $E^q(R,W)$. The reliability function with ML decoding metric is denoted $E(R,W)$.

Define the highest achievable exponent with $P$ constant composition codebooks of block length $n$ as
\begin{flalign}\label{eq: afuivudgfuvf}
e_n^q(R,P,W)&\triangleq 
\max_{\calC_n\subseteq\calT_n(P):\; |\calC_n|\geq e^{nR}}
-\frac{1}{n}\log P_e(q,W,\calC_n).
\end{flalign}
Using standard arguments that follow from the fact that the number of type-classes grows polynomially with $n$, it can be shown that
\begin{flalign}\label{eq: afuddivudgfuvf}
E^q(R,W)=\liminf_{n\rightarrow \infty}\max_{P_n\in\calP_n(\calX)} e_n^q(R,P_n,W),
\end{flalign}
and for this reason, the main focus of this paper is on analyzing constant composition codes.

\section{Main Results}\label{sc: NewResults} 

In this section we present new upper bounds on the mismatch capacity and the reliability function of the DMC when the decoder uses a mismatched decoding metric $q$. 
Before we present our new bounds, we describe the main idea behind our proof technique. 

\subsection{The Multicast Transmission with a Genie-Aided-Genie Proof Technique}\label{sc: GAG setup}

We refine our multicast transmission setup which was introduced in \cite{SomekhBaruch2020singleletter_part1}, 
that extends the single-user channel $W_{Y|X}$ to a two-output (broadcast) channel $W_{YZ|X}$ having an additional output $Z$ over some finite alphabet $\calZ$. An encoder uses a codebook $\calC_n=\{\overline{\bx}_i\}_{i=1}^M$ of size $\mathbb{M}_n=e^{nR}$ to transmit a message over the channel.

\begin{figure}[H]
	\resizebox{\columnwidth}{!}{	\tikzstyle{block} = [draw, fill=white!20, rectangle, 
	minimum height=3em, minimum width=6em]
	\tikzstyle{genblock} = [draw, fill=white!20, circle]
	
	\tikzstyle{block2} = [draw, fill=white!20, rectangle, 
	minimum height=3em, minimum width=3em]    
	\tikzstyle{sum} = [draw, fill=white!20, circle, node distance=1cm]
	\tikzstyle{input} = [coordinate]
	\tikzstyle{output} = [coordinate]
	\tikzstyle{pinstyle} = [pin edge={to-,thin,black}]
	
	\begin{tikzpicture}[auto, node distance=2cm]
	\node [input, name=input] {};

	\node [block, right of=input] (Encoder) at (-1.5,1) {Encoder} ;
	\node [block, right of=Encoder, node distance=3cm] (channel) at (1.5,1) {$\Large
	\substack{
	\normalsize \mbox{Channel}\\ \\
	W^n_{YZ|X}}$};
	\draw [thick,->] (Encoder) -- node[name=u] {$\bX$} (channel);
	\node [block, right of=channel, node distance=3cm] (Decoder) at (5.5,1) {$Y$-Decoder $q$};
	\draw [thick,->] (channel) -- node[name=y] {$\bY$}  (Decoder) ;  
	\node [output, right of=Decoder] (output) {};
	\node (M) (input)  at (-2.5,1) {$M$};
	\node (M2) (output)  at (11.5,1) {$\hat{M}$};
	\draw [thick](4.6,-2) --  node[name=z] {$\bZ$} (4.6, 0.4);
	\draw [thick](4.6,-4.5) --   (4.6, -2);
	\draw [thick,->](4.6,-2) --   (7.25, -2);
	\draw [thick,->](4.6,-4.5) --   (7.45, -4.5);
	
	\draw [thick](2.5,-5.5) -- (2.5, 1);
	\draw [thick,->](2.5,-5.5) --  node[name=jdfhv] {} (7.45, -5.5);

	\node [block, right of=channel, node distance=3cm] (Decoder2) at (5.5,-2){$\substack{$Z$-\mbox{Decoder}\\ \\   \mbox{List-Genie}}$};
	\node [genblock, below of=Decoder2, node distance=2cm] (TGenie) at (8.5,-3) {Type-Genie};
	\draw [thick,->] (TGenie) -- node[name=TTT] {$\hat{P}_{\bX\bZ}$}  (Decoder2) ; 
	
	\draw [thick,->] (Decoder2) --  node[right of= channel, name=L] {$\quad \quad \calL(\bZ,\hat{P}_{\bx\bz})=\{\bx_i(\bZ,\hat{P}_{\bx\bz})\}$}  (Decoder) ; 

	\draw [draw,thick,->] (input) -- (Encoder);
	
	\draw [thick,->] (Decoder) -- (output);

%
%
%
%

%
	
	\end{tikzpicture}
	
	}
	\caption{Transmission over a two-output (broadcast) channel with mismatched decoding and a Genie-Aided-Genie.}
	\label{Figure_BC_genie}
\end{figure}
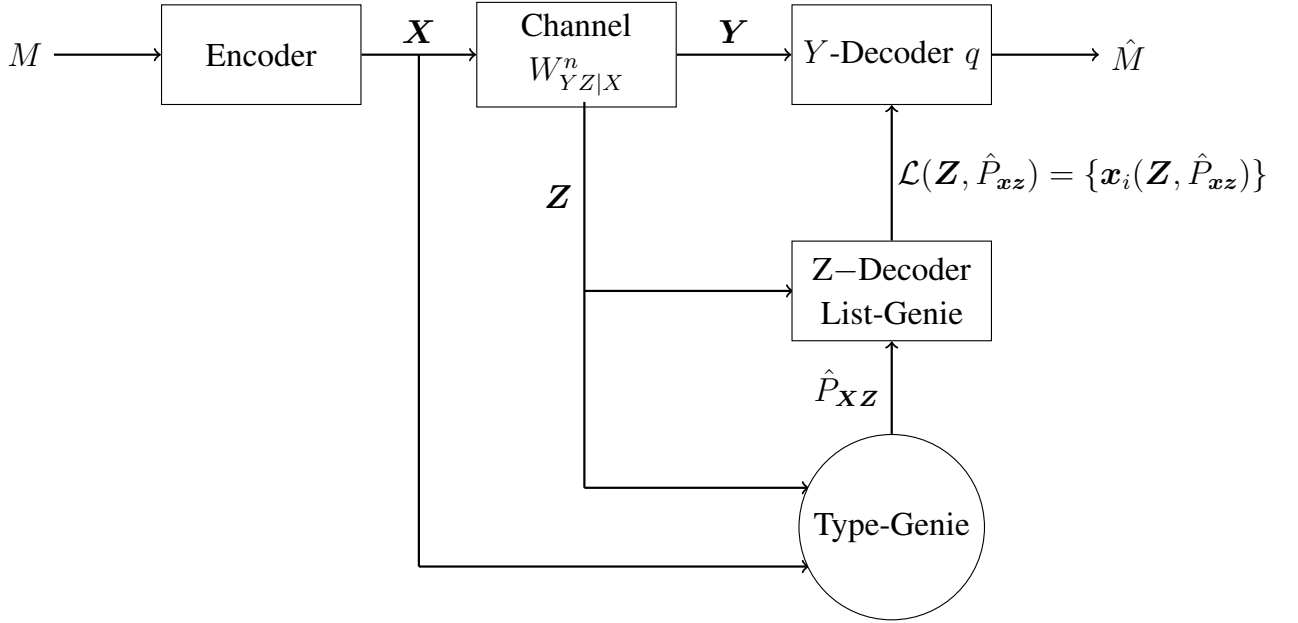

As in \cite[Theorem 2, Eq. (41)]{SomekhBaruch2020singleletter_part1}, we add a type-genie that informs the $Z$-receiver of the actual joint empirical distribution $\widehat{P}_{\bx\bz}$ of the input and output signals $(\bx,\bz)$. 
The refinement of the proof technique in this paper, is in that the $Z$-receiver, which observes $\bz$, serves as a genie to the $Y$-receiver by providing it with the list 
\begin{flalign}\label{eq: List dfn with type}
\calL(\bz,\widehat{P}_{\bx\bz})&\triangleq \calC_n\cap\calT_n(\widehat{P}_{\bx|\bz}|\bz)\nonumber\\
&\triangleq \big\{\bx_1(\bz,\widehat{P}_{\bx\bz}),\ldots,\bx_{|\calL(\bz,\widehat{P}_{\bx\bz})|}(\bz,\widehat{P}_{\bx\bz})\big\}
\end{flalign}
 of all the codewords, which lie in $\calT_n(\widehat{P}_{\bx|\bz}|\bz)$ (the conditional type-class given the received signal $\bz$). 
The $Y$-receiver compares the metrics of all the codewords in the list and outputs
\begin{flalign}\label{eq: genie aided decoder}
\widehat{m}&=\argmax_{i:\; \overline{\bx}_i\in \calL(\bz,\widehat{P}_{\bx\bz})} q(\overline{\bx}_i,\by),
\end{flalign}
where ties are broken uniformly between the maximizers. This setup is depicted in Fig.\  \ref{Figure_BC_genie}. 

 Since by definition the true codeword belongs to this narrowed down list, the error probability in mismatched decoding of the $Y$ receiver cannot exceed that of the original single-user setup.

The narrowing of the codebook to the list enables to lower bound the average pairwise error probability within the list since within the list, the joint pairwise empirical distribution of codewords is more constrained compared to the entire codebook (in particular, the joint empirical distribution of each codeword with $\bz$ is identical) this is beneficial for the bound. 
The actual size of the list does not play a major role in the proof, except for exceeding $1$; i.e., containing at least one additional message except the transmitted one. 
 
We show that for rates which exceed our upper bound on the mismatch capacity, $\overline{C}_q(W)$, the average error probability is bounded away from zero. As for rates below $\overline{C}_q(W)$, we upper bound the exponent of the average error probability.

Note that we use the terminology of multicast transmission over a {\it broadcast channel} similar to \cite{SomekhBaruch2020singleletter_part1}, and we also refer to it as a {\it two-output channel} in certain cases where this terminology is more suitable. 
We emphasize that this should not be confused with the 
ordinary broadcast channel over which two separate messages are transmitted, and each is intended to a different receiver.

A summary of the main differences between the proof technique of this work and previous ones can be found in Section \ref{sc: Major section comparison Cap}. 

\subsection{An Overview of the Main Results}\label{sc: An Overview of the Main Results}

In this section we present the new upper bounds on the achievable error exponent using decoding metric $q$, and the mismatch capacity $C_q(W)$. Throughout this paper, we adopt the shorthand notation that $W$ without subscript signifies the original single-user channel $W_{Y|X}$. Whenever we refer to other marginal distributions; i.e., $W_{Y|XZ}$, $W_{Z|XY}$ or $W_{YZ|X}$, the subscript is mentioned explicitly.

\subsubsection{\underline{The Main Upper Bound on $C_q(W)$}}

Consider the set of two-output (broadcast) 
channels: 
\begin{flalign}
&\calW_q(P_X)=\bigg\{
P_{YZ|X}:\; \min_{\substack{V_{U\widetilde{X}XZY}:V_{XZY}=P_{XYZ}\\
V_{U\widetilde{X}XZ}= V_{UX\widetilde{X}Z}\\
\widetilde{X}-(U,Z)-X\\
(\widetilde{X},U)-(X,Z)-Y
 }}\EE q(\widetilde{X},Y)\geq \EE q(X,Y) \bigg\}
\label{eq: calW q dfnasdfionidaf;i;}
\end{flalign}
where $U$ is an auxiliary RV\footnote{By inspecting (\ref{eq: calW q dfnasdfionidaf;i;}) it is evident that without loss of generality one can take $U$ such that $Z$ is a deterministic function of $U$ (perhaps with larger alphabet for $U$), because one can replace any $U$ by $U'=(U,Z)$. Therefore, one can add the constraint $H(Z|U)=0$ to the set of the minimization in (\ref{eq: calW q dfnasdfionidaf;i;}) without changing the resulting bound (\ref{eq: ;aifuhv;iuofv}). 
} with alphabet size $|\calU|\leq |\calX|^2|\calZ|$, and the condition $V_{U\widetilde{X}XZ}= V_{UX\widetilde{X}Z}$ signifies that for all $(u,x_1,x_2,z)$, $V_{U\widetilde{X}XZ}(u,x_1,x_2,z)= V_{U\widetilde{X}XZ}(u,x_2,x_1,z)$. 

Note that the RVs $(U,\widetilde{X},X,Y,Z)$ have the following intuitive explanation originating from the proofs: $X$- a typical channel input symbol, $(Y,Z)$ - typical channel output symbols, $\widetilde{X}$ - a worst-case competing codeword symbol, and $U$ - a time-sharing RV. 

Next, let
\begin{flalign}
\overline{C}_q(W)&\triangleq\max_{P_X} \min_{P_{YZ|X}\in \calW_q(P_X),\; P_{Y|X}=W}I(X;Z),\label{eq: ;aifuhv;iuofv}
\end{flalign}
where throughout this paper, we adopt the convention that the minimum and maximum over an empty set equal $\infty$ and $-\infty$, respectively. 

Our basic bound on $C_q(W)$ of the following theorem is proved in Section \ref{sc: Proof of Exponent Theorem }.
\begin{theorem}\label{th: thorem mismatch capacity}
For any $W$, additive metric $q\in\mathbb{R}\cup \{-\infty\}$, and finite alphabet $\calZ$, 
\begin{flalign}
C_q(W)&\leq\overline{C}_q(W).
\label{eq: dfhvdiudddsvhigdfddfiluhliufihi}\end{flalign}
\end{theorem}
The bound $\overline{C}_q(W)$ is tighter compared to previously known bounds, see Section \ref{sc: Comparison to Previous Results} for a comparison to previous works. 
Note that $\overline{C}_q(W)$ is quite difficult to compute, since in order to determine whether a two-output channel $P_{YZ|X}$ belongs to the set $\calW_q(P_X)$ or not, one needs to solve the minimization problem 
in (\ref{eq: Delta q definition a}). 
A similar problem arises with the computation of many of the previous bounds (e.g., those of \cite{SomekhBaruch_ISIT2021,KangarshahiGuilleniFabregas2021spherepackingArXiv}). 
For this reason, in the next section we present a few looser bounds that are easier to compute. 

It is worth mentioning though, that denoting
\begin{flalign}\label{eq: Delta q definition a}
&\Delta^q(P_{XZU},P_{Y|XZ})\triangleq \nonumber\\
&\sum_{x,z,u,\widetilde{x},y}P_{UZ}(u,z) P_{X|UZ}(x|u,z)P_{X|UZ}(\widetilde{x}|u,z) P_{Y|XZ}(y|x,z)[q(\widetilde{x},y)-q(x,y)],
\end{flalign}
the bound $\overline{C}_q(W)$ in (\ref{eq: ;aifuhv;iuofv}) can also be expressed as 
\begin{flalign}
\overline{C}_q(W)
&=\max_{P_X}\; \min_{\substack{P_{YZ|X}:\; \min_{P_{U|XZ} }\Delta^q(P_{XZU},P_{Y|XZ})\geq 0 ,\\ P_{Y|X}=W}}I(X;Z)\label{eq: aufdgvui}\\
&=\max_{(P_X,\; P_{U|XZ} )}\; \min_{\substack{P_{YZ|X}:\; \Delta^q(P_{XZU},P_{Y|XZ})\geq 0 ,\\ P_{Y|X}=W}}I(X;Z).
\end{flalign}
In this form of the bound, for every given $(P_X,P_{U|XZ})$, it is easy to determine whether the channel $P_{YZ|X}$ satisfies $\Delta^q(P_{XZU},P_{Y|XZ})\geq 0$ or not, nevertheless, one still needs to optimize over the pair $(P_X, P_{U|XZ} )$.

\subsubsection{\underline{Possibly Looser Easier to Compute Bounds on $C_q(W)$}}

We next present several sets\footnote{In fact, the sets, as well as $\calW_q(P)$ are also functions of the alphabet cardinality $|\calZ|$, but for the sake of simplicity we omit this dependence from our notation.}
 of two-output channels, 
 $\widetilde{\calW}^{sym}_q(P_X)$, and $\calW^{sym}_q(P_X)$, $\widetilde{\calW}_q(P_X)$, $\calW^{psd}_q(P_X)$, which are subsets of $\calW_q(P_X)$.  

Consider the following set of symmetric distributions:
\begin{flalign}\label{eq: sym dfn set}
\calP_{sym}(\calX^2\times \calZ)&\triangleq \{P_{\widetilde{X}ZX}\in \calP(\calX^2\times \calZ):\; \forall (x,\widetilde{x}, z),\; P_{\widetilde{X}ZX}(\widetilde{x},z,x)=P_{\widetilde{X}ZX}(x,z,\widetilde{x})\},
\end{flalign}
and define 
\begin{flalign}
&\widetilde{\calW}^{sym}_q(P_X)\triangleq \bigg\{P_{YZ|X}:\;\min_{\substack{V_{\widetilde{X}XZY}:\;V_{XYZ}=P_{XYZ}\\
V_{\widetilde{X}ZX}\in \calP_{sym}(\calX^2\times \calZ)\\
\widetilde{X}-(X,Z)-Y}}\EE q(\widetilde{X},Y)\geq \EE q(X,Y),\bigg\}\label{eq: cytctgyt} 
\\&{\calW}^{sym}_q(P_X)\triangleq \bigg\{P_{YZ|X}:\;\min_{\substack{V_{\widetilde{X}XZY}:\; V_{XYZ}=P_{XYZ}\\
V_{\widetilde{X}ZX}\in \calP_{sym}(\calX^2\times \calZ)
\\
\widetilde{X}-(X,Z)-Y,\;
 \forall (x,z),\; V_{\widetilde{X}|X,Z}(x|x,z)\geq P_{X|Z}(x|z) }}\EE q(\widetilde{X},Y)\geq \EE q(X,Y)\bigg\}. \label{eq: aiufhvuifhd}
\end{flalign}

Further, define the third set of two-output channels
\begin{flalign}
\widetilde{\calW}_q(P_X)&\triangleq 
\left\{ P_{YZ|X}
: \substack{\forall(z,V_{X|Z}):z\in\calZ,\\
 P_{Z}\times V_{X|Z}\ll P_{ZX} } ,\;\sum_{x,\widetilde{x},y} V(x|z)V(\widetilde{x}|z)P_{Y|XZ}(y|x,z)[q(\widetilde{x},y)-q(x,y)]\geq 0
\right\}
\\
&=
\left\{ P_{YZ|X}
: \substack{\forall(z,V_{XY|Z}):z\in\calZ,\\V_{Y|XZ}=P_{Y|XZ},\\
 P_{Z}\times V_{X|Z}\ll P_{XZ}}\;\EE_{V_{X|z}V_{Y|z}}q(X,Y)\geq \EE_{V_{XY|z}}q(X,Y)
\right\}
 \end{flalign}
where as mentioned in Section \ref{sc: Notation}, $\ll$ denotes absolute continuity.

Now, denote w.l.o.g.\ $\calX=\{1,...,|\calX|\}$, and consider the collection of symmetric $|\calX|\times |\calX|$ matrices $\{\calD^q(P_{Y|X,Z=z})\}$, indexed by $z\in\calZ$, whose $(i,j)$-th entries are given by:
\begin{flalign}
\{\calD^q(P_{Y|X,Z=z})\}_{i,j}&= 
\sum_y \left(P_{Y|XZ}(y|i,z)[q(j,y)-q(i,y)]+P_{Y|XZ}(y|j,z)[q(i,y)-q(j,y)]\right)\\
&=\sum_y[P_{Y|XZ}(y|i,z)-P_{Y|XZ}(y|j,z)]\cdot[q(j,y)-q(i,y)].\label{eq: ad;chdiuch}
\end{flalign}
Denoting that a matrix $\calD$ is positive semi-definite (p.s.d.) by $\calD\succeq 0$, we define the last set of two-output channels:
\begin{flalign}
\calW^{psd}_q(P_X)\triangleq 
 \left\{ P_{YZ|X}
:\;\forall z\in\calZ,\; \calD^q(P_{Y|X,Z=z})\succeq0\right\}.
\end{flalign}

Let \begin{flalign}
C^{sym}_q(W)&\triangleq\max_{P_X} \min_{P_{YZ|X}\in \calW^{sym}_q(P_X),\; P_{Y|X}=W}I(X;Z)\label{eq: Csym sihfi}\\
\widetilde{C}^{sym}_q(W)&\triangleq\max_{P_X} \min_{P_{YZ|X}\in \widetilde{\calW}^{sym}_q(P_X),\; P_{Y|X}=W}I(X;Z)\\
\widetilde{C}_q(W)&\triangleq\max_{P_X} \min_{ P_{YZ|X}\in \widetilde{\calW}_q(P_X),\; P_{Y|X}=W
}I(X;Z).\label{eq: oiafhovih222}\\
C^{psd}_q(W)&\triangleq\max_{P_X}\;  \min_{P_{YZ|X}\in \calW^{psd}_q(P_X)
:\; P_{Y|X}=W
}I(X;Z).\label{eq: oiafhovih222ad;ovugh;}
\end{flalign}

The possibly looser bounds compared to $\overline{C}_q(W)$ are presented in the following proposition, which is proved in Appendix \ref{sc: bounds sym ib}.

\begin{proposition}\label{pr: adh;oidvjhoi;hdjfoiv}
For any $P\in\calP(\calX)$, 
\begin{flalign}
\calW^{psd}_q(P)
\subseteq \widetilde{\calW}_q(P)& \subseteq \calW_q(P),\label{eq: oisfsdvijfdoijovdb}\\
\widetilde{\calW}^{sym}_q(P_X)\subseteq \calW^{sym}_q(P)&\subseteq\calW_q(P) \label{eq: oisfhbvufhdb}
\end{flalign}
and therefore
\begin{flalign}
\overline{C}_q(W)&\leq \widetilde{C}_q(W)\leq C^{psd}_q(W),\label{eq: dfhvdliufihi}\\
\overline{C}_q(W)&\leq C^{sym}_q(W)\leq \widetilde{C}^{sym}_q(W).
\end{flalign}
\end{proposition}

As mentioned before, the bound $C^{psd}_q(W)$ has a significant advantage over $\overline{C}_q(W)$, since it is easier to compute in the sense that determining whether a two-output channel $P_{YZ|X}$ belongs to the set $\calW^{psd}_q(P_X)$ requires a simple calculation. 
In particular, one needs to verify that $\forall z\in\calZ,\; \calD^q(P_{Y|X,Z=z})\succeq0$ by checking that the determinants of the  $2^{|\calX|}-1$ minors of each of these $|\calZ|$ matrices are all non-negative \cite{prussing1986principal}. 
This is in contrast to the calculation of $\overline{C}_q(W)$ and similarly several other previous bounds (e.g., those of \cite{SomekhBaruch_ISIT2021,KangarshahiGuilleniFabregas2021spherepackingArXiv}), which require to solve a certain minimization problem, such as the minimization in (\ref{eq: calW q dfnasdfionidaf;i;}), in order to determine whether a two-output channel belongs to the set $\calW_q(P_X)$ (see Section \ref{sc: Comparison to Previous Results}).

Furthermore, the bound $ C^{sym}_q(W)$ is also easier to compute numerically compared to $\overline{C}_q(W)$ and previous bounds, since the constraints in the definition of the set $\calW^{sym}_q(P)$ limit the range and the number of degrees of freedom of the solution of the optimization problem. 
In Section \ref{cs: Binary-Input-Any-Output Channels}, we analyze the various bounds for the binary-input channel case ($|\calX|=2$, $|\calY|<\infty$).

\subsubsection{\underline{Upper Bounds on the Reliability Function with Decoding Metric $q$}}

\vspace{0.1cm}

In this section we present our main bound $ E_{sp}^q(R,W)$ on the reliability function with mismatched decoding, $ E^q(R,W)$, and similar to the mismatch capacity, we present looser bounds that are easier to compute.

For $P\in \calP(\calX)$ define
\begin{flalign}
 E_{sp}^q(R,P,W)
&\triangleq 
\min_{\substack{P_{YZ|X}\in \calW_q(P):\;  I(X;Z)\leq R\\
}}
D(P_{Y|X}\|W|P)\label{eq: sp 1}\end{flalign}
Due to (\ref{eq: afuddivudgfuvf}), our main result concerning the reliability function is presented in terms of upper bounds on $e_n^q(R,P,W)$.
\begin{theorem}\label{th: Main Theorem }
Let $|\calZ|<\infty$, then 
for all $n$, and any $P\in \calP_n(\calX)$  
\begin{flalign}
e_n^q(R,P,W)
&\leq 
E_{sp}^{q}(R-\epsilon_{n,a},P,W)+\epsilon_{n,b},
\label{eq: dfhvdiudsvhidfiludffdhliufihi}
\end{flalign}
where $\epsilon_{n,a}=O(\frac{\log n}{n})$, and $\epsilon_{n,b}=O(\frac{\log n}{n})$.
\end{theorem}
The main idea of the proof of Theorem \ref{th: Main Theorem } was presented in Section \ref{sc: GAG setup}, the full proof can be found in Section \ref{sc: Proof of Exponent Theorem }, as well as the exact quantities $\epsilon_{n,a}$, and $\epsilon_{n,b}$. 
Note that the proof of Theorem \ref{th: Main Theorem } can be shortened to yield the same result as (\ref{eq: dfhvdiudsvhidfiludffdhliufihi}) with the exception that $E_{sp}^{q}$ is replaced\footnote{This is mentioned in Eq.\ (\ref{eq: advhof;idhvoisfbdfh}) which can replace the stronger result (\ref{Eq: a;ofhv;oifv}), which requires a lightly more involved proof. } by $\widetilde{E}_{sp}^{q,sym}$.

Now, define further
\begin{flalign}
\widetilde{E}_{sp}^{q}(R,P,W)&\triangleq
\min_{P_{YZ|X}\in \widetilde{\calW}_q(P):\; I(X;Z)\leq R}D(P_{Y|X}\|W|P)
\label{eq: sp 12}\\
E_{sp}^{q,psd}(R,P,W)&\triangleq
\min_{P_{YZ|X}\in \calW^{psd}_q(P):\; I(X;Z)\leq R}D(P_{Y|X}\|W|P)
\label{eq: sp 12sfoivozdfhjpv}\\
E_{sp}^{q,sym}(R,P,W)&\triangleq
\min_{P_{YZ|X}\in \calW^{sym}_q(P):\; I(X;Z)\leq R}D(P_{Y|X}\|W|P)
\label{eq: sp 12a'fhvpofdhovjpdfjopv}\\
\widetilde{E}_{sp}^{q,sym}(R,P,W)&\triangleq
\min_{P_{YZ|X}\in \widetilde{\calW}^{sym}_q(P):\; I(X;Z)\leq R}D(P_{Y|X}\|W|P).
\label{eq: sp 12a'fhsoihvodihiojpdfjopv}
 \end{flalign}
Further, let $E_{sp}^q(R,W), \widetilde{E}_{sp}^q(R,W), E_{sp}^{q,psd}(R,W)$, $E_{sp}^{q,sym}(R,W), \widetilde{E}_{sp}^{q,sym}(R,W)$ denote the maximum over $P\in\calP(\calX)$ of the above quantities (\ref{eq: sp 1})-(\ref{eq: sp 12a'fhsoihvodihiojpdfjopv}), respectively. 
The following theorem 
states the additional bounds. \begin{theorem}\label{th: advhodhvoid}
The following inequalities hold for all $(R,P,W)$
\begin{flalign}
E_{sp}^q(R,P,W)&\leq \widetilde{E}_{sp}^q(R,P,W)\leq E_{sp}^{q,psd}(R,P,W)
\label{eq: corollary proof tisssldejjj}\\
E_{sp}^q(R,P,W)&\leq E_{sp}^{q,sym}(R,P,W)\leq \widetilde{E}_{sp}^{q,sym}(R,P,W).
\label{eq: corollary proof tildejjj}
\end{flalign}
\end{theorem}
The 
inequalities of (\ref{eq: corollary proof tisssldejjj})-(\ref{eq: corollary proof tildejjj}) follow from (\ref{eq: oisfsdvijfdoijovdb})-(\ref{eq: oisfhbvufhdb}). 

The following corollary follows from (\ref{eq: afuddivudgfuvf}), and Theorems \ref{th: Main Theorem } and \ref{th: advhodhvoid}.
\begin{corollary}
For any $|\calZ|< \infty$, 
\begin{flalign}
E^q(R,W)&\leq 
E_{sp}^q(R,W)\leq \widetilde{E}_{sp}^q(R,W)\leq E_{sp}^{q,psd}(R,W)
\label{eq: sdfijj}\\
E^q(R,W)&\leq  E_{sp}^q(R,W)\leq  E_{sp}^{q,sym}(R,W)\leq \widetilde{E}_{sp}^{q,sym}(R,W).
\label{eq: afdjvpdofj}
\end{flalign}
\end{corollary}

The following corollary, which is easily verified, states that $\overline{C}_q(W)$ is a rate above which the average error probability in $q$ decoding cannot vanish exponentially fast. 
\begin{corollary}\label{eq: afuvg;oidfuhgv;oigh}
For all $R>\overline{C}_q(W)$, $E_{sp}^q(R,P,W)=0$.
\end{corollary}

\subsubsection{\underline{Upper Bound on the Reliability Function for Type-Dependent Metrics}}

Our next corollary extends the results of Theorems \ref{th: Main Theorem } and \ref{th: advhodhvoid} to the case of type-dependent metrics, similar to the extension in \cite{SomekhBaruch2020singleletter_part1}.
The class of type-dependent metrics generalizes additive metrics in the following manner. 
It is assumed that the decoding metric $q^{}(\bx,\by)$ depends on $\bx,\by$ solely via their joint empirical distribution; i.e., $q^{}(\bx,\by)=q(\hat P_{\bx,\by})$, so $q$ can be viewed as a mapping from the empirical distributions to the reals  
$
q:\;\calP_n(\calX\times\calY)\rightarrow\reals$.
More generally, in order not to restrict attention to a specific block-length $n$, we assume that it maps the simplex to a real number; i.e., 
\beq
q:\;\calP(\calX\times\calY)\rightarrow\reals.
\eeq
We refer to this class of metrics as type-dependent (formerly referred to as $\alpha$-decoders by Csisz\'{a}r and K\"{o}rner \cite{CsiszarKorner81graph}).
 In the case of type-dependent metrics, (\ref{eq:decoder}) becomes: 
\beq
\hat m = \argmax_{i\in \{1,\dotsc,\mathbb{M}_n\}} \,q(\widehat{P}_{\bx_i\by}).
\eeq
The equivalent of the set of two-output channels $\widetilde{\calW}^{sym}_q(P_X)$ in (\ref{eq: cytctgyt}) for type-dependent metrics is given by
\begin{flalign}
&\widetilde{\calW}^{sym}_q(P_X)\triangleq 
\bigg\{P_{YZ|X}:\;\min_{\substack{V_{\widetilde{X}XZY}:\; V_{\widetilde{X}ZX}\in \calP_{sym}(\calX^2\times \calZ),\\ 
V_{\widetilde{X}XZY}=P_{XYZ}\times V_{\widetilde{X}|XZ}}
}q(V_{\widetilde{X}Y})\geq q(P_{XY})\bigg\},\label{eq: cytctgytadvhi} 
\end{flalign}
yielding $\widetilde{E}_{sp}^{q,sym}(R,P,W)$ as defined in (\ref{eq: sp 12a'fhsoihvodihiojpdfjopv}).

The result pertaining to type-dependent metrics is the following. 
\begin{corollary}\label{cr: corollary Type Dependent}
Let $|\calZ|<\infty$, and let $q(P_{XY})$ be convex in $P_{Y|X}$ for fixed $P_{X}$, then 
for all $n$, and any $P\in \calP_n(\calX)$  
\begin{flalign}
e_n^q(R,P,W)
&\leq 
\widetilde{E}_{sp}^{q,sym}(R-\epsilon_{n,c},P,W)+\epsilon_{n,d},
\label{eq: dfhvdiudsvufihi}
\end{flalign}
where $\epsilon_{n,c}=O(\frac{\log n}{n})$ and $\epsilon_{n,d}=O(\frac{\log n}{n})$,
and consequently
\begin{flalign}
E^q(R,W)\leq \widetilde{E}_{sp}^{q,sym}(R,W).
\end{flalign}
\end{corollary}
The corollary is proved in Appendix \ref{ap: Cocorollary Type Dependent}.

\section{Binary-Input Channels}\label{cs: Binary-Input-Any-Output Channels}

As mentioned above, in \cite{CsiszarNarayan95}, a single-letter expression for the mismatch capacity in the case of the binary input binary output channel was derived. 
The single-letter converse result reported in \cite{Balakirsky95} for binary-input DMCs (with $2<|\calY|<\infty$) was disproved in \cite{ScarlettSomehkBaruchMartinezGuilleniFabregas2015}. Specifically, a rate based on superposition coding was shown to exceed the claimed mismatch capacity of \cite{Balakirsky95}.

The following lemma specifies simplified explicit expressions for $ C^{sym}_q(W), \widetilde{C}^{sym}_q(W)$, and $C^{psd}_q(W)$ for binary input DMCs, by explicitly solving the minimization problems in the definitions of ${\calW}^{sym}_q(P_X)$, $\widetilde{\calW}^{sym}_q(P_X)$, and ${\calW}^{psd}_q(P_X)$. 
Let 
\begin{flalign}
d_q(P_{YZ|X},z)= \sum_y [P(y|0,z)-P(y|1,z)][q(1,y)- q(0,y)].\label{eq: d_1 dfn}
\end{flalign}

\begin{lemma}\label{lm: aofihv;oidhfvo;idhfoivhdiof}
Let $W_{Y|X}$ be a DMC with a binary input alphabet $\calX=\{0,1\}$, and let $|\calZ|<\infty$, then\footnote{Since $P_{Y|XZ}(y|x,z) $ is not defined for $(x,z)$ such that $P_{XZ}(x,z)=0$, the condition $P_Z(z)P_{X|Z}(0|z)P_{X|Z}(1|z)d_q(P_{YZ|X},z)\geq 0$ should be understood as $P_Z(z)P_{X|Z}(0|z)P_{X|Z}(1|z)>0\Rightarrow d_q(P_{YZ|X},z)\geq 0$.} 
\begin{flalign}
 C^{sym}_q(W)
&= \max_{P_X}
\min_{\substack{P_{YZ|X}:\; \forall z:\;  P_Z(z)>0,\\ \EE_{P_{X|z}\times P_{Y|z}}
q(X,Y)\geq \EE_{P_{XY|z}}q(X,Y),\\
P_{Y|X}=W}} I(X;Z)
\label{eq: relma exampledacdscaadvds}\\
&= \max_{P_X}
\min_{\substack{P_{YZ|X}:\;\forall z,\; P_Z(z)P_{X|Z}(0|z)P_{X|Z}(1|z)d_q(P_{YZ|X},z)\geq 0,\\
P_{Y|X}=W}} I(X;Z).
\label{eq: relma example}
\end{flalign}
Consequently
\begin{flalign}\label{eq: conseahviodhsi}
C^{sym}_q(W)&\leq  
\min_{\substack{P_{YZ|X}:\;\forall z,\; P_{Z|X}(z|0)P_{Z|X}(z|1)d_q(P_{YZ|X},z)\geq 0,\\
P_{Y|X}=W}} C(P_{Z|X})
\end{flalign}
Further, 
\begin{flalign}
  C^{sym}_q(W)&=\widetilde{C}^{sym}_q(W)
\label{eq: relma example2}
\end{flalign}
and 
\begin{flalign}
C^{psd}_q(W)&= \max_{P_X}
\min_{\substack{P_{YZ|X}:\; \forall z,\; d_q(P_{YZ|X},z)= 0\\
P_{Y|X}=W}} I(X;Z)\label{eq: relma example3},
\end{flalign}
\end{lemma}
Lemma \ref{lm: aofihv;oidhfvo;idhfoivhdiof} is proved in Appendix \ref{sc: Proof of Binary Lemma}. 
For given $P_X$, it is easy to check whether the condition in the minimization in (\ref{eq: relma exampledacdscaadvds}) is satisfied for a suggested channel $P_{YZ|X}$, as well as the conditions in (\ref{eq: relma example})-(\ref{eq: relma example3}). Further, it is evident that in the binary-input case $ \widetilde{C}^{sym}_q(W)\leq C^{psd}_q(W)$.

\noindent{ \bf Example $1$):} Consider the example $\calX=\{0,1\}$, $\calY=\calZ=\{0,1,2\}$ (also studied in \cite{ScarlettSomehkBaruchMartinezGuilleniFabregas2015,KangarshahiGuilleniFabregas2021spherepackingArXiv}):
\begin{flalign}\label{eq: example parameters}
W_{Y|X}&=\left(\begin{array}{ccc} 0.97 & 0.03 & 0\\
0.1 & 0.1 & 0.8
\end{array}\right) ,\; 
 q(x,y)=\left(\begin{array}{ccc} 0 &0 &0\\
 0 &\log(0.5) &\log(1.36)
\end{array}\right).
\end{flalign}
It is easy to verify that the following two-output channel 
\begin{flalign}
\widetilde{W}_{YZ|X}(y,z|x)&= \left\{\begin{array}{ll}
0.77 &(x,y,z)=(0,0,1)\\
0.6 & (x,y,z)=(1,2,2)\\
0.2 & (x,y,z)\in\{(0,0,0),(1,2,1)\}\\
0.1 &  (x,y,z)\in\{ (1,0,0),(1,1,1)\}\\
0.03 & (x,y,z)=(0,1,1)\\
0& \mbox{ otherwise}
\end{array}\right.
\end{flalign}
satisfies 
\begin{flalign}
\widetilde{W}_{Y|X}&=W_{Y|X},\,
\widetilde{W}_{Z|X}=
\left(\begin{array}{ccc} 0.2 & 0.8 & 0\\
0.1 & 0.3 & 0.6
\end{array}\right) ,
\end{flalign} 
Assume w.l.o.g.\ that the maximizing $P_X$ is non-degenerate (as otherwise the mismatch capacity equals zero). It is easy to verify that $d_q(\widetilde{W}_{YZ|X},z)\geq 0$ for all $z$ such that $\widetilde{W}_{Z|X}(z|0)\widetilde{W}_{Z|X}(z|1)>0$, i.e., $z\in\{0,1\}$. 
Thus, 
\begin{flalign}
C_q^{sym}(W_{Y|X})\leq \max_{P_X}I(P_X\times \widetilde{W}_{Z|X})\approx 0.4081\, [\mbox{bits/channel use}],\label{eq; numerical calcu}
\end{flalign}
where $P_X\approx(0.59\, ,\, 0.41)$. 
For a comparison with previous results pertaining to this example and other cases, see the next Section \ref{sc: section comparison Cap}.

\section{A Comparison to Previous Results}\label{sc: Comparison to Previous Results}

We next discuss the relationship between our new results and some relevant previous converse results.
We begin by describing the major differences in the proof technique. 
Further, in Sections \ref{sc: section comparison Cap} and \ref{sc: Cpmparison eliability Function} we describe the differences between  the mismatch capacity and reliability function expressions, respectively.

\subsubsection{\underline{Main Differences Compared to Proof Techniques of Previous Works}}\label{sc: Major section comparison Cap}

The $Z$-receiver of our previous works \cite{SomekhBaruch2020singleletter_part1,SomekhBaruch_ISIT2021} 
was either genie-aided or not (depending on the specific theorem), but the major difference is that it did not serve as a genie to the $Y$-receiver in these works. These works relied on an argument that the broadcast channel $P_{YZ|X}$ is such that if the $Y$-receiver successfully decodes the message, then so does the $Z$ receiver (in the case of \cite{SomekhBaruch_ISIT2021}  this was relaxed to $Z$-receiver that is very likely to succeed). 
 
 As for \cite{Kangarshahi_GuilleniFabregas2020_IT,KangarshahiGuilleniFabregas2021spherepackingArXiv,KangarshahiGuilleniFabregasISIT2021spherepacking}, these works did not consider multicast transmission nor genies, the technique therein relied on a construction of a graph in the output space. Our bounding technique using a two-output channel, in which the $Z$-receiver serves as a genie to the $Y$-receiver is significantly simpler and does not involve the construction of a graph and graph-theoretic tools.

\vspace{0.2cm}

Another major difference in the proof technique between the current work and \cite{KangarshahiGuilleniFabregas2021spherepackingArXiv,KangarshahiGuilleniFabregasISIT2021spherepacking} is that in the current work we exploit the symmetric requirement from the codewords: 
For the reliability function, we analyze $\Pr(\calE_{ij}|\bx_i,\bz)$, the probability of the error event $\calE_{ij}$  that $q(\bx_j,\bY)$ exceeds $q(\bx_i,\bY)$ given that $i$ is transmitted and $\bz$ received, and lower bound the average of $\frac{1}{2}[\Pr(\calE_{ij}|\bx_i,\bz)+ \Pr(\calE_{ji}|\bx_j,\bz)]$; (see Eq.\ (\ref{eq: STAM})). 
For the mismatch capacity upper bound we state a condition such that for any pair of codewords $\bx_\ell,\bx_k$ in the list, the joint empirical distribution is such that either given $(\bx_\ell,\bz)$ the event $\calE_{\ell k}$ is very likely to occur or given $(\bx_k,\bz)$ the event $\calE_{k\ell }$ is very likely to occur (see Lemma \ref{lm: aifuvuidfv} Eqs.\ (\ref{eq: apd'ohvpodfjpvokq})-(\ref{eq: apd'ohvpodfjpvok})).

The upper bound of \cite{KangarshahiGuilleniFabregas2021spherepackingArXiv,KangarshahiGuilleniFabregasISIT2021spherepacking} implies that  the symmetry and between the possible events (of each of the messages being the transmitted one) was not exploited in the derivation.

\subsubsection{\underline{Reliability Function}}\label{sc: Cpmparison eliability Function}

The general upper bounds which hold for ML decoding are applicable for mismatched decoding as well, and in particular, 
the classical sphere-packing bound \cite{CsiszarKorner81}, which is given by:
\begin{flalign}
  E_{sp}(R,P,W)=& \min_{P_{Y|X}:\; I(P\times P_{Y|X})\leq R }D(P_{Y|X}\|W|P), \nonumber
\end{flalign}
as it holds for all metrics $q$ including the ML metric. 

Consider (\ref{eq: corollary proof tisssldejjj})-(\ref{eq: corollary proof tildejjj}), yielding that $E_{sp}^q(R,P,W)|_{q=\log W}$ essentially upper bounds $e_n^q(R,P,W)|_{q=\log W}$. 
Taking $Z=Y$ (instead of minimizing) in our bound (\ref{eq: sp 1}), $E_{sp}^q(R,P,W)|_{q=\log W}$, and noting that $P_{Y|X}\times \indicator_{\{Y=Z\}}\in \calW_q(P_X)$ (since $P_{XZ}=P_{\widetilde{X}Z}$ implies that $P_{XY}=P_{\widetilde{X}Y}$ and thus $\EE q(\widetilde{X},Y)-\EE q(X,Y)=0$) yields $E_{sp}^q(R,P,W)|_{q=\log W}\leq E_{sp}(R,P,W)$.

In \cite{KangarshahiGuilleniFabregas2021spherepackingArXiv}, 
an upper bound on the mismatched reliability function was derived, 
which we denote (to avoid confusion) as
\begin{flalign}
f_{sp}^q(R,P,W)\triangleq 
\min_{\substack{
P_{\widehat{Y}Y|X}:\;I_P(X;\widehat{Y})\leq  R, \\ \min_{\substack{P_{\widetilde{X}|X\widehat{Y}}:\; P_{X\widehat{Y}}=P_{\widetilde{X}\widehat{Y}},\\ \widetilde{X}-(X,\widehat{Y})-Y}} \EE q(\widetilde{X},Y)\geq \EE q(X,Y)}}D(P_{Y|X}\|W_{Y|X}|P_X).\nonumber
\end{flalign}
Note that we have (treating $\widehat{Y}$ in the role of $Z$)
\begin{flalign}
&\min_{\substack{P_{\widetilde{X}|XZ}:\; P_{XZ}=P_{\widetilde{X}Z},\\ \widetilde{X}-(X,Z)-Y}} \EE q(\widetilde{X},Y)\leq 
\min_{\substack{P_{\widetilde{X}|XZ}:\;P_{\widetilde{X}ZX}\in \calP^{sym} (\calX^2\times \calZ)}} \EE_{P_{XZ\widetilde{X}}\times P_{Y|XZ}} 
q(\widetilde{X},Y)\label{eq: aduvhoidhvi}
\end{flalign}
 where the r.h.s.\ is the minimization which 
appears in the definition of $\widetilde{\calW}^{sym}_q(P_X)$ (see (\ref{eq: cytctgyt})) and thus $\widetilde{E}_{sp}^{q,sym}(R,P,W)\leq f_{sp}^q(R,P,W)$.

In addition to being at least as tight, the bound $\widetilde{E}_{sp}^{q,sym}(R,P,W)$ has the advantage that due to the symmetry, solving the associated minimization problem on the r.h.s.\ of (\ref{eq: aduvhoidhvi}) involves fewer degrees of freedom to optimize over and narrower parameters range compared to the l.h.s.\ of (\ref{eq: aduvhoidhvi}). A similar comment holds even more so when $\widetilde{E}_{sp}^{q,sym}(R,P,W)$ is concerned, due to the constraint $\forall (x,z),\; P_{\widetilde{X}|X,Z}(x|x,z)\geq P_{X|Z}(x|z)$ in (\ref{eq: aiufhvuifhd}).

\subsubsection{\underline{Mismatch Capacity}}\label{sc: section comparison Cap}

As mentioned above, \cite{Kangarshahi_GuilleniFabregas2020_IT}, a single-letter bound was derived by forming a transformation
of the channel into another translated channel from $\calX$ to $\calY$ such that $q$-decoding error at the latter implies $q$-decoding error at the original channel. The idea was to connect the two channels by means
of a graph in the output space $\calY^n$.

In \cite{SomekhBaruch2020singleletter_part1}, the multicast transmission proof technique was proposed, which yielded a few tighter bounds. The main idea of the proof is that same message is transmitted simultaneously to two decoders over a two-output (broadcast) channel $W_{YZ|X}$ with two outputs $Y$ and $Z$.  
The $Z$-decoder employs an additive decoding metric $\rho$ that can be optimized, and the two-output channel belongs to the set  
\begin{flalign}\label{eq: Gamma q rho dfn}
&\Gamma(q,\rho)
\triangleq \nonumber\\
&\left\{\begin{array}{l}
P_{YZ|X}: \forall (x,y,z):\rho(x,z)-q(x,y)< \underset{x'\in\calX}{\max}\;[\rho(x',z)-q(x',y)]\\
\quad \quad \quad \quad  P_{YZ|X}(y,z|x) =0 \end{array}\right\}.\end{flalign}
It is easily verified (see details in \cite{SomekhBaruch2020singleletter_part1}), that any 
two-output channel in this class has the property that an error occurs at the $Z$ decoder, only if the $Y$-receiver makes an error. 
Thus, for any codebook $\calC_n$, we have 
\begin{flalign}\label{eq: THIS holds}
P_e(W_{Y|X},\calC_n,q)\geq P_e(W_{Z|X},\calC_n,\rho)
\end{flalign}
 and the following bound holds for any stationary memoryless channel. 
\begin{flalign}
C_q(W)\leq & \max_{P_X} \min_{P_{YZ|X}\in\Gamma(q,\rho):\; P_{Y|X}=W}I(X;Z)\label{eq: KG bound2 BB},
\end{flalign}
It can be shown that the bound of 
\cite{Kangarshahi_GuilleniFabregas2020_IT} can be expressed as a special case for the suboptimal choice of $\rho=q$ and $\calZ=\calY$. 
It is very easy to verify whether channel $P_{YZ|X}$ lies in the set of the minimization since calculating the marginal $P_{Y|X}$ involves a simple summation, and since $\Gamma(q,\rho)$ only dictates constraints on necessary zero values for $P_{YZ|X}$. 
Furthermore, this max-min problem is easily solvable numerically (see \cite[Section IV-C]{SomekhBaruch2020singleletter_part1}).

Another bound that was derived in \cite{SomekhBaruch2020singleletter_part1} and is applicable to DMCs, enlarges the set of channels to one that depends also on the input distribution $P_X$. This bound was obtained by including a genie that helps the $Z$-receiver by providing it with the actual value of the joint empirical distribution of the transmitted codeword and the output sequence $\bZ$. 
In \cite{SomekhBaruch_ISIT2021} the bound was improved to include a minimization over the larger class of channels
\begin{flalign}
&\Theta^*(q,P_X)\triangleq \nonumber\\ 
&\bigg\{P_{YZ|X}:
\min_{\substack{ V_{\widetilde{X}XYZ}:\;V_{XYZ}=P_{XYZ}, \\ V_{\widetilde{X}}= P_X,
V_{\widetilde{X}Z}=P_{XZ}}}
\EE q(\widetilde{X},Y)\geq \EE q(X,Y)
\bigg\}.\nonumber \end{flalign}
These channels satisfy the condition that if the $Y$-decoder successfully decodes the message, then with high probability (approaching $1$ as the block length tends to infinity), also does the genie-aided $Z$-decoder.

A further improved bound was derived in \cite{KangarshahiGuilleniFabregas2021spherepackingArXiv} 
where the set of channels is given by 
\begin{flalign}
&\calM_{max}(q,P_X)\triangleq \nonumber\\ 
&\bigg\{P_{YZ|X}:
\min_{\substack{ V_{\widetilde{X}XYZ}:\;V_{XYZ}=P_{XYZ}, \\ V_{\widetilde{X}}= P_X,
V_{\widetilde{X}Z}=P_{XZ}\\
 \widetilde{X}-(X,Z)-Y}}
\EE q(\widetilde{X},Y)\geq \EE q(X,Y)
\bigg\},\label{eq: Mmaxadhviuf} \end{flalign}
which includes the additional constraint $\widetilde{X}-(X,Z)-Y$. The resulting bound is denoted $\bar{R}(W,q)$. 

Note that 
\begin{flalign}
\left\{V_{\widetilde{X}ZX}\in \calP_{sym}(\calX^2\times \calZ):\; V_{XZ}=V_{\widetilde{X}Z}=P_{XZ} \right\}&\subseteq \left\{V_{\widetilde{X}ZX}:\; V_{XZ}=V_{\widetilde{X}Z}=P_{XZ} \right\}.\label{eq: compar cont}\end{flalign}
Therefore, from (\ref{eq: dfhvdiudddsvhigdfddfiluhliufihi}) 
we obtain
\begin{flalign}
\widetilde{C}^{sym}_q(W)\leq \bar{R}(W,q).\label{Eq: aifhjvoidhfo;}
\end{flalign}
In the next lemma we show that there are many cases for which (\ref{eq: compar cont}) is obtained in strict inclusion (for $|\calX|>2$). 
It is likely that there exist such cases where strict inclusion results in a strict inequality in (\ref{Eq: aifhjvoidhfo;}). 
This is reinforced by the fact that $\overline{C}_q(W)\leq C^{sym}_q(W)\leq \widetilde{C}^{sym}_q(W)$, and since the additional constraint which appears in $C^{sym}_q(W)$; i.e., $P_{X\widetilde{X}}(x,x|z)\geq P^2_{X|Z}(x|z)$ may be active. But, even more importantly, the above mentioned previous bounds except (\ref{eq: KG bound2}) are rather complicated to compute, compared to the bounds $C^{psd}_q(W)$ and $C^{sym}_q(W)$.
\begin{lemma}
For $|\calX|> 2$ the following inclusion is strict:
\begin{flalign}
\left\{V_{\widetilde{X}ZX}\in \calP_{sym}(\calX^2\times \calZ):\; V_{XZ}=V_{\widetilde{X}Z}=P_{XZ} \right\}&\subset \left\{V_{\widetilde{X}ZX}:\; V_{XZ}=V_{\widetilde{X}Z}=P_{XZ} \right\}.\label{eq: compar cont duavgiud}\end{flalign}
\end{lemma}
\begin{proof} 
It is sufficient to establish the result for $|\calX|=3$. 
Consider the following $3\times 3$ symmetric matrix: 
\begin{flalign}
A(z)=\left(\begin{array}{lll} a_1(z) & a_2(z) &a_3(z)\\
a_2(z) & a_4(z) & a_5(z) \\
a_3(z) & a_5(z) & a_6(z)
\end{array}\right),
\end{flalign}
where $a_i(z)\in(0,1)$. 
Next, consider the matrix $B(z)$ which is obtained by subtracting or adding a positive quantity which satisfies the condition  $\Delta\leq\min\{a_2(z),a_3(z),a_5(z),1-a_2(z),1-a_3(z),1-a_5(z)\}$ from $A(z)$'s entries as follows:
\begin{flalign}
B(z)=\left(\begin{array}{lll} a_1(z) & a_2(z)+\Delta &a_3(z)-\Delta\\
a_2(z)-\Delta & a_4(z) & a_5(z)+\Delta \\
a_3(z)+\Delta & a_5(z)-\Delta & a_6(z)
\end{array}\right).,
\end{flalign}
Note that the condition guarantees that $0\leq B_{i,j}(z)\leq 1$. 
A more general choice is also applicable
with the appropriate conditions on $k_1,k_2\neq k_3,k_4$ (which are not all necessarily positive) ensuring that $0\leq K_{i,j}(z)\leq 1$ where:
\begin{flalign}
K(z)=\left(\begin{array}{lll} a_1(z)+k_1 & a_2(z)+k_2 &a_3(z)-k_1-k_2\\
a_2(z)+k_3 & a_4(z)+k_4 & a_5(z)-k_3-k_4 \\
a_3(z)-k_1-k_3 & a_5(z)-k_2-k_4& a_6(z)+k_1+k_2+k_3+k_4
\end{array}\right)..
\end{flalign}
Clearly, the sums of corresponding columns and rows of $A(z)$ and $K(z)$ are identical (e.g. $\sum_j A_{i,j}(z)=\sum_j K_{i,j}(z)$), but while $A(z)$ is symmetric, $K(z)$ is not (since $k_2\neq k_3$). 
Now, let $V_{XZ\widetilde{X}}\in \calP_{sym}(\calX^2\times \calZ)$ which satisfies $V_{XZ}=V_{\widetilde{X}Z}=P_{XZ}$ be given, and think of $A_{i,j}(z)$ as $V_{X\widetilde{X}|Z=z}(i,j|z)$ and define 
$V'_{X\widetilde{X}|Z=z}(i,j|z)=K_{i,j}(z)$ which is clearly non symmetric. Thus $P_Z\times V'_{\widetilde{X}X|Z}$ belongs to the set on the r.h.s.\ of (\ref{eq: compar cont duavgiud}) but does not belong to the set on the l.h.s.\ of (\ref{eq: compar cont duavgiud}), and the lemma follows.
\end{proof}
It is easy to realize that for $|\calX|=2$ the inclusion (\ref{eq: compar cont}) is {\it not} strict; i.e., (\ref{eq: compar cont}) is obtained with equality. 
Combining this with (\ref{eq: relma example2}), we obtain 
\begin{flalign}
\mbox{If $|\calX|=2$, then }\bar{R}(W,q)=\widetilde{C}_q^{sym}(W)={C}_q^{sym}(W)\geq \overline{C}_q(W).
\end{flalign}
Further, unlike our Lemma \ref{lm: aofihv;oidhfvo;idhfoivhdiof}, the optimization problem over $V_{\widetilde{X}XYZ}$ in (\ref{eq: Mmaxadhviuf}) associated with binary input channels was not solved explicitly in \cite{KangarshahiGuilleniFabregas2021spherepackingArXiv}, and therefore our expression (\ref{eq: relma example}) is significantly simpler, and easier to compute.

Consequently, for the binary input ternary output Example 1) (see (\ref{eq: example parameters})) it was established numerically in \cite{KangarshahiGuilleniFabregas2021spherepackingArXiv} that \begin{flalign}
\bar{R}(W,q)\leq 0.4999,\end{flalign} 
while we rigorously established (\ref{eq; numerical calcu}); i.e., 
\begin{flalign}
C^{sym}_q(W)\leq \max_{P_X}I(P_X\times \widetilde{W}_{Z|X})\approx0.4081.\end{flalign} 
Note that it is not evident whether the bound $0.4081$ may be improved by calculating $\overline{C}_q(W)$ rather than $C^{sym}_q(W)$, or by another choice of channel instead of $\widetilde{W}_{YZ|X}$.

\section{Proof of Theorem \ref{th: Main Theorem }}\label{sc: Proof of Exponent Theorem }

In this section we prove Theorem \ref{th: Main Theorem } based on the outline described in Section \ref{sc: GAG setup}.

Let a DMC $W=W_{Y|X}$ be given. 
Fix $P_n\in\calP_n(\calX)$, and let $\calC_n=\{\overline{\bx}_i\}_{i=1}^{\mathbb{M}_n}$ be a $P_n$-constant composition codebook of rate $R$ for the channel $W_{Y|X}$. 
Consider another channel from $\calX\times\calY$ to a finite set $\calZ$ denoted by $W_{Z|XY}$, which along with $W_{Y|X}$ constitutes a two-output channel $W_{YZ|X}$. For technical reasons, we assume that $W_{Z|XY}$ takes on the following form:
\begin{flalign}\label{eq: ahfuv W star}
\forall (x,y,z),\; W_{Z|XY}(z|x,y)&= (1-\overline{\epsilon}_n) \cdot W^*_{Z|XY}(z|x,y)+\overline{\epsilon}_n\cdot \frac{1}{|\calZ|},
\end{flalign}
where $\overline{\epsilon}_n=\frac{1}{n}$, and $W^*_{Z|XY}(z|x,y)$ is some conditional distribution from $\calX\times\calY$ to $\calZ$.
Note that this implies that 
\begin{flalign}\label{eq: ahfuv W starert7eft}
\forall (x,y,z),\; W_{Y|XZ}(y|x,z)&= \frac{W(y|x)W(z|x,y)}{W(z|x)}\geq W_{Y|X}(y|x)\cdot \frac{\overline{\epsilon}_n}{|\calZ|}. 
\end{flalign}
Denote
\begin{flalign}
w_{min}&\triangleq \min_{(x,y):\; W(y|x)>0} W_{Y|X}(y|x),\label{eq: W min dfn 1} \\
t_{n,min}&\triangleq \frac{\overline{\epsilon}_n}{|\calZ|}\cdot w_{min}.\label{eq: W min dfn} 
\end{flalign}
Given the channel input $\bX\in \calT_n(P_n)$, and the joint type-class of $(\bX,\bZ)$, $\widehat{P}_{\bX\bZ}=\widehat{P}_{XZ}$, we clearly have that $\bZ$ is distributed uniformly over $\calT_n(\widehat{P}_{Z|X}|\bX)$; i.e.,
\begin{flalign}\label{eq: mu channel}
\Pr(\bZ=\bz|\bX=\bx,\widehat{P}_{\bX\bZ}=\widehat{P}_{XZ})&=\frac{\indicator_{\{\bz\in \calT_n(\widehat{P}_{Z|X}|\bx)\}}}{|\calT_n(\widehat{P}_{Z|X}|\bx)|}.
\end{flalign}
Recall the definition of $\calL(\bz,\widehat{P}_{\bx\bz})$, and $\bx_i(\bz,\widehat{P}_{\bx\bz})$ in (\ref{eq: List dfn with type}), and note that by definition, $|\calL(\bz,\widehat{P}_{XZ})| \geq 1$ for any $\widehat{P}_{XZ} $ which is a possible joint empirical distribution of a channel input-output sequences pair $\bx,\bz$. 
Further recall the assumption that the $Y$-decoder is informed of the list $\calL(\bZ,\widehat{P}_{XZ})=\{\bx_i(\bZ,\widehat{P}_{XZ})\}$ and employs the decoding rule (\ref{eq: genie aided decoder}). 

It is easily verified that for any possible channel output $\bz$ such that $\widehat{P}_{\bz}=\widehat{P}_Z$, it holds that 
$\{\bx_i(\bz,\widehat{P}_{XZ})\}$ are equiprobable given $\{\bZ=\bz,\widehat{P}_{\bX\bZ}=\widehat{P}_{XZ}\}$; that is, 
\begin{flalign}\label{eq: idfuhvilufg}
 P(\bX=\bx_i(\bz,\widehat{P}_{XZ})|\bZ=\bz,\widehat{P}_{\bX\bZ}=\widehat{P}_{XZ})=\frac{1}{|\calL(\bz,\widehat{P}_{XZ})|}.
\end{flalign}
To see this, note that by applying Bayes' law twice we have 
\begin{flalign}
 &\Pr(\bX=\bx_i(\bz,\widehat{P}_{XZ}),\bZ=\bz|\widehat{P}_{\bX\bZ}=\widehat{P}_{XZ})\nonumber\\
&= \Pr(\bZ=\bz|\widehat{P}_{\bX\bZ}=\widehat{P}_{XZ})\cdot  \Pr(\bX=\bx_i(\bz,\widehat{P}_{XZ})|\bZ=\bz,\widehat{P}_{\bX\bZ}=\widehat{P}_{XZ})\\
&= \Pr(\bX=\bx_i(\bz,\widehat{P}_{XZ})|\widehat{P}_{\bX\bZ}=\widehat{P}_{XZ})\cdot \Pr(\bZ=\bz|\bX=\bx_i(\bz,\widehat{P}_{XZ}),\widehat{P}_{\bX\bZ}=\widehat{P}_{XZ}).
\end{flalign}
Now, since the code is constant composition, the actual joint type-class $\widehat{P}_{\bX\bZ}$ does not depend on the codeword $\bX$, and hence $\Pr(\bX=\bx_i(\bz,\widehat{P}_{XZ})|\widehat{P}_{\bX\bZ}=\widehat{P}_{XZ})=\frac{1}{\mathbb{M}_n}$, and further we have $\Pr(\bZ=\bz|\bX=\bx_i(\bz,\widehat{P}_{XZ}),\widehat{P}_{\bX\bZ}=\widehat{P}_{XZ})=\frac{1}{|\calT_n(\widehat{P}_{Z|X})|}$, this yields
\begin{flalign}
  \Pr(\bX=\bx_i(\bz,\widehat{P}_{XZ})|\bZ=\bz,\widehat{P}_{\bX\bZ}=\widehat{P}_{XZ})&=\frac{1}{|\calT_n(\widehat{P}_{Z|X})|\cdot \mathbb{M}_n\cdot \Pr(\bZ=\bz|\widehat{P}_{\bX\bZ}=\widehat{P}_{XZ})}.
\end{flalign}
Since the r.h.s.\ does not depend on $i$ we obtain the desired result (\ref{eq: idfuhvilufg}).

Next, let 
\begin{flalign}
\calE_{ij}\triangleq \calE_{ij}(\bz,\widehat{P}_{XZ})\triangleq \{\by:\; q(\bx_j(\bz,\widehat{P}_{XZ}),\by)\geq q(\bx_i(\bz,\widehat{P}_{XZ}),\by)\},
\end{flalign}
and adopt the shorthand notation
\begin{flalign}
\bx_i\triangleq \bx_i(\bz,\widehat{P}_{XZ}),\; \calL\triangleq \calL(\bz,\widehat{P}_{XZ}). 
\end{flalign}
Since $\Pr(error|\bx_i,\bz)= \Pr(\cup_{j\neq i}\calE_{ij}|\bx_i,\bz)$, 
we have the following lower bound on the average error probability in $q$-mismatched decoding at the $Y$-receiver given that $\bZ=\bz$ and $\widehat{P}_{\bX\bZ}=\widehat{P}_{XZ}$, 
\begin{flalign}\label{eq: STAM}
\Pr(error|\bz,\widehat{P}_{\bX\bZ}=\widehat{P}_{XZ})&\geq\left\{\begin{array}{ll} \frac{1}{|\calL|(|\calL|-1)}\sum_{i,j\in\calL,\; j\neq i}
\Pr(\calE_{ij}|\bx_i,\bz)&|\calL|\geq 2\\
0&|\calL|=1\end{array}\right. .
\end{flalign}
Evidently, we shall focus on the case where $(\bz,\widehat{P}_{XZ})$ are such that $|\calL|\geq 2$ to obtain a meaningful lower bound on the error probability. 
Note that
\begin{flalign}
\Pr(\calE_{ij}|\bx_i,\bz)=& \sum_{\by:\; q(\bx_j,\by)\geq q(\bx_i,\by)}W_{Y|XZ}^n(\by|\bx_i,\bz)\\
&\geq \frac{1}{(n+1)^{|\calX|^2|\calZ||\calY|}}\cdot 
\sum_{
\substack{V_{\widetilde{X}XZY}\in\calP_n(\calX^2\times\calZ\times\calY):\; \\
V_{\widetilde{X}ZX}=\widehat{P}_{\bx_i \bz \bx_j},\\ q(V_{\widetilde{X}Y})\geq q(V_{XY})}}e^{-n\cdot D(V_{Y|XZ\widetilde{X}}\|W_{Y|XZ}|\widehat{P}_{\bx_i \bz \bx_j})
}\\
&\geq \frac{1}{(n+1)^{|\calX|^2|\calZ||\calY|}}\cdot e^{-n\Omega_n(\widehat{P}_{\bx_i\bz\bx_j}, W_{Y|XZ})}.\label{eq: afhjviodfhvaifudhi}
\end{flalign}
where for an empirical distribution $P_{\widetilde{X}ZX}\in\calP_n(\calX^2\times\calZ)$ satisfying $P_{\widetilde{X}Z}=P_{XZ}$ we define
\begin{flalign}
&\Omega_n(P_{XZ\widetilde{X}}, W_{Y|XZ})\triangleq 
\min_{\substack{V_{\widetilde{X}XYZ}\in\calP_n(\calX^2\times\calZ\times\calY)\cap \calS(P_{XZ\widetilde{X}})}} D(V_{Y|XZ\widetilde{X}}\|W_{Y|XZ}|P_{XZ\widetilde{X}})\label{eq: Ialpha dfndfugilugi}\\
&\calS(P_{XZ\widetilde{X}})\triangleq \left\{V_{\widetilde{X}XYZ}\in\calP(\calX^2\times\calZ\times\calY):
V_{XZ\widetilde{X}}=P_{XZ\widetilde{X}},\; 
q(V_{\widetilde{X}Y})\geq q(V_{XY}) \right\}\label{eq: calSdfn}.
\end{flalign}

Consider the function $\Omega(P_{XZ\widetilde{X}}, W_{Y|XZ})$ which extends $\Omega_n(P_{XZ\widetilde{X}}, W_{Y|XZ})$ in a twofold manner: (a) it is defined for $P_{\widetilde{X}ZX}\in \calP(\calX^2\times \calZ)$ which need not necessarily be an empirical distribution of order $n$, and (b) the minimization is over the simplex $\calP(\calX^2\times\calZ\times\calY)$ rather than empirical distributions; that is, 
\begin{flalign}
\Omega(P_{XZ\widetilde{X}}, W_{Y|XZ})
&\triangleq 
\min_{V_{\widetilde{X}XYZ}\in\calS(P_{XZ\widetilde{X}})
 } D(V_{Y|XZ\widetilde{X}}\|W_{Y|XZ}|P_{XZ\widetilde{X}}) 
 .\label{eq: Ialpha dfndfugilugi11}
\end{flalign}
We present the following approximation lemma whose proof appears in Appendix \ref{ap: approx 1 elmm}.
\begin{lemma}\label{lm: aiuvdgiudg}
For all $P_{XZ\widetilde{X}}\in  \calP_n(\calX^2\times\calZ)$, 
\begin{flalign}\label{eq: asdcoi}
\Omega_n(P_{XZ\widetilde{X}}, W_{Y|XZ})&\leq \Omega(P_{XZ\widetilde{X}}, W_{Y|XZ})+\delta_n, 
\end{flalign}
where $\delta_n= 2\cdot \frac{|\calX|^2|\calZ||\calY|}{n}\log \frac{n^2|\calZ|}{w_{min}}$,  
with $w_{min}$ defined in (\ref{eq: W min dfn}). 
\end{lemma}
Thus, 
\begin{flalign}
\Pr(\calE_{ij}|\bx_i,\bz)&\geq
 \frac{1}{(n+1)^{|\calX|^2|\calZ||\calY|}}\cdot e^{-n[\Omega(\widehat{P}_{\bx_i\bz\bx_j}, W_{Y|XZ})+\delta_n]}.\label{eq: afhjviodfhvaifudhi111}
\end{flalign}
Further, 
\begin{flalign}
&\sum_{\bx_i,\bx_j\in\calL,\; j\neq i}e^{-n\Omega(\widehat{P}_{\bx_i\bz\bx_j}, W_{Y|XZ})}\nonumber
\\
&=
\sum_{\bx_i,\bx_j\in\calL,\; j\neq i}\frac{1}{2}\cdot 
\left[e^{-n\Omega(\widehat{P}_{\bx_i\bz\bx_j}, W_{Y|XZ})}+e^{-n\Omega(\widehat{P}_{\bx_j\bz\bx_i}, W_{Y|XZ})}\right]\label{eq: a;uf}\\
&\geq \sum_{\bx_i,\bx_j\in\calL,\; j\neq i}\frac{1}{2}\cdot 
e^{-n\min\left\{\Omega(\widehat{P}_{\bx_i\bz\bx_j}, W_{Y|XZ}),\Omega(\widehat{P}_{\bx_j\bz\bx_i}, W_{Y|XZ})\right\}},\label{eq: afviudfh}
\end{flalign}
where (\ref{eq: a;uf}) follows by switching the roles of the summation indices and multiplying and dividing by $2$, (\ref{eq: afviudfh}) follows since for positive $A,B$, ,we have, $A+B\geq\max\{A,B\}$, and since $f(t)=e^{-t}$ is a monotonically decreasing function. 

Now, in analogy to the definition of $\calS(P_{\widetilde{X}ZX})$ in (\ref{eq: calSdfn}), define the following set of conditional distributions rather than joint distributions
\begin{flalign}
\calS^{cond}(P_{XZ\widetilde{X}})&\triangleq \left\{V_{Y|XZ\widetilde{X}}:\; 
P_{XZ\widetilde{X}}\times V_{Y|XZ\widetilde{X}}\in \calS(P_{XZ\widetilde{X}})
\right\}.\label{eq: calS cond dfn}
\end{flalign}

Next, observe that since $\Omega(\widehat{P}_{XZ\widetilde{X}}, W_{Y|XZ})=\min_{V_{Y|XZ\widetilde{X}}\in\calS^{cond}(\widehat{P}_{XZ\widetilde{X}}) } D(V_{Y|XZ\widetilde{X}}\|W_{Y|XZ}|\widehat{P}_{XZ\widetilde{X}})$, 
\begin{flalign}
&\min\left\{\Omega(\widehat{P}_{\bx_i\bz\bx_j}, W_{Y|XZ}),\Omega(\widehat{P}_{\bx_j\bz\bx_i}, W_{Y|XZ})\right\}\nonumber\\
&\leq \min\left\{\substack{\min_{\substack{V_{Y|XZ\widetilde{X}}\in\calS^{cond}(\widehat{P}_{\bx_i\bz\bx_j})  :\\ I_V(\widetilde{X};Y|X,Z)=0}} D(V_{Y|XZ}\|W_{Y|XZ}|\widehat{P}_{XZ})
,\\\min_{\substack{V_{Y|XZ\widetilde{X}}\in\calS^{cond}(\widehat{P}_{\bx_j\bz\bx_i}):\\ I_V(\widetilde{X};Y|X,Z)=0 }} D(V_{Y|XZ}\|W_{Y|XZ}|\widehat{P}_{XZ})}\right\}\label{eq: ;fiuhbvifdu}
\\
&=\min_{\substack{V_{Y|XZ\widetilde{X}}\in\calS^{cond}(\widehat{P}_{\bx_i\bz\bx_j})\cup  \calS^{cond}(\widehat{P}_{\bx_j\bz\bx_i})
:\\ I_V(\widetilde{X};Y|X,Z)=0
}} D(V_{Y|XZ}\|W_{Y|XZ}|\widehat{P}_{XZ})\label{eq: a;ofihvoifhv}
\\
&\leq\min_{\substack{V_{Y|XZ\widetilde{X}}\in\calS^{cond}(\frac{1}{2}[\widehat{P}_{\bx_i\bz\bx_j}+\widehat{P}_{\bx_j\bz\bx_i}]):\\ I_V(\widetilde{X};Y|X,Z)=0
}} D(V_{Y|XZ}\|W_{Y|XZ}|\widehat{P}_{XZ})\label{eq: afiuvufgv},
\end{flalign}

where (\ref{eq: a;ofihvoifhv}) holds since $D(V_{Y|XZ\widetilde{X}}\|W_{Y|XZ}|\widehat{P}_{\bx_j\bz\bx_i})=D(V_{Y|XZ}\|W_{Y|XZ}|\widehat{P}_{\bx_i\bz})+I_V(\widetilde{X};Y|X,Z)$, and since $\forall i$, $\widehat{P}_{\bx_i\bz}=\widehat{P}_{XZ}$, (\ref{eq: a;ofihvoifhv}) holds since $\min\{\min_{t\in\calS_1}f(t),\min_{t\in\calS_2}f(t)\}=\min_{t\in\calS_1\cup \calS_2}f(t)$, (\ref{eq: afiuvufgv}) follows since $\calS^{cond}(\frac{1}{2}[\widehat{P}_{\bx_i\bz\bx_j}+\widehat{P}_{\bx_j\bz\bx_i}])\subseteq \calS^{cond}(\widehat{P}_{\bx_i \bz \bx_j})\cup  \calS^{cond}(\widehat{P}_{\bx_j \bz \bx_i})$ 
which follows from the linearity of the expectation; that is,
\begin{flalign}
&\calS^{cond}(\widehat{P}_{\bx_i \bz \bx_j})\cup  \calS^{cond}(\widehat{P}_{\bx_j \bz \bx_i})\triangleq \nonumber\\
&=\left\{V_{Y|XZ\widetilde{X}}:
\EE_{\widehat{P}_{\bx_i \bz \bx_j}\times V_{Y|XZ\widetilde{X}}}[q(\widetilde{X},Y)-q(X,Y)]\geq 0\mbox{ or }
\EE_{\widehat{P}_{\bx_j \bz \bx_i}\times V_{Y|XZ\widetilde{X}}}[q(\widetilde{X},Y)-q(X,Y)]\geq 0
\right\}\\
&\supseteq\left\{V_{Y|XZ\widetilde{X}}:
\EE_{\frac{1}{2}[\widehat{P}_{\bx_i \bz \bx_j}+\widehat{P}_{\bx_j \bz \bx_i}]\times V_{Y|XZ\widetilde{X}}}[q(\widetilde{X},Y)-q(X,Y)]\geq 0\right\}
\label{eq: calSdfnzd v;vknc kn}.
\end{flalign}

Gathering (\ref{eq: STAM}), (\ref{eq: afhjviodfhvaifudhi111}), (\ref{eq: afviudfh}), (\ref{eq: afiuvufgv}), and denoting 
$k_n\triangleq \delta_n+\frac{|\calX|^2|\calZ||\calY|}{n} \log (n+1)+\frac{\ln(2)}{n} $
 we obtain that whenever $|\calL(\bz,\widehat{P}_{XZ})|\geq 2$,
\begin{flalign}
&\Pr(error|\bz,\widehat{P}_{\bX\bZ}=\widehat{P}_{XZ})\nonumber\\
&\geq \min_{\bx_i\in\calL ,\bx_j\in\calL,j\neq i}\exp\bigg\{-n \bigg[\min_{\substack{V_{\widetilde{X}XYZ}\in\calS(\frac{1}{2}[\widehat{P}_{\bx_i \bz \bx_j}+\widehat{P}_{\bx_j \bz \bx_i}]): \\I_V(\widetilde{X};Y|XZ)=0}
} D(V_{Y|XZ}\|W_{Y|XZ}|\widehat{P}_{XZ})+k_n\bigg]\bigg\}.\label{eq: advhof;idhvoidfh}
\end{flalign}
Since $\frac{1}{2}[\widehat{P}_{\bx_i \bz \bx_j}+\widehat{P}_{\bx_j \bz \bx_i}]\in \calP_{sym}(\calX^2\times \calZ)$ (see (\ref{eq: sym dfn set})), it straightforwardly follows that 
\begin{flalign}
&\min_{\substack{V_{\widetilde{X}XYZ}\in\calS(\frac{1}{2}[\widehat{P}_{\bx_i \bz \bx_j}+\widehat{P}_{\bx_j \bz \bx_i}]):\\ I_V(\widetilde{X};Y|XZ)=0
}} D(V_{Y|XZ}\|W_{Y|XZ}|\widehat{P}_{XZ})\nonumber
\\
&\leq \min_{V_{XYZ}:\; V_{YZ|X}\in
\widetilde{\calW}^{sym}_q(\widehat{P}_X):\; V_{XZ}=\widehat{P}_{XZ}
}  D(V_{Y|XZ}\|W_{Y|XZ}|\widehat{P}_{XZ}),\label{eq: advhof;idhvoisfbdfh}
\end{flalign}
where $\widetilde{\calW}^{sym}_q(\widehat{P}_X)$ is defined in (\ref{eq: cytctgyt}). 

We next establish a stronger result with $\calW_q(\widehat{P}_X)$ in lieu of $\widetilde{\calW}^{sym}_q(\widehat{P}_X)$, but note that the proof can be shortened if we wish to prove (\ref{eq: dfhvdiudsvhidfiludffdhliufihi}) with $E_{sp}^{q}$ replaced by $\widetilde{E}_{sp}^{q,sym}$. 
\begin{claim}\label{cl: calWs stronger claim}
For any $\bx_i,\bx_j\in\calL$
\begin{flalign}
&\min_{\substack{V_{\widetilde{X}XYZ}\in\calS(\frac{1}{2}[\widehat{P}_{\bx_i \bz \bx_j}+\widehat{P}_{\bx_j \bz \bx_i}]):\\ I_V(\widetilde{X};Y|XZ)=0
}} D(V_{Y|XZ}\|W_{Y|XZ}|\widehat{P}_{XZ})\nonumber
\\
&\leq \min_{V_{XYZ}:\; V_{YZ|X}\in
\calW_q(\widehat{P}_X):\; V_{XZ}=\widehat{P}_{XZ}
} D(V_{Y|XZ}\|W_{Y|XZ}|\widehat{P}_{XZ}),\label{Eq: a;ofhv;oifv}
\end{flalign}
\end{claim}
Claim \ref{cl: calWs stronger claim} is proved in Appendix \ref{sc: calWs stronger claim}.

Hence, from (\ref{eq: advhof;idhvoidfh}) and (\ref{Eq: a;ofhv;oifv}), we obtain that whenever $|\calL(\bz,\widehat{P}_{XZ})|\geq 2$,
\begin{flalign}
\Pr(error|\bz,\widehat{P}_{\bX\bZ}=\widehat{P}_{XZ})&\geq 
 e^{-n\left[\min_{V_{YZ|X}\in
\calW_q(\widehat{P}_X):\; V_{XZ}=\widehat{P}_{XZ}
} D(V_{Y|XZ}\|W_{Y|XZ}|\widehat{P}_{XZ})+k_n\right]}\label{eq: afiud;hvv}
\end{flalign}

We next present a lemma which enables to assess the size of the list $\calL(\bZ, \widehat{P}_{XZ})$ defined in (\ref{eq: List dfn with type}). 
\begin{lemma}\label{eq: List size lemma}
Let a codebook $\calC_n=\{\overline{\bx}_i\}_{i=1}^{\mathbb{M}_n}$ be given, let $\bX$ denote the random codeword (distributed uniforly over $\calC_n$), and let $\bZ$ denote the output of the channel $W_{Z|X}^n$ when fed by $\bX$. 
For any $\tau\geq 0$, 
\begin{flalign}\label{eq: sdhvodishfiov}
&\Pr\left(
 |\calL(\bZ, \widehat{P}_{XZ})|\geq e^{n\tau}\big|\widehat{P}_{\bX\bZ}=\widehat{P}_{XZ}\right)\geq 1-(n+1)^{|\calX||\calZ|-1}\cdot e^{-n[R-I(\widehat{P}_{XZ})-\tau]}.
\end{flalign}
\end{lemma}
Lemma \ref{eq: List size lemma} is proved in Appendix \ref{sc: List size lemma}.

Note that Lemma \ref{eq: List size lemma}, implies that 
for any $\widehat{P}_{XZ}$, $\widetilde{\epsilon}_n>0$, and $\epsilon_n>0$, such that 
\begin{flalign}\label{eq: ilsufgviludf}
R\geq I(\widehat{P}_{XZ})+\epsilon_n+\frac{|\calX||\calZ|-1}{n}\log(n+1)+\widetilde{\epsilon}_n,
\end{flalign}
it holds that
\begin{flalign}
\Pr\left(|\calL(\bZ,\widehat{P}_{XZ})|<e^{n\epsilon_n}\big|\widehat{P}_{\bX\bZ}=\widehat{P}_{XZ}\right)\leq  e^{-n\widetilde{\epsilon}_n} .
\end{flalign}
Consequently, if $\widehat{P}_{XZ}$ is a possible joint empirical distribution of a codeword and a channel output $\bZ$ such that (\ref{eq: ilsufgviludf}) holds, we have for $\widetilde{\epsilon}_n>1/n$ that
\begin{flalign}
\Pr(error|\widehat{P}_{XZ})&\geq \Pr(error, \calL(\bZ,\widehat{P}_{XZ})\geq e^{n\epsilon_n}|\widehat{P}_{\bX\bZ}=\widehat{P}_{XZ})\label{eq: aiufviufgv}\\
&\geq \left(1-e^{-n\widetilde{\epsilon}_n}\right)\cdot 
\Pr(error|\widehat{P}_{\bX\bZ}=\widehat{P}_{XZ},\calL(\bZ,\widehat{P}_{XZ})\geq e^{n\epsilon_n})\\
&\geq \left(1-e^{-n\widetilde{\epsilon}_n}\right)\cdot 
\min_{\bz\in\calT_n(\widehat{P}_Z):\; |\calL|
\geq e^{n\epsilon_n}}\Pr(error|\widehat{P}_{\bX\bZ}=\widehat{P}_{XZ},\bZ=\bz)\label{eq: udfhv0}\\
&\geq \left(1-e^{-n\widetilde{\epsilon}_n}\right)\cdot e^{-n[\min_{V_{YZ|X}\in
\calW_q(\widehat{P}_X):\;V_{XZ} =\widehat{P}_{XZ}
} D(V_{Y|XZ}\|W_{Y|XZ}|\widehat{P}_{XZ})+k_n]}
,\label{eq: udfhv}
\end{flalign}
where the last step follows from (\ref{eq: afiud;hvv}).

Now, let 
\begin{flalign}
\Psi\left(P_{XZ},  W_{Y|XZ}\right)&\triangleq 
\min_{V_{XYZ}\in
\calW_q(\widehat{P}_X):\;V_{XZ} =P_{XZ}
} D(V_{Y|XZ}\|W_{Y|XZ}|P_{XZ}).\label{eq: aofhvdufh}
\end{flalign}
Since (\ref{eq: udfhv}) holds for every $ \widehat{P}_{XZ}\in\calP_n(\calX\times \calZ)$ such that $\widehat{P}_{X}=P_n$, which is a possible empirical distribution of $\bX,\bZ$, and which satisfies (\ref{eq: ilsufgviludf}), 
and since 
\begin{flalign}
\Pr(error)&=\sum_{\widehat{P}_{XZ}\in\calP_n(\calX\times\calZ):\; \widehat{P}_X=P_n}\Pr(\widehat{P}_{\bX\bZ}=\widehat{P}_{XZ})\cdot \Pr(error|\widehat{P}_{\bX\bZ}=\widehat{P}_{XZ})\nonumber\\
&\geq\sum_{\substack{
\widehat{P}_{XZ}\in\calP_n(\calX\times \calZ):\; \widehat{P}_{X}=P_n,\\
I(\widehat{P}_{XZ})\leq R-\epsilon_n''}} \frac{1}{(n+1)^{|\calX||\calZ|}}e^{-nD(\widehat{P}_{Z|X}\|W_{Z|X}|\widehat{P}_X)}\cdot \Pr(error|\widehat{P}_{\bX\bZ}=\widehat{P}_{XZ}) ,\label{eq: aiuvgiufgv}
\end{flalign} 
denoting $\delta_n''=k_n-\frac{1}{n}\log (1-e^{-n\widetilde{\epsilon}_n}) +\frac{|\calX||\calZ|}{n}\log(n+1)$, we get
\begin{flalign}\label{eq: idfugvsliudfg}
&-\frac{1}{n}\log \Pr(error)\leq \nonumber\\
&\min_{\substack{
\widehat{P}_{XZ}\in\calP_n(\calX\times \calZ):\; \widehat{P}_{X}=P_n,\\
I(\widehat{P}_{XZ})\leq R-\epsilon_n''}}D(\widehat{P}_{Z|X}\|W_{Z|X}|\widehat{P}_X)+ \Psi\left(\widehat{P}_{XZ},  W_{Y|XZ}\right)+\delta''_n,
\end{flalign}
where $\epsilon_n''\triangleq \epsilon_n+\frac{|\calX||\calZ|-1}{n}\log(n+1)+\widetilde{\epsilon}_n$.
The following lemma shows that the minimization over empirical conditional distributions $\calP_n(\calX\times \calZ)$ can be approximated by minimization over the simplex $\calP(\calX\times \calZ)$. 
\begin{lemma}\label{lm: APPPROX five}
For any $P_n\in\calP_n(\calX)$,  and any $P_{YZ|X}\in\calW_q(P_n)$, there exists $\widehat{P}_{XZ}\in\calP_n(\calX\times\calZ)$ such that $\widehat{P}_X=P_n$, $\widehat{P}_{Z|X}\times P_{Y|XZ}\in\calW_q(P_n)$,
\begin{flalign}
&\|P_n\times P_{Z|X}-\widehat{P}_{XZ}\|\leq\frac{|\calX||\calZ|}{n},\; \mbox{ and }
P_{Z|X}(z|x)=0\Rightarrow \widehat{P}_{Z|X}(z|x)=0
.\label{ADDeq: auadfhiov;hfdoihv;oifdh;iv}
\end{flalign} \end{lemma}
Lemma \ref{lm: APPPROX five} is proved in Appendix \ref{ap: Proof of Lemma app}.

Next, let $P_{XYZ}^*$ be the minimizer of (\ref{eq: aofhvdufh}). 
Note that $\|P_{XZ}^*-\widehat{P}_{XZ}\|\leq\frac{|\calX||\calZ|}{n}$ implies 
 $|D(P_{YZ|X}^*\|W_{YZ|X}|P_X)- D(\widehat{P}_{Z|X}\times P_{Y|XZ}^*\|W_{YZ|X}|P_X)|\leq 2\frac{|\calX||\calY||\calZ|}{n}\log n+\frac{|\calX||\calY||\calZ|}{n}\log\frac{n|\calZ|}{w_{min}}\triangleq \overline{\delta}_n$, where $w_{min}$ is defined in 
 (\ref{eq: W min dfn})
 and further, $|I(P_{XZ}^*)-I(P_n\times\widehat{P}_{Z|X})|\leq 2 \frac{|\calX||\calZ|}{n}\log n\triangleq \epsilon_{1,n}$.

This implies that for any $P_n\in\calP_n(\calX)$, 
\begin{flalign}
&\min_{\substack{
\widehat{P}_{XZ}\in\calP_n(\calX\times \calZ):\; \widehat{P}_{X}=P_n,\\
I(\widehat{P}_{XZ})\leq R}}D(\widehat{P}_{Z|X}\|W_{Z|X}|\widehat{P}_X)+ \Psi\left(\widehat{P}_{XZ},  W_{Y|XZ}\right)\nonumber\\
& \leq \min_{\substack{
P_{XZ}\in\calP(\calX\times \calZ):\; P_{X}=P_n,\\
I(P_{XZ})\leq R-\epsilon_{1,n}}}D(P_{Z|X}\|W_{Z|X}|\widehat{P}_X)+ \Psi\left(P_{XZ}, W_{Y|XZ}\right)+\overline{\delta}_n.
\end{flalign}

Thus, we obtained
\begin{flalign}
-\frac{1}{n}\log \Pr(error)&\leq \min_{
P_{YZ|X}\in
\calW_q(P_n):\; 
I(P_n\times P_{Z|X})\leq R-\epsilon_{1,n}-\epsilon_n''}D(P_{YZ|X}\|W_{YZ|X}|P_n)+\overline{\delta}_n+\delta''_n.
\end{flalign}
Now, since $W_{Z|XY}$ can be optimized, with the exception that (\ref{eq: ahfuv W star}) 
must hold, we can choose in (\ref{eq: ahfuv W star}) $W^*_{Z|XY}=P_{Z|XY}$, yielding $\|W_{Z|XY}-P_{Z|XY}\|\leq \frac{\overline{\epsilon}_n}{|\calZ|}$. Since we also have $P_{Z|XY}\ll W_{Z|XY}$, from Pinsker's inverse inequality (see e.g.\ \cite[Lemma 4.1]{GotzeSabaleSinulis2019}), we get $D(P_{Z|XY}\|W_{Z|XY}|P_n\times P_{Z|X})\leq
2\frac{\overline{\epsilon}_n}{|\calZ|}$, and consequently
\begin{flalign}
-\frac{1}{n}\log \Pr(error)&\leq\min_{
P_{YZ|X}\in
\calW_q(P_n):\; 
I(P_n\times P_{Z|X})\leq R-\epsilon_{1,n}-\epsilon_n''}D(P_{Y|X}\|W_{Y|X}|P_n)+\overline{\delta}_n+\delta''_n+2\frac{\overline{\epsilon}_n}{|\calZ|}.
\end{flalign}
Now, taking for example $\epsilon_n=\widetilde{\epsilon}_n=\frac{\log n}{n}$, $\overline{\epsilon}_n=\frac{1}{n}$, and recalling that $t_{n,min}=w_{min}\frac{1}{n|\calZ|}$ we obtain 
\begin{flalign}
-\frac{1}{n}\log \Pr(error)&\leq\min_{
P_{YZ|X}\in
\calW_q(P_n):\; 
I(P_n\times P_{Z|X})\leq R-\epsilon_{n,a}}D(P_{Y|X}\|W_{Y|X}|P_n)+\epsilon_{n,b},
\end{flalign}
where 
$\epsilon_{n,b}=2\frac{|\calX||\calY||\calZ|}{n}\log n+\frac{|\calX||\calY||\calZ|}{n}\log\frac{n|\calZ|}{w_{min}}
+ 2\cdot \frac{|\calX|^2|\calZ||\calY|}{n}\log \frac{n^2|\calZ|}{w_{min}}
+\frac{|\calX||\calZ|[|\calX||\calY|+1]}{n} \log (n+1)+\frac{\ln(2)}{n} 
-\frac{1}{n}\log (1-\frac{1}{n}) 
+2\frac{1}{n|\calZ|}$ 
and $
\epsilon_{n,a}= 2 \frac{(|\calX||\calZ|+1)}{n}\log n+
\frac{|\calX||\calZ|-1}{n}\log(n+1)$.

\section{Proof of Theorem \ref{th: thorem mismatch capacity}}\label{sc: Proof of Theorem Capacity}

Let a DMC $W=W_{Y|X}$ be given. 
As in the proof of Theorem \ref{th: Main Theorem }, 
consider another channel from $\calX\times\calY$ to a finite set $\calZ$ denoted by $W_{Z|XY}$, which along with $W_{Y|X}$ constitutes a two-output channel $W_{YZ|X}$. 
Fix $P_n\in\calP_n(\calX)$, and let $\calC_n=\{\overline{\bx}_i\}_{i=1}^{\mathbb{M}_n}$ be a $P_n$-constant composition codebook of rate $R$ for the channel $W_{Y|X}$. Consider the genie-aided-genie setup depicted in Fig.\ \ref{Figure_BC_genie}.

First, we treat the case of $q=-\infty$: note that if there exists a pair of symbols $(x_0,y_0)$ such that $q(x_0,y_0)=-\infty$ and $W(y_0|x_0)>0$, then $x_0$ must be absent from $\calC_n$ for all $n$ sufficiently large, as otherwise the maximal error probability would not vanish as $n$ tends to infinity. Hence assume w.l.o.g. 
\begin{flalign}\label{eq infinite q case }
\forall n,\; P_n(x)=0 \mbox{ if }\sum_{y:\; q(x,y)=-\infty}W(y|x)>0.
\end{flalign}
Recall the definition of the list $\calL(\bz, \widehat{P}_{XZ})$ (see (\ref{eq: List dfn with type})). 
We use the shorthand notation $\calL=\calL(\bz, \widehat{P}_{XZ})$, and $\bx_i=\bx_i(\bz,\widehat{P}_{XZ})$. 
Further recall the notation of 
$\Delta^q(P_{XZU},P_{Y|XZ})$ and $\calW_q(P_X)$ defined in (\ref{eq: Delta q definition a}), (\ref{eq: calW q dfnasdfionidaf;i;}), respectively, (see also (\ref{eq: aufdgvui})), and let
\begin{flalign}
&\calA_q(W_{Y|XZ},P_n)\triangleq\nonumber\\
& \left\{P_{XZ}:\; P_X=P_n,\;\min_{P_{U|XZ}} 
\Delta^q(P_{XZU},W_{Y|XZ})\geq 0\right\}\label{eq: calA dfn dnf}\\
&=\left\{P_{XZ}:\; P_X=P_n,\;P_{Z|X}\times W_{Y|XZ}\in\calW_q(P_X) \right\}.\label{eq: calA dfn dnfsdcihjsdih}
\end{flalign}
For any collection of codewords $\calL'\subseteq\calL$ such that $|\calL'|\geq 2$, let 
\begin{flalign}
P^{avg, \calL'}_{XZ\widetilde{X}}(x,z,\widetilde{x})&\triangleq \frac{1}{|\calL'|(|\calL'|-1)}\sum_{i,j\in\calL':\;  j\neq i}\widehat{P}_{\bx_j \bz \bx_i}(x,z,\widetilde{x}). \label{eq: Pavg dfnsuhfv;oiudhf;oho;idhvf}
\end{flalign}
\begin{lemma}\label{lm: aifuvuidfv}
If $\widehat{P}_{XZ}\in\calA_q(W_{Y|XZ},P_n)\cap \calP_n(\calX\times\calZ)$, then for any subset $\calL'\subseteq\calL(\bz,\widehat{P}_{XZ})$ such that $|\calL'|\geq 2$, 
\begin{flalign}\label{eq: iudshgviudg1}
 \sum_{x,\widetilde{x},z,y} P^{avg, \calL'}_{XZ\widetilde{X}}(x,z,\widetilde{x}), 
 \cdot W(y|x,z)\left[q(\widetilde{x},y)-q(x,y)\right]\geq 0,
 \end{flalign}
 and consequently, for any pair of codewords $\bx_\ell,\bx_k$ in $\calL$, either
 \begin{flalign}\label{eq: apd'ohvpodfjpvokq}
 \sum_{x,\widetilde{x},z,y} \widehat{P}_{\bx_\ell\bz\bx_k}(\widetilde{x},z,x) 
 \cdot W(y|x,z)\left[q(\widetilde{x},y)-q(x,y)\right]\geq 0,
 \end{flalign}
 or
 \begin{flalign}\label{eq: apd'ohvpodfjpvok}
 \sum_{x,\widetilde{x},z,y} \widehat{P}_{\bx_k\bz\bx_\ell}(x,z,\widetilde{x}) 
 \cdot W(y|x,z)\left[q(\widetilde{x},y)-q(x,y)\right]\geq 0.
 \end{flalign}
\end{lemma}
\begin{proof}
Recall the definition of  
$\overline{P}_{XZ\widetilde{X}}^{\calL'}$ in 
(\ref{eq: iusdgicugd}), hence similar to (\ref{eq: iudsdvdssdjv'osidjvvsdshgviudg}),
 \begin{flalign}
&\frac{1}{|\calL'|(|\calL'|-1)}\sum_{i,j\in\calL':\;  j\neq i}\;
\sum_{x,\widetilde{x},z,y} \widehat{P}_{\bx_i \bz \bx_j}(x,z,\widetilde{x}) \cdot W(y|x,z)\left[q(\widetilde{x},y)-q(x,y)\right]\label{eq: aliufdsdfsddf}\\
&=\sum_{x,\widetilde{x},z,y} P^{avg, \calL'}_{XZ\widetilde{X}}(x,z,\widetilde{x})\cdot W(y|x,z)\left[q(\widetilde{x},y)-q(x,y)\right]\label{eq: aliufdgviludf}\\
  &=\frac{|\calL'|}{|\calL'|-1}\sum_{x,\widetilde{x},z,y} \overline{P}_{XZ\widetilde{X}}(x,z,\widetilde{x})
 \cdot W(y|x,z)\left[q(\widetilde{x},y)-q(x,y)\right]\label{eq: ai;udvgh;iudhvoidhio}\\
   &\geq
   \min_{\substack{P_{UXZ}:\\P_{XZ}=\widehat{P}_{XZ}} }\frac{|\calL'|}{|\calL'|-1}\sum_{u,x,\widetilde{x},z,y} 
   P_{UZ}(u,z)P_{X|UZ}(x|u,z)P_{X|UZ}(\widetilde{x}|u,z) \cdot W(y|x,z)\left[q(\widetilde{x},y)-q(x,y)\right]\label{eq: ifuav}\\
   &\geq 0,\label{eq: iudsdvdsvsdshgviudg}
\end{flalign}
where we abbreviate $W=W_{Y|XZ}$, (\ref{eq: ai;udvgh;iudhvoidhio})
follows (\ref{eq: iudfh}), (\ref{eq: ifuav}) follows by definition of $\overline{P}_{XZ\widetilde{X}}$ (see (\ref{eq: iusdgicugd})), 
and the last step is by definition of $\calA_q(P_n,W_{Y|XZ})$.
\end{proof}

The next lemma 
shows that for at least half of the pairs of distinct codewords, the pairwise error probability is bounded away from zero. 
It is based on \cite[Lemma 3]{CsiszarNarayan95}, which was used by Csisz\'{a}r and Narayan to establish a necessary and sufficient condition for the positivity of the mismatch capacity. 
 \begin{lemma}\label{lm: nu lemma}
Assume (\ref{eq infinite q case }) holds. There exists $\nu>0$, such that for all $n$ sufficiently large, for any $P_n\in\calP_n(\calX)$, any $\widehat{P}_{XZ}\in\calA_q(W_{Y|XZ},P_n)\cap \calP_n(\calX\times\calZ)$ 
 and any $\bz\in\calT_n(\widehat{P}_Z)$ which satisfies $|\calL(\bz,\widehat{P}_{XZ})|\geq 2$, any pair of codewords $\bx_\ell,\bx_k$ in $\calL(\bz,\widehat{P}_{XZ})$ such that $\bz$ is a possible output of $W_{Z|X}^n$ when fed by $\bx_\ell$ or $\bx_k$ satisfies either 
\begin{flalign}
 W_{Y|XZ}^n\left(q(\bx_k,\bY)\geq q(\bx_\ell,\bY)\big|\bX=\bx_\ell,\bZ=\bz\right)>\nu,
 \end{flalign}
 or 
 \begin{flalign}
 W_{Y|XZ}^n\left(q(\bx_\ell,\bY)\geq q(\bx_k,\bY)\big|\bX=\bx_k,\bZ=\bz\right)>\nu.
 \end{flalign}
\end{lemma}
\begin{proof}
From Lemma \ref{lm: aifuvuidfv}, we know that
for any pair of codewords $\bx_\ell,\bx_k$ in $\calL(\bz,\widehat{P}_{XZ})$ we have either (\ref{eq: apd'ohvpodfjpvokq}) or (\ref{eq: apd'ohvpodfjpvok}). Assume w.l.o.g.\ that (\ref{eq: apd'ohvpodfjpvokq}) holds, then clearly, \begin{flalign}
 &W_{Y|XZ}^n\left(q(\bx_k,\bY)\geq q(\bx_\ell,\bY)\big|\bX=\bx_\ell,\bZ=\bz\right)\nonumber\\
 &\geq W_{Y|XZ}^n\left(q(\bx_k,\bY)\geq q(\bx_\ell,\bY)+
 \bigg[
 \sum_{x,\widetilde{x},z,y} \widehat{P}_{\bx_\ell\bz\bx_k}(x,z,\widetilde{x}) 
  W(y|x,z)\left[q(\widetilde{x},y)-q(x,y)\right]\
 \bigg]
 \big|\bX=\bx_\ell,\bZ=\bz\right)\\
 &\triangleq 
 W_{Y|XZ}^n\left(\sum_{i=1}^nS_{i}\geq 0\big|\bX=\bx_\ell,\bZ=\bz\right),
 \end{flalign}
  where $\{ S_{i}\}$ are the following RVs satisfying $\EE( S_{i}\big | \bx_\ell,\bz)=0$, 
  \begin{flalign}
  S_{i,\ell,k}&=[q(x_{\ell,i},Y_i)-q(x_{k,i},Y_i) ]\nonumber\\
  &-
  \EE[q(x_{\ell,i},Y_i)-q(x_{k,i},Y_i) |Z_i=z_i,X_i=x_{\ell,i}].\end{flalign} 
Note also that the distribution of $S_i$ given $(\bx_\ell,\bz)$ depends only on $(x_{\ell,i},x_{k,i},z_i)$, and therefore, 
there is a finite set, denoted $\widetilde{\calP}$ (of size not exceeding $|\calZ||\calX|^2$) of conditional distributions that $S_i$ can have given $(\bx_\ell,\bz)$. 
Hence, it follows from \cite[Lemma 3]{CsiszarNarayan95} that if the $q(\cdot)$ values are bounded, there exists a constant $\nu >0$ such that for all $n$, 
\begin{flalign}\label{eq: qfwtrqtwyqwty}
 W_{Y|XZ}^n\left(\sum_iS_i\geq 0\bigg|\bx_\ell, \bz\right)>\nu .
\end{flalign}
It remains to treat infinite values of $\EE\left(q(\bx_k, \bY) |\bx_\ell,\bz\right)$ and $\EE\left(q(\bx_\ell, \bY)\big |\bx_\ell,\bz\right)$:
Since we assume metric values $q\in\mathbb{R}\cup\{-\infty\}$, the cases of interest for their values are limited to $(c,-\infty)$, $(-\infty,c)$, $(-\infty,-\infty)$, where $c$ represents a finite constant, respectively. The case $(-\infty,c)$ yields the inequality (\ref{eq: qfwtrqtwyqwty}) trivially. 

Since we assume that $\bz$ is a possible output of $W_{Z|X}^n$ when fed by $\bx_\ell$ (or $\bx_k$), the cases $(c,-\infty)$ and $(-\infty,-\infty)$; i.e., $\EE\left(q(\bx_\ell, \bY)\big |\bx_\ell,\bz\right)=-\infty$, imply that there exists a symbol $x_0$ in $\bx_\ell$ and a pair $(x_0,y_0)$ such that $W(y_0|x_0)>0$ and $q(x_0,y_0)=-\infty$, in contradiction to assumption (\ref{eq infinite q case }).

\end{proof}

 Now, since the members of $\calL(\bz,\widehat{P}_{XZ})$ are equiprobable given $(\bz,\widehat{P}_{XZ})$ (see (\ref{eq: idfuhvilufg})), we obtain from Lemma \ref{lm: nu lemma} that if 
$|\calL(\bz,\widehat{P}_{XZ})|\geq 2$, there exists $\nu >0$ such that 
\begin{flalign}
&P(error|\widehat{P}_{\bX\bZ}=\widehat{P}_{XZ},\bZ=\bz)\nonumber\\
&\geq \frac{1}{2}
\nu\cdot \indicator\{\widehat{P}_{XZ}\in\calA_q(W_{Y|X,Z},P_n)\}. \label{eq: laifulvdfg}
\end{flalign}

Next recall (\ref{eq: ilsufgviludf})-(\ref{eq: udfhv0}), stating that 
for any $\widehat{P}_{XZ}$, $\epsilon_n>0$, and $\widetilde{\epsilon}_n>0$ such that 
\begin{flalign}\label{eq: ilsufgviludf777}
R_n\geq I(\widehat{P}_{XZ})+\epsilon_n+\frac{|\calX||\calZ|-1}{n}\log(n+1)+\widetilde{\epsilon}_n,
\end{flalign}
it holds that
$\Pr\left(|\calL(\bZ,\widehat{P}_{XZ})|<e^{n\epsilon_n}\big|\widehat{P}_{\bX\bZ}=\widehat{P}_{XZ}\right)\leq  e^{-n\widetilde{\epsilon}_n}$,
and also 
\begin{flalign}
\Pr(error|\widehat{P}_{\bX\bZ}=\widehat{P}_{XZ})
&\geq \left(1-e^{-n\widetilde{\epsilon}_n}\right)\cdot \min_{\bz\in\calT_n(\widehat{P}_Z):\; |\calL(\bz,\widehat{P}_{XZ})|
\geq e^{n\epsilon_n}}\Pr(error|\widehat{P}_{\bX\bZ}=\widehat{P}_{XZ},\bZ=\bz).\label{eq: aofhv;oiudhfov;uihdfov}
\end{flalign}
and thus, we get from (\ref{eq: laifulvdfg}) and (\ref{eq: aofhv;oiudhfov;uihdfov}) that if (\ref{eq: ilsufgviludf777}) holds, 
\begin{flalign}
\Pr(error|\widehat{P}_{\bX\bZ}=\widehat{P}_{XZ})
&\geq \left(1-e^{-n\widetilde{\epsilon}_n}\right)\cdot
\frac{1}{2}\nu\cdot \indicator\{\widehat{P}_{XZ}\in\calA_q(W_{Y|X,Z},P_n)\}
 .\label{eq: aofhv;oiudhfov;uihdfdfdfov}
\end{flalign}

Next, take a vanishing sequence $c_n$ that satisfies $\lim_{n\rightarrow\infty}n[c_n-\frac{1}{n}|\calX||\calZ|\log(n+1) ]=\infty$;  we have
\begin{flalign}
&\Pr(D(\widehat{P}_{\bZ|\bX}\|W_{Z|X}|P_n)> c_n|\bX=\bx)\nonumber\\
&= 
\sum_{\bz:\;D(\widehat{P}_{\bz|\bx}\|W_{Z|X}|P_n)> c_n}W_{Z|X}^n(\bz|\bx)\\
&\leq (n+1)^{|\calX||\calZ|-1}\max_{\widehat{P}_{Z|X}:\;D(\widehat{P}_{Z|X}\|W_{Z|X}|P_n)> c_n}e^{-nD(\widehat{P}_{Z|X}\|W_{Z|X}|P_n)}\\
&\leq e^{-n[c_n-\frac{1}{n}|\calX||\calZ|\log(n+1) ]}.
\end{flalign}
denote $f_n=c_n-\frac{1}{n}|\calX||\calZ|\log(n+1) $, and  $d_n=\epsilon_n+\frac{|\calX||\calZ|-1}{n}\log(n+1)+\widetilde{\epsilon}_n$, we have \begin{flalign}
 &\Pr(error) = \sum_{\widehat{P}_{XZ}\in\calP_n(\calX\times\calZ):\; \widehat{P}_X=P_n}\Pr(\widehat{P}_{\bX\bZ}=\widehat{P}_{XZ})\cdot \Pr(error|\widehat{P}_{\bX\bZ}=\widehat{P}_{XZ})\nonumber\\
  & \geq  \sum_{\substack{\widehat{P}_{XZ}\in \calA_q(W_{Y|X,Z},P_n)\cap \calP_n(\calX\times\calZ):\\ \widehat{P}_X=P_n,\;  R\geq I(\widehat{P}_{XZ})+d_n\\
  D(\widehat{P}_{Z|X}\|W_{Z|X}|P_n)\leq f_n
  }}\Pr(\widehat{P}_{\bX\bZ}=\widehat{P}_{XZ})\cdot \Pr(error|\widehat{P}_{\bX\bZ}=\widehat{P}_{XZ})\\
&\geq (1-e^{-nf_n})\cdot\left(1-e^{-n\widetilde{\epsilon}_n}\right)\cdot \frac{1}{2}\nu\cdot \indicator\left\{\exists 
\widehat{P}_{XZ}\in
\calK_q(R,P_n ,W_{YZ|X})
 \right\},\label{eq: lilili}
\end{flalign}
where 
\begin{flalign}\label{eq: akduvudh}
&\calK_q(R,P_n, W_{YZ|X})\triangleq 
\left\{\widehat{P}_{XZ}\in\calP_n(\calX\times\calZ):\; \substack{ D(\widehat{P}_{Z|X}\|W_{Z|X}|P_n)\leq f_n,\\R_n\geq I(\widehat{P}_{XZ})+d_n,\\
 \widehat{P}_{XZ}\in\calA_q(W_{Y|X,Z},P_n)}
 \right\}.
\end{flalign}
The following lemma concludes the proof of Theorem \ref{th: thorem mismatch capacity}.

\begin{lemma}\label{ADDlm: non emptiness calG}
For any $P_n\in\calP_n(\calX)$ and 
$W_{YZ|X}\in \calW_q(P_n)$ if $R>I(P_n\times W_{Z|X})+\epsilon$, then the set $\calK_q(R,P_n, W_{YZ|X})$ is non-empty, and consequently 
there exists $\nu>0$ (which does not depend on $P_n$), such that for any $\calC_n\subseteq\calT_n(P_n)$, such that $|\calC_n|>e^{nR}$, we have $P_e(W,\calC_n,q)>(1-e^{-nf_n})\cdot\left(1-e^{-n\widetilde{\epsilon}_n}\right)\cdot \frac{1}{2}\nu$.
\end{lemma}
Lemma \ref{ADDlm: non emptiness calG} is proved in Appendix \ref{ap: non emptiness calG proof}, and this concludes the proof of Theorem \ref{th: thorem mismatch capacity}.

\section{Simpler Bounds on the Reliability Function and Sufficient Conditions for Tightness}\label{sc: aifugviudfg}

We next show how looser yet simpler bounds on the reliability function using the method of \cite{SomekhBaruch2020singleletter_part1} can be derived, and these bounds provide sufficient conditions for tightness for certain ranges of rates.

As mentioned at the beginning of Section \ref{sc: section comparison Cap}, any 
two-output channel which belongs to $\Gamma(q,\rho)$ has the property that an error occurs at the $Z$ decoder, only if the $Y$-receiver makes an error, and thus for any codebook $\calC_n$, (\ref{eq: THIS holds}) holds which implies the bound (\ref{eq: KG bound2 BB}) on $C_q(W)$. 

For exactly the same reason, the following bound on the reliability function can be deduced.
\begin{theorem}\label{th: simple theorem}
For all $\calZ$, additive metrics $q,\rho$,  
and a 
stationary memoryless channel $W$
\begin{flalign}
E^q(R,W)&\leq 
\min_{P_{YZ|X}\in\Gamma(q,\rho):\; P_{Y|X}=W}E(R,P_{Y|X})\label{eq: KG bound2}.
\end{flalign}
\end{theorem}
Additionally, as mentioned above, in \cite{SomekhBaruch2020singleletter_part1} 
equivalence classes of isomorphic channel-metric pairs $(W,q)$ that share the same mismatch capacity for additive metrics were introduced. The following definition of \cite{SomekhBaruch2020singleletter_part1} is repeated here for completeness.
\begin{definition}\label{df: isomorph}
A channel-metric pair $(P_{Z|X},\rho)$ is superior to the channel-metric pair $(P_{Y|X},q)$ if there exists a joint conditional distribution $P_{YZ|X}\in\Gamma(q,\rho)$, 
whose marginal conditional distributions are $P_{Y|X}$ and $P_{Z|X}$. The superiority relation is denoted by
$(P_{Y|X},q)\rightarrowtriangle (P_{Z|X},\rho)$. 
If both $(P_{Y|X},q)\rightarrowtriangle (P_{Z|X},\rho)$ and $(P_{Z|X},\rho) \rightarrowtriangle (P_{Y|X},q)$, denote $(P_{Y|X},q)\leftrightarrowtriangle (P_{Z|X},\rho)$, and the pairs are called isomorphic. 
\end{definition}
It was proved that if one of the pairs in the class is matched, then the mismatch capacity of the entire class is fully characterized and equal to 
that of the matched pair. The following theorem follows straightforwardly for exactly the same reason of superiority/equivalence. 
\begin{theorem}\label{lm:ldfslvdfhgivhfiuhsss}
If $(W,q)\rightarrowtriangle (P_{Z|X},\rho)$ then
\begin{flalign}\label{eq: adhvgiludhviuh}
E^q(R,W)&\leq E^{\rho}(R,P_{Z|X}),
\end{flalign}
and consequently, if $(W,q)\leftrightarrowtriangle (P_{Z|X},\rho)$ then
\begin{flalign}\label{eq: trivya}
E^q(R,W)&= E^{\rho}(R,P_{Z|X}).
\end{flalign}
If there exists a matched channel-metric pair $(\widetilde{P}_{Z|X},\widetilde{q}_{ML})$ where $\widetilde{q}_{ML}=\log \widetilde{P}_{Z|X}$ is the maximum likelihood metric w.r.t.\ $\widetilde{P}_{Z|X}$ such that $(W,q)\leftrightarrowtriangle (\widetilde{P}_{Z|X},\widetilde{q}_{ML})$ then  
\begin{flalign}
E^q(R,W)&=
E(R,\widetilde{P}_{Z|X}).
\end{flalign}
\end{theorem}
Note that the theorem implies that for the range of rates such that $E(R,\widetilde{P}_{Z|X})$ is known; e.g., at $R=0^+$ or above the critical rate where the tangential straight line bound meets the sphere packing bound, the $E^q(R,W)$ is known as well. 

Theorem \ref{th: simple theorem} can be extended to yield tighter bounds for larger classes of channels that depend also on the codebook composition $P$ using a similar approach. For example, using the set (see \cite{SomekhBaruch2020singleletter_part1})
 \begin{flalign}
&\Gamma(q,\rho,P)=\nonumber\\
&\Bigg\{P_{YZ|X}:\; 
\min_{\substack{ V_{XYZ\widetilde{X}}:V_{X}=V_{\widetilde{X}}=P, \\V_{XYZ}\ll P\times P_{YZ|X}
,\\ \rho(V_{XZ})\leq \rho(V_{\widetilde{X}Z})}}\left[ q(V_{\widetilde{X}Y})-q(V_{XY})\right]\geq 0\Bigg\}\label{eq: Gamma q rho P dfn s;kdfjbv ;ibf}.
\end{flalign}

\section{Concluding Remarks}\label{sc: Concluding Remarks}

The new technique presented in the paper yields the tightest bounds known to date on the mismatch capacity and the reliability function with mismatched decoding for $R>0$. One of the main contributions of our work is the derivation of bounds that are easier to compute compared to previous bounds, either by reducing the number of degrees of freedom of the parameters that are optimized in the calculation of the bounds (such as in $C^{sym}_q(W)$), or by providing a looser bound that is considerably easier to compute, 
$C^{psd}_q(W)$ in addition to our previous bound (\ref{eq: KG bound2 BB}) of \cite{SomekhBaruch2020singleletter_part1}.

It would be interesting to see whether there are cases for which $\overline{C}_q(W)$ can be strictly tighter than $C^{sym}_q(W)$ or not; another interesting question is what are the exact relations between the bounds $\widetilde{C}_q(W)$, $\widetilde{C}^{sym}_q(W)$, $C^{sym}_q(W)$ and $C^{psd}_q(W)$. A partial answer to this question was given by analyzing the binary input channels. 

Similar questions apply also to the bounds on the reliability function.

\appendix

\subsection{Proof of Proposition \ref{pr: adh;oidvjhoi;hdjfoiv}}\label{sc: bounds sym ib}

\vspace{0.2cm}

\noindent{\underline{Proof of $\overline{C}_q(W) \leq C^{sym}_q(W)$ }}:

Clearly, the marginal $P_{\widetilde{X}ZX}$ distribution of $P_{\widetilde{X}XZU}$ where $\widetilde{X}-(U,Z)-X$ and $P_{XZU}=P_{\widetilde{X}ZU}$ is symmetric, and thus satisfies $P_{\widetilde{X}ZX}\in \calP_{sym}(\calX^2\times \calZ)$.

Secondly, note that any distribution $P_{\widetilde{X}XZU}$ where $\widetilde{X}-(U,Z)-X$ and $P_{XZU}=P_{\widetilde{X}ZU}$ satisfies for any $(x,z)\in \calX\times\calZ$:
\begin{flalign}
P_{\widetilde{X}X|Z}(x,x|z)& = \sum_uP_{U|Z}(u|z)\left[P_{X|UZ}(x|u,z)\right]^2\nonumber\\
&\geq \left[\sum_uP_{U|Z}(u|z)P_{X|UZ}(x|u,z)\right]^2\label{eq: aiugviudgv}\\
&=\left[P_{X|Z}(x|z)\right]^2.
\end{flalign}
where (\ref{eq: aiugviudgv}) follows from Jensen's inequality. Hence, $P_{\widetilde{X}|XZ}(x|x,z)\geq P_{X|Z}(x|z)$. 

Thus the claim $\overline{C}_q(W) \leq C^{sym}_q(W)$ follows since 
\begin{flalign}
&\min_{\substack{P_{\widetilde{X}|XZ}:\; P_{\widetilde{X}ZX}\in \calP_{sym}(\calX^2\times \calZ)\\
\forall (x,z),\; P_{\widetilde{X}|XZ}(x|x,z)\geq P_{X|Z}(x|z)}}
\sum_{x,\widetilde{x},z,y}P_{XZ}(x,z)P_{\widetilde{X}|XZ}(\widetilde{x}|x,z)P_{Y|XZ}(y|x,z)[q(\widetilde{x},y)-q(x,y)]\nonumber\\
&\leq 
\min_{P_{U|XZ}}
\sum_{u,x,\widetilde{x},z,y}P_{XZ}(x,z,u)P_{X|UZ}(\widetilde{x}|u,z)P_{Y|XZ}(y|x,z)[q(\widetilde{x},y)-q(x,y)].
\end{flalign}

\noindent{\underline{Proof of $\overline{C}_q(W)\leq \widetilde{C}_q(W)$}}:

We show that $\widetilde{\calW}_q(P_X)\subseteq \calW_q(P_X)$. Fix $P_X$, and let $P_{YZ|X}\in \widetilde{\calW}_q(P_X)$ be given, which implies that 
for all $(z,V_{X|Z})$ such that $z\in\calZ$, and $V_{X|Z}$ such that $P_Z\times V_{X|Z}\ll P_{ZX}$, 
\begin{flalign}
 &\sum_{x,\widetilde{x},y} V(x|z)V(\widetilde{x}|z)P_{Y|XZ}(y|x,z)[q(\widetilde{x},y)-q(x,y)]\geq 0.\label{eq: ozfihv}
\end{flalign}

Next fix $P_{U|XZ}$. Along with $P_{XZ}$ (the marginal of $P_X\times P_{YZ|X}$) this also induces $P_{XZU}=P_{XZ}\times P_{U|XZ}$. 
From (\ref{eq: ozfihv}) it follows that for all $(z,u)\in\calZ\times \calU$ such that either $P_{X|U=u,Z=z}\ll  P_{X|Z=z}$ or $P_{UZ}(u,z)=0$, it must hold that 
\begin{flalign}
 &
P_{UZ}(u,z)\cdot \sum_{x,\widetilde{x},y} P_{X|UZ}(x|u,z)P_{X|UZ}(\widetilde{x}|u,z)P_{Y|XZ}(y|x,z)[q(\widetilde{x},y)-q(x,y)]\geq 0.\label{eq: ozfihv2}
\end{flalign}

Since for any $P_{U|XZ}$ it holds that for all $(z,u)\in\calZ\times \calU$ either $P_{X|U=u,Z=z}\ll  P_{X|Z=z}$ or $P_{UZ}(u,z)=0$, summing over $(u,z)$ we obtain
\begin{flalign}\label{eq: Delta q definition sdu;hc;isdpckopkusdha}
\sum_{x,z,u,\widetilde{x},y}P_{UZ}(u,z) P_{X|UZ}(x|u,z)P_{X|UZ}(\widetilde{x}|u,z) P_{Y|XZ}(y|x,z)[q(\widetilde{x},y)-q(x,y)]\geq 0,
\end{flalign}
and this yields that $P_{YZ|X}\in \calW_q(P_X)$, because the above holds for any $P_{U|XZ}$.

\noindent{\underline{Proof of $\widetilde{C}_q(W)\leq C^{psd}_q(W)$}}:

Observe that 
\begin{flalign}
&\sum_{x,\widetilde{x},y} V(x|z)V(\widetilde{x}|z)P_{Y|XZ}(y|x,z)[q(\widetilde{x},y)-q(x,y)]\nonumber\\
&=\frac{1}{2}\sum_{x}\sum_{\widetilde{x},y}V(x|z)V(\widetilde{x}|z)\left(
P_{Y|XZ}(y|x,z)[q(\widetilde{x},y)-q(x,y)]+P_{Y|XZ}(y|\widetilde{x},z)[q(x,y)-q(\widetilde{x},y)]\right)\label{eq: adch;diohc}\\
&=\frac{1}{2}\sum_{x}\sum_{\widetilde{x}}V(x|z)V(\widetilde{x}|z)\calD^q(P_{Y|X,Z=z})_{x,\widetilde{x}}.
\end{flalign}

Hence, we get for any $z\in\calZ$, 
\begin{flalign}
&\calD^q(P_{Y|X,Z=z})\succeq 0  \Rightarrow \nonumber\\ &\forall V_{X|Z}:\; P_Z\times V_{X|Z}\ll P_{XZ},\; 
 \sum_{x,\widetilde{x},y} V(x|z)V(\widetilde{x}|z)P_{Y|XZ}(y|x,z)[q(\widetilde{x},y)-q(x,y)]\geq 0,\label{eq: ozfihv3}
\end{flalign}
because the requirement $\sum_{x}\sum_{\widetilde{x}}V(x|z)V(\widetilde{x}|z)\calD^q(P_{Y|X,Z=z})_{x,\widetilde{x}}\geq 0$ for any
$\left(V(1|z),...,V(|\calX||z)\right)$ which is a probability vector (that has non-negative entries) is looser than the same requirement for any real vector $\underline{V}\in\mathbb{R}$, which is nothing but the definition of a p.s.d. matrix.

\subsection{Proof of Lemma \ref{lm: aofihv;oidhfvo;idhfoivhdiof}}\label{sc: Proof of Binary Lemma}

Consider a binary input DMC, whose transition probability distribution is given by
\begin{flalign}
\{W(y|x)\}_{x\in\calX, y\in\calY},\; \calX=\{0,1\}, |\calY|<\infty.\end{flalign}

Let $\calZ$ be a finite set, and consider the bound $C^{sym}_q(W)$ in (\ref{eq: Csym sihfi}), where ${\calW}^{sym}_q(P_X)$ is defined in (\ref{eq: aiufhvuifhd}), and can be expressed as 
\begin{flalign}
&{\calW}^{sym}_q(P_X)\triangleq \bigg\{P_{YZ|X}:\;\min_{\substack{V_{\widetilde{X}ZX}\in \calP_{sym}(\calX^2\times \calZ):\; V_{XZ}=P_{XZ}
,\;\\ \forall (x,z),\; V_{\widetilde{X}X|Z}(x,x|z)\geq P_{X|Z}^2(x|z) }}\EE_{V_{\widetilde{X}ZX}P_{Y|XZ}} [q(\widetilde{X},Y)- q(X,Y)]\geq 0\bigg\}. \label{eq: aiufhvuifhddasvl`dih}
\end{flalign}

We next solve the minimization problem of (\ref{eq: aiufhvuifhddasvl`dih}) explicitly. We have from symmetry of $V$
\begin{flalign}
&\EE_{V_{\widetilde{X}ZX}P_{Y|XZ}} [q(\widetilde{X},Y)- q(X,Y)]\nonumber\\
&= \sum_{z,x,\widetilde{x}}P_Z(z)V_{X\widetilde{X}|Z}(x,\widetilde{x}|z)P(y|x,z)[q(\widetilde{x},y)- q(x,y)]\\
&=\sum_{z,(x,\widetilde{x})\in\{0,1\}^2 ,\widetilde{x}\neq x}P_Z(z)V_{X\widetilde{X}|Z}(x,\widetilde{x}|z)\sum_y P(y|x,z)[q(\widetilde{x},y)- q(x,y)]\\
&= \frac{1}{2}\sum_{z,(x,\widetilde{x})\in\{0,1\}^2 ,\widetilde{x}\neq x}P_Z(z)V_{X\widetilde{X}|Z}(x,\widetilde{x}|z)d_q(P_{YZ|X},z)\\
&= \sum_{z}P_Z(z)a(z)d_q(P_{YZ|X},z),\label{eq:doihoihi}
\end{flalign}
where $d_q(P_{YZ|X},z)$ is defined in (\ref{eq: d_1 dfn}), and $a(z)\triangleq V_{X\widetilde{X}|Z}(0,1|z)=V_{X\widetilde{X}|Z}(1,0|z)$. 
Note that the constraint $V_{\widetilde{X},X|Z}(0,0|z)\geq P_{X|Z}^2(0|z) $ combined with 
$V_{\widetilde{X},X|Z}(0,0|z)+V_{\widetilde{X},X|Z}(0,1|z)= P_{X|Z}(0|z) $
becomes $a(z)\leq P_{X|Z}(0|z)P_{X|Z}(1|z) $. Therefore, the minimizing $a(z)$ equals
\begin{flalign}
a_{opt}^{sym}(z)&= \left\{\begin{array}{ll}
0 & d_q(P_{YZ|X},z)\geq 0\\
P_{X|Z}(0|z)P_{X|Z}(1|z) & d_q(P_{YZ|X},z)< 0
\end{array}\right..
\end{flalign}
This yields
\begin{flalign}
&\min_{\substack{V_{\widetilde{X}ZX}\in \calP_{sym}(\calX^2\times \calZ):\; V_{XZ}=P_{XZ}
,\;\\ \forall (x,z),\; V_{\widetilde{X},X|Z}(x,x|z)\geq P_{X|Z}^2(x|z) }}\EE_{V_{\widetilde{X}ZX}P_{Y|XZ}} [q(\widetilde{X},Y)- q(X,Y)]\nonumber\\
&= \sum_zP_Z(z)a_{opt}^{sym}(z)d_q(P_{YZ|X},z)\\
&=\sum_{z:\; d_q(P_{YZ|X},z)< 0}
P_Z(z)P_{X|Z}(0|z)P_{X|Z}(1|z)d_q(P_{YZ|X},z).
\end{flalign}
Now, this quantity is non-negative iff for all $z$
\begin{flalign}
 d_q(P_{YZ|X},z)< 0\Rightarrow P_Z(z)P_{X|Z}(0|z)P_{X|Z}(1|z)=0.\label{eq: aofhvodifhv}
\end{flalign}
Note that the condition (\ref{eq: aofhvodifhv}) can also be rewritten as:
\begin{flalign}
\forall z,\; P_Z(z)P_{X|Z}(0|z)P_{X|Z}(1|z)d_q(P_{YZ|X},z)\geq 0,\label{eq: ;afhiodho;ivhdfoiv}
\end{flalign}
which yields (\ref{eq: relma example}), and since \begin{flalign}
&P_{X|Z}(0|z)P_{X|Z}(1|z)d_q(P_{YZ|X},z)\nonumber\\
&=\frac{1}{2}\sum_{(x,\widetilde{x})\in\{0,1\}^2}P_{X|Z}(x|z)P_{X|Z}(\widetilde{x}|z)d_q(P_{YZ|X},z)\\
&=\sum_{(x,\widetilde{x})\in\{0,1\}^2}P_{X|Z}(x|z)P_{X|Z}(\widetilde{x}|z)\sum_y P_{Y|XZ}(y|x,z)[q(\widetilde{x},y)-q(x,y)]\\
&=\sum_{\widetilde{x},y}P_{X|Z}(\widetilde{x}|z)P_{Y|Z}(y|z)q(\widetilde{x},y)-\sum_{x,y} P_{XY|Z}(x,y|z)q(x,y)
,\end{flalign} we get the equivalent expression (\ref{eq: relma exampledacdscaadvds}).

To obtain (\ref{eq: conseahviodhsi}), assume w.l.o.g.\ that the maximizing $P_X$ is non-degenerate; i.e., $P_X(0)\in(0,1)$, as otherwise $C_q(W)=0$. In this case, the coupling between $P_X$ and the set of the minimization is removed since $P_Z(z)P_{X|Z}(0|z)P_{X|Z}(1|z)d_q(P_{YZ|X},z)\geq 0$ is equivalent to $P_{Z|X}(z|0)P_{Z|X}(z|1)d_q(P_{YZ|X},z)\geq 0$, and thus\begin{flalign}
&\max_{P_X}
\min_{\substack{P_{YZ|X}:\;\forall z,\; P_Z(z)P_{X|Z}(0|z)P_{X|Z}(1|z)d_q(P_{YZ|X},z)\geq 0,\\
P_{Y|X}=W}} I(X;Z)\nonumber\\
&= \max_{P_X}
\min_{\substack{P_{YZ|X}:\;\forall z,\; P_{Z|X}(z|0)P_{Z|X}(z|1)d_q(P_{YZ|X},z)\geq 0,\\
P_{Y|X}=W}} I(X;Z)\\
&\leq \min_{\substack{P_{YZ|X}:\;\forall z,\; P_{Z|X}(z|0)P_{Z|X}(z|1)d_q(P_{YZ|X},z)\geq 0,\\
P_{Y|X}=W}} \max_{P_X}I(X;Z)\label{eq: maxmin nimmax}\\
&=\min_{\substack{P_{YZ|X}:\;\forall z,\; P_{Z|X}(z|0)P_{Z|X}(z|1)d_q(P_{YZ|X},z)\geq 0,\\
P_{Y|X}=W}}C(P_{Z|X}),
\end{flalign}
where (\ref{eq: maxmin nimmax}) follows since $\max_{a\in\calA}\min_{b\in\calB}f(a,b)\leq  \min_{b\in\calB}\max_{a\in\calA}f(a,b)$, and the last step follows from Shannon's capacity formula.

The bound (\ref{eq: relma example2}) follows similarly to (\ref{eq: relma example}), but without the constraint $V_{\widetilde{X},X|Z}(x,x|z)\geq P_{X|Z}^2(x|z) $. The optimal $a(z)$ of (\ref{eq:doihoihi}) in this case is given by
\begin{flalign}
\widetilde{a}_{opt}^{sym}(z)&= \left\{\begin{array}{ll}
0 & d_q(P_{YZ|X},z)\geq 0\\
\min\{P_{X|Z}(0|z),P_{X|Z}(1|z)\} & d_q(P_{YZ|X},z)< 0
\end{array}\right.,
\end{flalign}
yielding the condition
\begin{flalign}
& \sum_{z}P_Z(z)\widetilde{a}_{opt}^{sym}(z)d_q(P_{YZ|X},z)\nonumber\\
&=\sum_{z:\; d_q(P_{YZ|X},z)< 0}\min\{P_{XZ}(0,z),P_{XZ}(1,z)\}\cdot d_q(P_{YZ|X},z)  \geq 0
\end{flalign}
which is equivalent to (\ref{eq: ;afhiodho;ivhdfoiv}), proving (\ref{eq: relma example2}).

As for (\ref{eq: relma example3}),
note that
\begin{flalign}
\calD^q(P_{Y|X,Z=z})=\left(\begin{array}{cc}0 & d_q(P_{YZ|X},z)\\d_q(P_{YZ|X},z) & 0\end{array}\right), 
\end{flalign}
hence in this case, $\calD^q(P_{Y|X,Z=z})\succeq 0$ is equivalent to $d_q(P_{YZ|X},z)=0$, which yields (\ref{eq: relma example3}) so in this case we have $ C^{sym}_q(W)\leq C^{psd}_q(W)$. 

\subsection{Proof of Lemma \ref{lm: aiuvdgiudg}}\label{ap: approx 1 elmm}

Consider the argument of the minimization in (\ref{eq: Ialpha dfndfugilugi11}); i.e., 
\begin{flalign}
& D(V_{Y|XZ\widetilde{X}}\|W_{Y|XZ}|P_{XZ\widetilde{X}})\\
&= \sum_{x,\widetilde{x},z}P_{\widetilde{X}ZX}(\widetilde{x},x,z)\sum_y V_{Y|XZ\widetilde{X}}(y|x,z,\widetilde{x})\log \frac{V_{Y|XZ\widetilde{X}}(y|x,z,\widetilde{x})}{ W_{Y|XZ}(y|x,z)}\label{eq:eruhfoihr},
\end{flalign}
and recall the definition of $\calS^{cond}(P_{\widetilde{X}ZX})$ in (\ref{eq: calS cond dfn}).

Clearly, if $P_{\widetilde{X}ZX}$ is such that $\Omega(P_{\widetilde{X}ZX}, W_{Y|XZ})=\infty$, the inequality (\ref{eq: asdcoi}) holds trivially, thus assume $\Omega(P_{\widetilde{X}ZX}, W_{Y|XZ})<\infty$, which implies that $\calS^{cond}(P_{\widetilde{X}ZX})$ is non-empty, and that a minimizer satisfies  $V_{Y|XZ\widetilde{X}}(y|x,z,\widetilde{x})=0$ whenever $W(y|x,z)=0$. 
Let $V_{Y|XZ\widetilde{X}}^*\in \calS^{cond}(P_{\widetilde{X}ZX})$ be such a distribution. 

We next show that $V_{Y|XZ\widetilde{X}}^*$ can be approximated by an empirical distribution $V_{Y|XZ\widetilde{X}}^{(n)}(y|x,z,\widetilde{x})\in\calS^{cond}(P_{\widetilde{X}ZX})$. 

To this end, we introduce two additional technical lemmas. The first lemma is obtained as a special case of Krein-Milman Theorem. It asserts that any distribution in $\calP(\calA)$ where $\calA$ is finite, can be expressed as a convex combination of no more than $2^{|\calA|}$ empirical distributions in $\calP_{\ell}(\calA)$, which are all $1/\ell$ close to it in the $L_{\infty}$-sense:
\begin{lemma}\label{lm: Krein Milman}
Let $\xi=(\xi_1,\xi_2,...,\xi_{|\calA|})\in\calP(\calA)$ be a given distribution, let $\ell\geq 1$ be an integer, and consider the following convex subset of $\calP(\calA)$
\begin{flalign}
\Pi(\xi,\ell)&\triangleq \left\{V\in \calP(\calA):\; \forall j\in\calA,\; \frac{\lfloor \ell \xi_j\rfloor }{\ell}\leq V(j)\leq \frac{\lceil \ell \xi_j\rceil }{\ell}\right\}.
\end{flalign}
There exist $K\leq 2^{|\calA|}$ empirical distributions $\{P^{(i)}\}_{i=1}^{K}$ in $\Pi(\xi,\ell)\cap \calP_{\ell}(\calA)$ such that any $P\in\Pi(\xi,\ell)$ (and in particular, $\xi$) can be expressed as:
\begin{flalign}
P&=\sum_{i=1}^{K}\alpha_i \cdot P^{(i)}
\end{flalign}
for some $\{\alpha_i\}$, such that $\alpha_i\in[0,1]$ and $\sum_{i=1}^{K}\alpha_i=1$. 
\end{lemma}
\begin{proof}
By Krein-Milman's Theorem, $\Pi(\xi,\ell)$ is the closed convex hull of its extreme points. The extreme points of $\Pi(\xi,\ell)$ are empirical distributions of order $\ell$. There are no more than $2^{|\calA|}$ empirical distributions in $\Pi(\xi,\ell)\cap \calP_{\ell}(\calA)$, because each entry $V(j)$ can only take $2$ values.
\end{proof}

The next lemma is a straightforward consequence of 
Lemma \ref{lm: Krein Milman}.

\begin{lemma}\label{lm: Krein Milman 2}
Let $\calA,\calB$ be finite sets, and $\ell$ an integer. 
For any $Q_A\in\calP_\ell(\calA)$ and $V_{B|A}\in\calP(\calB|\calA)$, there exist empirical distributions $P^{(j)}_{AB}\in\calP_\ell(\calA\times\calB)$, $i=1,2,..K$, with 
$K\leq|\calA|\cdot 2^{|\calB|}$, such that
\begin{flalign}\label{eq: a;fiuhvoifho;dv}
Q_A\times V_{B|A}&=\sum_{j=1}^{K} \beta_j P^{(j)}_{AB}
\end{flalign}
and
\begin{flalign}
&\forall j,\; P^{(j)}_{A}=Q_A, \|Q_A\times V_{B|A}-P_{AB}^{(j)}\|\leq\frac{|\calA||\calB|}{\ell},\; \mbox{ and }
V_{B|A}(b|a)=0\Rightarrow P_{B|A}^{(j)}(b|a)=0
.\label{eq: ahv;oifdh;iv}
\end{flalign} 
\end{lemma}
\begin{proof}
For each $a\in\calA$, we use Lemma \ref{lm: Krein Milman} to express $V_{B|A=a}$ as a convex combination of empirical distributions, \begin{flalign}
V_{B|A=a}(b|a)&=\sum_{i_a=1}^{K_a} \alpha_{i_a} V_{B|A}^{(i_a)}(b|a), 
\end{flalign}
where $K_a\leq 2^{|\calB|}$, and naturally, if $V_{B|A=a}=0$, we take  $V_{B|A}^{(i_a)}(b|a)=0$ for all $i_a$, 
and by definition we have $V_{B|A=a}^{(i_a)}\in\calP_{\ell_a}(\calA)$, where $\ell_a=\ell\cdot Q(a)$, and for all $i_a$, 
$\|V_{B|A=a}-V_{B|A=a}^{(i_a)}\|\leq\frac{|\calB|}{\ell_a}$. 

Now, let $j=(i_1,...,i_{|\calA|})$ denote the index that takes $K=\sum_{x=1}^{|\calA|} K_a$ values, denote further
\begin{flalign}
P^{(j)}_{AB}(a,b)= Q(a)\cdot V_{B|A}^{(i_a)}(b|a),
\end{flalign}
thus, we obtain (\ref{eq: a;fiuhvoifho;dv}), 
where for $j=(i_1,...,i_{|\calA|})$ we have $\beta_j=\alpha_{i_j}$, and clearly (\ref{eq: ahv;oifdh;iv}) holds. 
\end{proof}

Next, we invoke Lemma \ref{lm: Krein Milman 2} with 
$\left(n, \calX^2\times\calZ, P_{\widetilde{X}ZX}, \calY, V_{Y|XZ\widetilde{X}}^*\right)$ in the roles of 
$\left(\ell, \calA, Q_A, \calB, P_{B|A}\right)$, respectively, and we let $ V_{Y|XZ\widetilde{X}}^{(j,n)}$ denote $P_{B|A}^{(j)}$. 

By this construction we have for all $j$, 
$
\|P_{XZ\widetilde{X}}\times V_{Y|XZ\widetilde{X}}^{*}-P_{XZ\widetilde{X}}\times V_{Y|XZ\widetilde{X}}^{(j,n)}\|\leq\frac{|\calX|^2|\calZ||\calY|}{n}$, 
and by affinity of $q(\cdot)$, we have that there must exist at least one index $j_0$ such that  
$q(V_{\widetilde{X}Y}^{(j_0,n)})-q(V_{XY}^{(j_0,n)})\geq q(V_{\widetilde{X}Y}^*)-q(V_{XY}^*)\geq 0$, and hence $V_{\widetilde{X}XYZ}^{(j_0,n)}\in \calS(P_{\widetilde{X}ZX})$.

Recall again that the argument of the minimization is $D(V_{Y|XZ\widetilde{X}}\|W_{Y|XZ}|P_{Xz\widetilde{X}})$, 
which is equal to $-H(V_{Y|XZ\widetilde{X}})-\EE_{P_{\widetilde{X}ZX}\times V_{Y|XZ\widetilde{X}}}\log W_{Y|XZ}$.

Hence, denoting $A=|\calX|^2|\calZ||\calY|$, and $c_n=\frac{|\calX|^2|\calZ||\calY|}{n}$ from \cite[Lemma 2.7]{CsiszarKorner81}, if follows that 
\begin{flalign}
&|D(V_{Y|XZ\widetilde{X}}^{(j_0,n)}\|W_{Y|XZ}|P_{\widetilde{X}ZX})-D(V_{Y|XZ\widetilde{X}}\|W_{Y|XZ}|P_{\widetilde{X}ZX})|\nonumber\\
&\leq - 2\cdot c_n\log\frac{c_n}{A}+c_n\log \frac{1}{t_{n,min}},
\end{flalign}
 where $t_{n,min}$ is defined (\ref{eq: W min dfn}) 
and this concludes the proof of (\ref{eq: asdcoi}).

\subsection{Proof of Corollary \ref{cr: corollary Type Dependent}}\label{ap: Cocorollary Type Dependent}

The proof is similar to that of Theorem \ref{th: Main Theorem }, so it is advisable to read the latter up to (\ref{eq: advhof;idhvoidfh}), before reading this proof. 
For convenience, we introduce the following notation for a joint distribution $V_{\widetilde{X}XZY}$; Let 
\begin{flalign}
r_q(V_{\widetilde{X}XZY})\triangleq q(V_{\widetilde{X}Y}). 
\end{flalign}
Denote also
\begin{flalign}
\calA^*\triangleq \{V_{Y|XZ\widetilde{X}}:\; I_V(\widetilde{X};Y|XZ)=0\}. 
\end{flalign}
We follow the steps of the proof of Theorem \ref{th: Main Theorem } up to (\ref{eq: advhof;idhvoidfh}) where step (\ref{eq: afiuvufgv}) follows since 
\begin{flalign}
&\calS^{cond}(\frac{1}{2}[\widehat{P}_{\bx_i \bz \bx_j}+\widehat{P}_{\bx_j \bz \bx_i}])\cap \calA^*\nonumber\\
&\subseteq \left(\calS^{cond}(\widehat{P}_{\bx_i \bz \bx_j})\cap \calA^*\right) \cup  \left(\calS^{cond}(\widehat{P}_{\bx_j \bz \bx_i})\cap \calA^*\right)
\end{flalign}
 is also valid for convex type-dependent metrics. To realize this, note that since $\widehat{P}_{\bx_j\bz}=\widehat{P}_{\bx_i\bz}$, 
\begin{flalign}
&(\calS^{cond}(\widehat{P}_{\bx_i \bz \bx_j})\cap \calA^*)\cup ( \calS^{cond}(\widehat{P}_{\bx_j \bz \bx_i})\cap\calA^*) \nonumber\\
&=\left\{V_{Y|XZ}: r_q\left(\widehat{P}_{\bx_i \bz \bx_j}\times V_{Y|XZ}\right)\geq q(V_{XY})
\mbox{ or }
r_q\left(\widehat{P}_{\bx_j \bz \bx_i}\times V_{Y|XZ}\right)\geq q(V_{XY})
\right\}\\
&=\left\{V_{Y|XZ}: \frac{1}{2}\left(r_q\left(\widehat{P}_{\bx_i \bz \bx_j}\times V_{Y|XZ}\right)+r_q\left(\widehat{P}_{\bx_j \bz \bx_i}\times V_{Y|XZ}\right)\right)\geq q(V_{XY})
\right\}\\
&\supseteq\left\{V_{Y|XZ}:r_q\left(\frac{1}{2}[\widehat{P}_{\bx_i \bz \bx_j}+\widehat{P}_{\bx_j \bz \bx_i}]\times V_{Y|XZ}\right)\geq q(V_{XY})\right\}
\label{eq: calSd`cdcsdfnzd v;vknc kn},
\end{flalign}
where the last step follows from convexity of $q(P_{XY})$ in $P_{Y|X}$ for fixed $P_X$.

Thus, we have (\ref{eq: advhof;idhvoidfh}), i.e.,  \begin{flalign}
&\Pr(error|\bz,\widehat{P}_{\bX\bZ}=\widehat{P}_{XZ})\nonumber\\
&\geq \exp\bigg\{-n \max_{\bx_i\in\calL ,\bx_j\in\calL,j\neq i}\bigg(\min_{\substack{V_{\widetilde{X}XYZ}\in\calS(\frac{1}{2}[\widehat{P}_{\bx_i \bz \bx_j}+\widehat{P}_{\bx_j \bz \bx_i}]): \\I_V(\widetilde{X};Y|XZ)=0}
} D(V_{Y|XZ}\|W_{Y|XZ}|\widehat{P}_{XZ})+k_n\bigg)\bigg\}\\
&\geq \exp\bigg\{-n \left(\max_{P_{XZ\widetilde{X}}:\; P_{XZ}=P_{\widetilde{X}Z}=\widehat{P}_{XZ}}F(P_{XZ\widetilde{X}}, W_{Y|XZ})+k_n\right)\bigg\},\label{eq: advhofadsochoiudhc;idhvoidfh}
\end{flalign}
where
\begin{flalign}
F(P_{XZ\widetilde{X}}, W_{Y|XZ})&\triangleq\min_{\substack{V_{\widetilde{X}XYZ}\in\calS(\frac{1}{2}[P_{XZ\widetilde{X}}+P_{\widetilde{X}ZX}]): \\I_V(\widetilde{X};Y|XZ)=0}
} D(V_{Y|XZ}\|W_{Y|XZ}|\widehat{P}_{XZ}).
\label{eq: o;fibhoissdcdcsfhgb}
\end{flalign}
Next, we use (\ref{eq: aiufviufgv})-(\ref{eq: udfhv0}) and this gives
for $\widehat{P}_{XZ}$ which is a possible joint empirical distribution of a codeword and a channel output $\bZ$ such that (\ref{eq: ilsufgviludf}) holds, for $\widetilde{\epsilon}_n>1/n$ \begin{flalign}
\Pr(error|\widehat{P}_{XZ})&\geq \left(1-e^{-n\widetilde{\epsilon}_n}\right)\cdot e^{-n \left[\max_{\bx_i\in\calL,\bx_j\in\calL:\; i\neq j}F(\widehat{P}_{\bx_i \bz \bx_j}, W_{Y|XZ})+k_n\right]}\\
&\geq \left(1-e^{-n\widetilde{\epsilon}_n}\right)\cdot e^{-n \left[\max_{P_{XZ\widetilde{X}}:\; P_{XZ\widetilde{X}}=P_{\widetilde{X}ZX},\; P_{XZ}=\widehat{P}_{XZ}}F(P_{XZ\widetilde{X}}, W_{Y|XZ})+k_n\right]}.\end{flalign}
And, similar to (\ref{eq: aiuvgiufgv})-(\ref{eq: idfugvsliudfg}) this gives
\begin{flalign}\label{eq: idfugvsliudadvfadvfg}
&-\frac{1}{n}\log \Pr(error)\leq \nonumber\\
&\min_{\substack{
\widehat{P}_{XZ}\in\calP_n(\calX\times \calZ):\; \widehat{P}_{X}=P_n,\\
I(\widehat{P}_{XZ})\leq R-\epsilon_n''}}D(\widehat{P}_{Z|X}\|W_{Z|X}|\widehat{P}_X)+ \max_{\substack{P_{XZ\widetilde{X}}:\; P_{XZ\widetilde{X}}=P_{\widetilde{X}ZX},\\ P_{XZ}=\widehat{P}_{XZ}}}F(P_{XZ\widetilde{X}}, W_{Y|XZ})+\delta''_n,
\end{flalign}
where $\epsilon_n''\triangleq \epsilon_n+\frac{|\calX||\calZ|-1}{n}\log(n+1)+\widetilde{\epsilon}_n$.
It remains to show that the minimization over empirical conditional distributions $\calP_n(\calX\times \calZ)$ can be approximated by a minimization over the simplex $\calP(\calX\times \calZ)$, as asserted in the following lemma whose proof appears in Appendix \ref{ap: aiudvgiudfgviugdfv}. 

\begin{lemma}\label{lm: udvagiudfv}
\begin{flalign}
&\min_{\substack{
\widehat{P}_{XZ}\in\calP_n(\calX\times \calZ):\;\widehat{P}_{X}=P_n,\\
I(\widehat{P}_{XZ})\leq R}}D(\widehat{P}_{Z|X}\|W_{Z|X}|\widehat{P}_X)+ \max_{\substack{\mu_{XZ\widetilde{X}}:\; \mu_{XZ\widetilde{X}}=\mu_{\widetilde{X}ZX},\\ \mu_{XZ}=\widehat{P}_{XZ}}}F(\mu_{XZ\widetilde{X}}, W_{Y|XZ})\nonumber\\
&\leq \min_{\substack{
P_{XZ}\in\calP(\calX\times \calZ):\;P_{X}=P_n,\\
I(P_{XZ})\leq R-\epsilon_{1,n}}}D(P_{Z|X}\|W_{Z|X}|P_X)+ \max_{\substack{\mu_{XZ\widetilde{X}}:\; \mu_{XZ\widetilde{X}}=\mu_{\widetilde{X}ZX},\\ \mu_{XZ}=P_{XZ}}}F(\mu_{XZ\widetilde{X}}, W_{Y|XZ})+\overline{\delta}_n+\epsilon_{1,n},
 \end{flalign}
 where $\epsilon_{1,n}=2 \frac{|\calX||\calZ|}{n}\log n$ and $\overline{\delta}_n=2\frac{|\calX||\calY||\calZ|}{n}\log n+\frac{|\calX||\calY||\calZ|}{n}\log\frac{n|\calZ|}{w_{min}}$, where  
$w_{min}$ is defined in (\ref{eq: W min dfn}). 
\end{lemma}

Thus, denoting $a_n= \overline{\delta}_n+\epsilon_{1,n}+\delta_n''$ we proved that 
\begin{flalign}\label{eq: idfugvfzdbdfsliudadvfadvfg}
&-\frac{1}{n}\log \Pr(error)\leq \nonumber\\
&\min_{\substack{
P_{XZ}\in\calP(\calX\times \calZ):\\P_{X}=P_n,\\
I(P_{XZ})\leq R -\epsilon_n''-\epsilon_{1,n}}}D(P_{Z|X}\|W_{Z|X}|P_X)+ \max_{\substack{\mu_{XZ\widetilde{X}}:\\ \mu_{XZ\widetilde{X}}=\mu_{\widetilde{X}ZX},\\ \mu_{XZ}=P_{XZ}}}
\min_{\substack{V_{\widetilde{X}XYZ}:\; 
V_{XZ\widetilde{X}}=\mu_{XZ\widetilde{X}}\\
q(V_{\widetilde{X}Y})\geq q(V_{XY}), \\I_V(\widetilde{X};Y|XZ)=0}
} D(V_{Y|XZ}\|W_{Y|XZ}|P_{XZ})+a_n,
\end{flalign}
and since $D(V_{YZ|X}\|W_{YZ|X}|P_Z)=D(P_{Z|X}\|W_{Z|X}|P_X)+D(V_{Y|XZ}\|W_{Y|XZ}|P_{XZ})$ and $W_{Z|XY}$ can be optimized, we can choose in (\ref{eq: ahfuv W star}) $W^*_{Z|XY}=P_{Z|XY}$, yielding $\|W_{Z|XY}-P_{Z|XY}\|\leq \frac{\overline{\epsilon}_n}{|\calZ|}$. Since we also have $P_{Z|XY}\ll W_{Z|XY}$, from Pinsker's inverse inequality (as mentioned above, see e.g.\ \cite[Lemma 4.1]{GotzeSabaleSinulis2019}), we get $D(P_{Z|XY}\|W_{Z|XY}|P_n\times P_{Z|X})\leq
2\frac{\overline{\epsilon}_n}{|\calZ|}$, yielding
\begin{flalign}
&-\frac{1}{n}\log \Pr(error)\leq
\widetilde{E}_{sp}^{q,sym}(R-\epsilon_n''-\epsilon_{1,n},P_n,W)+a_n+2\frac{\overline{\epsilon}_n}{|\calZ|}.
\end{flalign}
which concludes the proof.

\subsection{Proof of Claim \ref{cl: calWs stronger claim}}\label{sc: calWs stronger claim}
We prove (\ref{Eq: a;ofhv;oifv}) 
by establishing that if $V_{XYZ}$ is such that $V_{YZ|X}\in
\calW_q(\widehat{P}_X)$ and $ V_{XZ}=\widehat{P}_{XZ}$, then for any $\bz$ and $\bx_i,\bx_j\in \calL(\bz,\widehat{P}_{XZ})$, we have
\begin{flalign}
\EE_{
\frac{1}{2} [\widehat{P}_{\bx_i \bz \bx_j}(\widetilde{x},z,x)+\widehat{P}_{\bx_j \bz \bx_i}(\widetilde{x},z,x)]
\times V_{Y|XZ}}[q(\widetilde{X},Y)-q(X,Y)]\geq 0.\label{eq: fivudiv}
\end{flalign}
To this end we present a few definitions and two lemmas. 

Recall the definition of $P^{avg, \calL'}_{XZ\widetilde{X}}$ (see (\ref{eq: Pavg dfnsuhfv;oiudhf;oho;idhvf})) defined for any collection of codewords $\calL'\subseteq\calL$ such that $|\calL'|\geq 2$, 
\begin{flalign}
P^{avg, \calL'}_{XZ\widetilde{X}}(x,z,\widetilde{x})&\triangleq \frac{1}{|\calL'|(|\calL'|-1)}\sum_{i,j\in\calL':\;  j\neq i}\widehat{P}_{\bx_j \bz \bx_i}(x,z,\widetilde{x}). \label{eq: Pavg dfnsuhfv;oiudhf;oho;idhvf}
\end{flalign}
We next define an additional distribution $\overline{P}^{\calL'}_{TZX\widetilde{X}}$. Let $T$ be a RV uniformly distributed over $\{1,...,n\}$ and define
\begin{flalign}
\overline{P}^{\calL'}_{TZ}(t,z)&\triangleq \frac{1}{n}\cdot \indicator_{\{z=\bz(t)\}}\\
\overline{P}^{\calL'}_{X|T}(x|t)&\triangleq  \frac{\sum_{j\in \calL'}
\indicator_{\{\bx_j(t)=x\}}}{ |\calL'|}\\
\overline{P}^{\calL'}_{TZX\widetilde{X}}(t,z,x,\widetilde{x})&\triangleq \overline{P}^{\calL'}_{TZ}(t,z)\overline{P}^{\calL'}_{X|T}(x|t)\overline{P}^{\calL'}_{X|T}(\widetilde{x}|t).\label{eq: iusdgicugd}
\end{flalign}
The following lemma detects a few relations between $P^{avg, \calL'}_{XZ\widetilde{X}}$ and $\overline{P}^{\calL'}_{XZ\widetilde{X}}$. \begin{lemma}\label{lm: avg overline lemma}
For any collection of codewords $\calL'\subseteq\calL$ such that $|\calL'|\geq 2$ and any $z,x\neq \widetilde{x}$ 
\begin{flalign}
P^{avg, \calL'}_{X
Z\widetilde{X}}(x,z,\widetilde{x})=\frac{|\calL'|}{|\calL'|-1} \cdot \overline{P}^{\calL'}_{XZ\widetilde{X}}(x,z,\widetilde{x}).\label{eq: iudfh}
\end{flalign}

Further, 
for any given $P_{Y|XZ\widetilde{X}}$ letting $\overline{P}^{\calL'}_{X\widetilde{X}ZY}=\overline{P}^{\calL'}_{XZ\widetilde{X}}\times P_{Y|XZ\widetilde{X}}$, and 
$P^{avg, \calL'}_{X\widetilde{X}ZY}=P^{avg, \calL'}_{XZ\widetilde{X}}\times P_{Y|XZ\widetilde{X}}$, one has 
\begin{flalign}
q(P^{avg, \calL'}_{\widetilde{X}Y})-q(P^{avg, \calL'}_{XY})\geq 0\mbox{ iff }q(\overline{P}^{\calL'}_{\widetilde{X}Y})-q(\overline{P}^{\calL'}_{XY})\geq 0\label{eq: afuhv;iuhdf}
\end{flalign}
\end{lemma}
\begin{proof}
Recall that all members of $\calL'$ lie in $\calT_n(P_{X|Z}|\bz)$. 
Next, based on Plotkin's counting idea (similar to \cite{SHANNONGallagerBerlekamp1967522}), we obtain that for all $(x,\widetilde{x},z)\in\calX^2\times\calZ$ such that $x\neq \widetilde{x}$, 
\begin{flalign}
&|\calL'|(|\calL'|-1)\cdot P^{avg, \calL'}_{XZ\widetilde{X}}(x,z,\widetilde{x})\nonumber\\
&= \sum_{i,\;j\neq i\in \calL'}
\widehat{P}_{\bx_j \bz \bx_i}(x,\widetilde{x},z)\\
& =\sum_{j\in \calL',i\in \calL'}
\widehat{P}_{\bx_j \bz \bx_i}(x,z,\widetilde{x})- \sum_{j\in \calL'}
\widehat{P}_{\bx_j\bz\bx_j}(x,z,\widetilde{x})\label{eq: iudfgivgv122}\\
& =\frac{1}{n}\sum_{t=1}^n\left[\sum_{j\in \calL',i\in \calL'}
\indicator_{\{\bx_j(t)=x,\bz(t)=z,\bx_i(t)=\widetilde{x}\}}\right]\label{eq: iudfgivgv1}\\
& =\frac{1}{n}\sum_{t=1}^n\indicator_{\{\bz(t)=z\}} \left[\sum_{j\in \calL',i\in \calL'}
\indicator_{\{\bx_j(t)=x\}}\cdot \indicator_{\{\bx_i(t)=\widetilde{x}\}}\right]\label{eq: iudfgivgv}\\
& =\frac{1}{n}\sum_{t=1}^n\indicator_{\{\bz(t)=z\}}\left[\sum_{j\in \calL'}
\indicator_{\{\bx_j(t)=x\}}\right]\cdot \left[\sum_{i\in \calL'}\indicator_{\{\bx_i(t)=\widetilde{x}\}}\right]\label{eq: iudfgivgvMM}\\
&\triangleq  |\calL'|^2 
\sum_{t=1}^n \overline{P}^{\calL'}_{TZ}(t,z)\cdot \overline{P}^{\calL'}_{X|T}(x|t)\cdot \overline{P}^{\calL'}_{X|T}(\widetilde{x}|t),
\end{flalign}
where (\ref{eq: iudfgivgv1}) follows since $\sum_{j\in \calL'}
\widehat{P}_{\bx_j\bx_j\bz}(x,\widetilde{x},z)=0$ whenever $x\neq \widetilde{x}$, and 
(\ref{eq: iudfgivgvMM}) follows since the indicator $\indicator_{\{\bx_j(t)=x\}}$ does not depend on $i$. 
This yields (\ref{eq: iudfh}).
To prove (\ref{eq: afuhv;iuhdf}), for given $P_{Y|XZ\widetilde{X}}$ denote $\overline{P}^{\calL'}_{X\widetilde{X}ZY}=\overline{P}^{\calL'}_{XZ\widetilde{X}}\times P_{Y|XZ\widetilde{X}}$, and 
$P^{avg, \calL'}_{X\widetilde{X}ZY}=P^{avg, \calL'}_{XZ\widetilde{X}}\times P_{Y|XZ\widetilde{X}}$, and note that from (\ref{eq: iudfh}) it follows that 
\begin{flalign}
q(P^{avg, \calL'}_{\widetilde{X}Y})-q(P^{avg, \calL'}_{XY})&= \frac{|\calL'|}{|\calL'|-1}\cdot \left(q(\overline{P}^{\calL'}_{\widetilde{X}Y})-q(\overline{P}^{\calL'}_{XY})\right),
\end{flalign}
and thus (\ref{eq: afuhv;iuhdf}) follows. 
\end{proof}

The following lemma shows that the random variable $T$ (uniformly distributed over $\{1,...,n\}$) can be replaced by another random variable $U$ of finite alphabet cardinality that does not increase with $n$. 
\begin{lemma}\label{lm: Caratheodory}
There exists a random variable $U$ whose alphabet size is $|\calU|\leq |\calX|^2|\calZ|$, and a joint distribution $P_{UXZ}\in\calP(\calU\times \calX\times\calZ)$ such that for any $(x,\widetilde{x},z)$, 
\begin{flalign}\label{eq: iuafgivufigvdfiu}
\sum_u P_{UZ}(u,z)P_{X|UZ}(x|u,z)P_{X|UZ}(\widetilde{x}|u,z)&=\overline{P}_{XZ\widetilde{X}}(x,\widetilde{x},z).
\end{flalign}
There also exists a joint distribution $P_{UXZ}\in\calP(\calU\times \calX\times\calZ)$ such that for any $(x,\widetilde{x},z)$ (\ref{eq: iuafgivufigvdfiu}) holds, $Z$ is a deterministic function of $U$, and $|\calU|\leq |\calX|^2|\calZ|+1$.
\end{lemma}
Lemma \ref{lm: Caratheodory} is proved in Appendix \ref{ap: Caratheodory}. 
It remains to show that (\ref{eq: fivudiv}) holds for any $V_{XYZ}$ such that $V_{YZ|X}\in
\calW_q(\widehat{P}_X)$ and $ V_{XZ}=\widehat{P}_{XZ}$. 
Let $\widetilde{\calL}=\{\bx_i,\bx_j\}$. Obviously, $\widetilde{\calL}\subseteq\calL$, and thus we can invoke (\ref{eq: afuhv;iuhdf}) to obtain
 \begin{flalign}
 &\EE_{\frac{1}{2}[\widehat{P}_{\bx_i \bz \bx_j}+\widehat{P}_{\bx_j \bz \bx_i}]\times V_{Y|XZ}}[q(\widetilde{X},Y)-q(X,Y)]
 \nonumber\\
&=\sum_{x,\widetilde{x},z,y} P^{avg, \widetilde{\calL}}_{XZ\widetilde{X}}(x,z,\widetilde{x})\cdot V(y|x,z)\left[q(\widetilde{x},y)-q(x,y)\right]\label{eq: aliufdgvdsoijv'osidjviludf}\\
  &=\frac{|\widetilde{\calL}|}{|\widetilde{\calL}|-1}\sum_{x,\widetilde{x},z,y} \overline{P}^{\widetilde{\calL}}_{XZ\widetilde{X}}(x,z,\widetilde{x})
 \cdot V(y|x,z)\left[q(\widetilde{x},y)-q(x,y)\right]\label{eq: ai;udvghodsijviojsidj;iudhvoidhio}\\
   &\geq
   \min_{\substack{P_{UXZ}:\\P_{XZ}=\widehat{P}_{XZ}} }\frac{|\widetilde{\calL}|}{|\widetilde{\calL}|-1}\sum_{u,x,\widetilde{x},z,y} 
   P_{UZ}(u,z)P_{X|UZ}(x|u,z)P_{X|UZ}(\widetilde{x}|u,z) \cdot V(y|x,z)\left[q(\widetilde{x},y)-q(x,y)\right]\label{eq: ifuadsijv'osdjv}\\
   &\geq 0,\label{eq: iudsdvdssdjv'osidjvvsdshgviudg}
\end{flalign}
where (\ref{eq: aliufdgvdsoijv'osidjviludf}) follows by definition of $P^{avg, \widetilde{\calL}}_{XZ\widetilde{X}}$, (\ref{eq: ai;udvghodsijviojsidj;iudhvoidhio}) follows from (\ref{eq: iudfh}), (\ref{eq: ifuadsijv'osdjv}) follows by definition of $\overline{P}^{\widetilde{\calL}}_{XZ\widetilde{X}}$ and from Lemma \ref{lm: Caratheodory}, and  (\ref{eq: iudsdvdssdjv'osidjvvsdshgviudg}) follows since
$V_{YZ|X}\in
\calW_q(\widehat{P}_X)$ and $ V_{XZ}=\widehat{P}_{XZ}$. 
Thus, we have shown that (\ref{eq: fivudiv}) holds and consequently also (\ref{Eq: a;ofhv;oifv}).

\subsection{Proof of Lemma \ref{eq: List size lemma}}\label{sc: List size lemma}

From the law of total probability
\begin{flalign}
&\Pr\left(
 |\calL(\bZ,\widehat{P}_{XZ})|<e^{n\tau}\big|\widehat{P}_{\bX\bZ}=\widehat{P}_{XZ}\right)\\
&=\frac{1}{\mathbb{M}_n}\sum_{i=1}^{\mathbb{M}_n} \Pr\left(
 |\calL(\bZ,\widehat{P}_{XZ})|<e^{n\tau}\big|\bX=\overline{\bx}_i,\widehat{P}_{\bX\bZ}=\widehat{P}_{XZ}\right)\label{eq: X1}\\
&=\frac{1}{\mathbb{M}_n}\sum_{i=1}^{\mathbb{M}_n} \left|\{\bz\in \calT_n(\widehat{P}_{Z|X}|\overline{\bx}_i):\;|\calL| <e^{n\tau}\}\right|\cdot \frac{1}{|\calT_n(\widehat{P}_{Z|X}|\overline{\bx}_i)|}\label{eq: X2}\\
&=\frac{1}{\mathbb{M}_n}\sum_{\bz\in\calT_n(\widehat{P}_Z):\; |\calL| <e^{n\tau}} |\calL|\cdot \frac{1}{|\calT_n(\widehat{P}_{Z|X}|\overline{\bx}_i)|}\label{eq: X4}\\
&\leq\frac{1}{\mathbb{M}_n}\cdot e^{n\tau} \cdot |\calT_n(\widehat{P}_Z)| \cdot \frac{1}{|\calT_n(\widehat{P}_{Z|X}|\overline{\bx}_i)|}\label{eq: X3}\\
&\leq (n+1)^{|\calX||\calZ|-1}\cdot e^{-n[R-I(\widehat{P}_{XZ})-\tau]}\label{eq: X5}\\
&=e^{-n[R-I(\widehat{P}_{XZ})-\tau-\frac{|\calX||\calZ|-1}{n}\log(n+1)]},
\end{flalign}
where (\ref{eq: X2}) follows since $\bZ$ is uniform over $\calT_n(\widehat{P}_{Z|X}|\overline{\bx}_i)$ given $\overline{\bx}_i$, (\ref{eq: X3}) holds by replacing the count over codewords by a count over sequences $\bz$, and (\ref{eq: X5}) follows by a standard bound on the size of a type-class.

\subsection{Proof of Lemma \ref{lm: Caratheodory}}\label{ap: Caratheodory}
We prove that while $T$ is a RV uniformly distributed over $\{1,...,n\}$, indeed the cardinality of the alphabet of the random variable $U$ can
be limited without loss of generality by $|\calX|^2\cdot |\calZ|$. 
This is done by an application of
the support Lemma (Caratheodory's Theorem).
Note that by Bayes' Law $\overline{P}_{XZ\widetilde{X}}(x,\widetilde{x},z)= \sum_t \overline{P}_T(t) \frac{\overline{P}_{ZX|T}(z,x|t)\overline{P}_{ZX|T}(z,\widetilde{x}|t)}{V_{Z|T}(z|t)}$, and thus
there exists a distribution $V_{UXZ}$ such that the expectations (w.r.t.\ $U$) of the following $|\calU|\leq |\calX|^2\cdot |\calZ|$ functionals of $V_{XZ|U=u}$:
\begin{flalign}
&\left\{ \frac{V_{ZX|U}(z,x|u)V_{ZX|U}(z,\widetilde{x}|u)}{V_{Z|U}(z|u)}\right\}_{(z,x,\widetilde{x})\in\calZ  \times\calX^2 }, 
\end{flalign}
preserve those of $\overline{P}_{TXZ}$; i.e., 
\begin{flalign}
\forall (z,x,\widetilde{x}),\; \sum_u V_U(u) \frac{V_{ZX|U}(z,x|u)V_{ZX|U}(z,\widetilde{x}|u)}{V_{Z|U}(z|u)}&= 
\sum_t \overline{P}_{T} (t)\frac{\overline{P}_{ZX|T}(z,x|T=t) \overline{P}_{ZX|T}(z,\widetilde{x}|T=t)}{\overline{P}_{Z|T}(z|T=t)}
\end{flalign}
because there are in fact $|\calZ | \cdot|\calX|^2-1$ degrees of freedom in $\overline{P}_{ZX\widetilde{X}}(z,x,\widetilde{x})$, it suffices to preserve only $|\calX|^2\cdot |\calZ|-1$ of the functionals. 
 
As for the last assertion of the lemma, note that preserving the expectation of one additional functional $H_V(Z|U=u) =-\sum_{x,z}V_{ZX|U}(z,x|u)\log \sum_{x'}V_{ZX|U}(z,x'|u)$ yields 
\begin{flalign}
\sum_u V_U(u) H_{\overline{T}}(Z|U=u) &= 
\sum_t \overline{P}_{T} (t)
H_{\overline{P}_{Z|T}}(Z|T=t),
\end{flalign}
since $H_{\overline{P}}(Z|T) =0$, $Z$ is also  deterministic function of $U$, where the alphabet cardinality increase is $1$.

\subsection{Proof of Lemma \ref{lm: APPPROX five}}\label{ap: Proof of Lemma app}

Let $P_{XZ}=P_n\times P_{Z|X}$,  
recall the definition of $\Delta^q(P_{XZU},P_{Y|XZ})$ in (\ref{eq: Delta q definition a}), and let
\begin{flalign}
&\calQ_{diff}(P_{XZ}, P_{Y|XZ})\triangleq \min_{P_{U|XZ}}\Delta^q(P_{XZU},P_{Y|XZ}).\label{eq: a;duhvu}
\end{flalign}

We use Lemma \ref{lm: Krein Milman 2} to express $P_{XZ}=P_n\times P_{Z|X}$ as a convex combination of empirical distributions, with $(\calX,\calZ,P_{XZ},n)$ in the roles of $(\calA,\calB,P_{AB},\ell)$, respectively, to obtain 
 \begin{flalign}
P_{XZ}&= \sum_i \beta_i \cdot P^{(i)}_{XZ}
\end{flalign}
where 
\begin{flalign}
&\forall i,\; P^{(i)}_{X}=P_X=P_n,\; \|P_{XZ}-P_{XZ}^{(i)}\|\leq\frac{|\calX||\calZ|}{n},\; \mbox{ and }
P_{Z|X}(z|x)=0\Rightarrow P_{Z|X}^{(i)}(z|x)=0
.\label{ADDeq: ahv;oifdh;iv}
\end{flalign} 
Denote by $\{P_{U|XZ}^{(i)}\}$ the minimizers corresponding to $\{P_{XZ}^{(i)}\}$ in (\ref{eq: a;duhvu}), respectively; that is, those $\{P_{U|XZ}^{(i)}\}$ which satisfy
\begin{flalign}
\calQ_{diff}(P_{XZ}^{(i)}, P_{Y|XZ})&\triangleq \Delta^q(P_{XZ}^{(i)}\times P_{U|XZ}^{(i)},P_{Y|XZ}).
\end{flalign}
Let $\{P_{\widetilde{X}XZU}^{(i)}\}$ be the corresponding induced  probabilities (by Bayes' Law); i.e., 
\begin{flalign}
P_{\widetilde{X}XZU}^{(i)}(\widetilde{x},x,z,u)&= P_{XZ}^{(i)}(x,z)P_{U|XZ}^{(i)}(u|x,z)
\cdot \frac{P_{U|XZ}^{(i)}(u|\widetilde{x},z)\cdot P_{XZ}^{(i)}(\widetilde{x},z)}{P_{UZ}^{(i)}(u,z)}.\label{ADDeq: foduiagv1111111}
\end{flalign}
Let $\overline{T}$ be the RV whose distribution is $[\beta_1,...,\beta_K]$, and signifies the value of $"i"$ in (\ref{ADDeq: foduiagv1111111}), we have $P_{\widetilde{X}XZ\overline{T}}(\widetilde{x},x,z,i)=\beta_i P_{\widetilde{X}ZX}^{(i)}(\widetilde{x},x,z)$, and therefore 
\begin{flalign}
&\max_i  \calQ_{diff}(P_{XZ}^{(i)}, P_{Y|XZ})\label{ADDeq: yujopjnodskjbavkj}\\
&\geq \sum_i \beta_i \calQ_{diff}(P_{XZ}^{(i)}, P_{Y|XZ}\label{ADDeq: yujopjno})\\
&=\sum_i\beta_i \EE_{P_{XZ\widetilde{X}}^{(i)}\times P_{Y|XZ}} [q(\widetilde{X},Y)-q(X,Y)]\\
&\triangleq \Delta^q(P_n\times P_{Z|X}\times  P_{U,\overline{T}|XZ},P_{Y|XZ})\\
&\triangleq \Delta^q(P_n\times P_{Z|X}\times  P_{\widetilde{U}|XZ},P_{Y|XZ}\label{ADDeq: laidfhs})\\
&\geq \calQ_{diff}(P_n\times P_{Z|X}, P_{Y|XZ})\label{ADDeq: laidfhv}\\
&\geq 0,\label{ADDeq: laidfhvdfavhopi}
\end{flalign}
where (\ref{ADDeq: yujopjno}) follows since the average is upper bounded by the maximal value, in (\ref{ADDeq: laidfhs}) we define $\widetilde{U}=(U,\overline{T})$, and (\ref{ADDeq: laidfhv}) follows by 
definition of $\calQ_{diff}$ as the minimum over $P_{\widetilde{U}|XZ}$ of $\Delta^q(P_n\times P_{Z|X}\times  P_{\widetilde{U}|XZ},P_{Y|XZ})$. The fact that the alphabet of $\widetilde{U}$ is larger than that of $\calU$ does not impair the derivation as by Caratheodory's Theorem, any $\widetilde{U}$ can be substituted with $U$ of alphabet size not larger than $|\calX|^2|\calZ|$ as asserted in Lemma \ref{lm: Caratheodory}. The last step (\ref{ADDeq: laidfhvdfavhopi}) follows since $P_{YZ|X}\in \calW_q(P_X)$. 

Finally, let $i^*$ be the maximizer of (\ref{ADDeq: yujopjnodskjbavkj}), thus $P_{XZ}^{(i^*)}$ serves as the empirical distribution whose existence is claimed in Lemma \ref{lm: APPPROX five}.

\subsection{Proof of Lemma \ref{ADDlm: non emptiness calG}}\label{ap: non emptiness calG proof}

 We continue the proof of Lemma \ref{lm: APPPROX five} (in Appendix \ref{ap: Proof of Lemma app}) applied to $W_{YZ|X}$ in the role of $P_{YZ|X}$. 
 
To conclude the proof of Lemma \ref{ADDlm: non emptiness calG}, we verify that $P_{XZ}^{(i^*)}\in \calK_q(R,P_n, W_{YZ|X})$ by checking that the $3$ requirements in its definition (\ref{eq: akduvudh}) are met. 
\begin{enumerate}
\item 
From (\ref{ADDeq: yujopjnodskjbavkj})-(\ref{ADDeq: laidfhvdfavhopi}) we have that $P_{XZ}^{(i^*)}\in \calA_q(W_{Y|XZ},P_n)$. 
\item 
From (\ref{ADDeq: ahv;oifdh;iv}) we have 
$D(P_{Z|X}^{(i^*)}\|W_{Z|X}|P_n)\leq 2\frac{(|\calX||\calZ|)^2}{n^2 \cdot \min_{z,x:\; W_{Z|X}(z|x)>0}W_{Z|X}(z|x)}=O(1/n^2)$, by the inverse Pinsker's inequality (see e.g.\ \cite[Lemma 4.1]{GotzeSabaleSinulis2019}). 
Since $f_n=c_n-\frac{1}{n}|\calX||\calZ|\log(n+1) $, where $c_n\gg \frac{1}{n}$ (for example $c_n=n^{-1/2}$), for $n$ sufficiently large, we clearly have $D({P}_{Z|X}^{(i^*)}\|W_{Z|X}|P_n)\leq f_n$. 
\item 
By continuity of the mutual information, inequality (\ref{ADDeq: ahv;oifdh;iv}) also implies that $|I(P_n\times W_{Z|X})- I(P_{XZ}^{(i^*)})|\leq |\calX||\calZ|\cdot \frac{\log n}{n}$ (see \cite[Lemma 2.7]{CsiszarKorner81}). 
Since $I(P_n\times W_{Z|X})\leq R-\epsilon$, this gives $I(P_{XZ}^{(i^*)})\leq R - \epsilon+|\calX||\calZ|\cdot \frac{\log n}{n}$. 
Since $d_n=\epsilon_n+\frac{|\calX||\calZ|-1}{n}\log(n+1)+\widetilde{\epsilon}_n$, where both $\epsilon_n,\widetilde{\epsilon}_n$ are vanishing sequences, for $n$ sufficiently large we thus have $I(P_{XZ}^{(i^*)})\leq R -d_n$. 
\end{enumerate}
Hence, $P_{XZ}^{(i^*)}\in \calK_q(R,P_n, W_{YZ|X})$, and this concludes the proof of Lemma \ref{ADDlm: non emptiness calG}. 

\subsection{Proof of Lemma \ref{lm: udvagiudfv}}\label{ap: aiudvgiudfgviugdfv}

Let $P_{XZ}^*$ denote the minimizer of the following function
\begin{flalign}
&\min_{\substack{
P_{XZ}\in\calP(\calX\times \calZ):\;P_{X}=P_n,\\
I(P_{XZ})\leq R}}D(P_{Z|X}\|W_{Z|X}|P_X)+ \max_{\substack{\mu_{XZ\widetilde{X}}:\; \mu_{XZ\widetilde{X}}=\mu_{\widetilde{X}ZX},\\ \mu_{XZ}=P_{XZ}}}F(\mu_{XZ\widetilde{X}}, W_{Y|XZ})\nonumber\\
&\triangleq 
D(P_{Z|X}^*\|W_{Z|X}|P_X)+ \max_{\mu_{XZ\widetilde{X}}:\; \mu_{XZ\widetilde{X}}=\mu_{\widetilde{X}ZX},\; \mu_{XZ}=P_{XZ}^*}F(\mu_{XZ\widetilde{X}}, W_{Y|XZ}).
\label{Eq: davij'dfjvo}
\end{flalign}
Next, we use Lemma \ref{lm: Krein Milman 2} to express $P_{XZ}^*$ as a convex combination of empirical distributions $P_{XZ}^*=\sum_i\beta_i P_{XZ}^{(i)}$, with $\left(n, \calX, P_n,\calZ, P_{Z|X}^*\right)$ in the roles of 
$\left(\ell, \calA, Q_A,\calB, P_{B|A}\right)$, respectively, and we let $ P_{Z|X}^{(i)}$ denote $P_{B|A}^{(i)}$. 

By this construction we have for all $i$, 
$
\|P_n\times P_{Z|X}^{*}-P_n\times  P_{Z|X}^{(i)}\|\leq\frac{|\calX||\calZ|}{n}$. 
Now, for each $i$, denote by $P_{\widetilde{X}|XZ}^{(i)}$ the maximizing $\mu_{\widetilde{X}|XZ}^{(i)}$ of:
$
\max_{\mu_{XZ\widetilde{X}}:\; \mu_{XZ\widetilde{X}}=\mu_{\widetilde{X}ZX},\; \mu_{XZ}=P_{XZ}^{(i)}}F(\mu_{XZ\widetilde{X}}, W_{Y|XZ})
$, and 
this gives
\begin{flalign}
& \max_{\mu_{XZ\widetilde{X}}:\; \mu_{XZ\widetilde{X}}=\mu_{\widetilde{X}ZX},\; \mu_{XZ}=P_{XZ}^*}F(\mu_{XZ\widetilde{X}}, W_{Y|XZ})\nonumber\\
&\geq
F(\sum_i\beta_i P_{XZ\widetilde{X}}^{(i)}, W_{Y|XZ})\label{Eq: ;afhv}\\
&=\min_{\substack{V_{Y|XZ\widetilde{X}}\in\calS^{cond}(\frac{1}{2}\sum_i\beta_i[P_{XZ\widetilde{X}}^{(i)}+P_{\widetilde{X}ZX}^{(i)}]): \\I_V(\widetilde{X};Y|XZ)=0}
} D(V_{Y|XZ}\|W_{Y|XZ}|\sum_i \beta_iP_{XZ}^{(i)})\label{eq: aifuhv;ud}.\end{flalign}
where (\ref{Eq: ;afhv}) follows since by definition, the marginal $(X,Z)$ distribution of $\sum_i\beta_i P_{XZ\widetilde{X}}^{(i)}$ is equal to $\sum_i\beta_i P_{XZ}^{(i)}$ which is equal to $P_{XZ}^*$, and thus  $\sum_i\beta_i P_{XZ\widetilde{X}}^{(i)}$ yields an $F(\cdot)$ value that cannot exceed the maximal possible value. Step (\ref{eq: aifuhv;ud}) is by definition of $F(\cdot)$. Now, by affinity of the expectation, there must exist $i_0$ such that the minimizing $V_{Y|XZ\widetilde{X}}$ belongs to $\calS^{cond}(\frac{1}{2}[P_{XZ\widetilde{X}}^{(i_0)}+P_{\widetilde{X}ZX}^{(i_0)}])$, and hence
\begin{flalign}
&\min_{\substack{V_{Y|XZ\widetilde{X}}\in\calS^{cond}(\frac{1}{2}\sum_i\beta_i[P_{XZ\widetilde{X}}^{(i)}+P_{\widetilde{X}ZX}^{(i)}]): \\I_V(\widetilde{X};Y|XZ)=0}
} D(V_{Y|XZ}\|W_{Y|XZ}|\sum_i \beta_iP_{XZ}^{(i)})\nonumber\\
&\geq \min_{\substack{V_{Y|XZ\widetilde{X}}\in\calS^{cond}(\frac{1}{2}[P_{XZ\widetilde{X}}^{(i_0)}+P_{\widetilde{X}ZX}^{(i_0)}]): \\I_V(\widetilde{X};Y|XZ)=0}
} D(V_{Y|XZ}\|W_{Y|XZ}|\sum_i \beta_iP_{XZ}^{(i)})\\
&\geq \min_{\substack{V_{Y|XZ\widetilde{X}}\in\calS^{cond}(\frac{1}{2}[P_{XZ\widetilde{X}}^{(i_0)}+P_{\widetilde{X}ZX}^{(i_0)}]): \\I_V(\widetilde{X};Y|XZ)=0}
} D(V_{Y|XZ}\|W_{Y|XZ}|P_{XZ}^{(i_0)})-\overline{\delta}_n\label{eq: aifudvfgv}\\
&= \max_{\mu_{XZ\widetilde{X}}:\; \mu_{XZ\widetilde{X}}=\mu_{\widetilde{X}ZX},\; \mu_{XZ}=P_{XZ}^{(i_0)}}F(\mu_{XZ\widetilde{X}}, W_{Y|XZ}) -\overline{\delta}_n,
\end{flalign}
where (\ref{eq: aifudvfgv}) follows since by definition $\sum_i \beta_i P_{XZ}^{(i)}=P_{XZ}^*$, and since $\|P_{XZ}^*-P_{XZ}^{(i_0)}\|\leq\frac{|\calX|\calZ|}{n}$, which implies that for any $P_{Y|XZ}$, 
 $|D(P_{Y|XZ}\|W_{YZ|X}|P_{XZ}^{(i_0)})- D( P_{Y|XZ}\|W_{YZ|X}|P_{XZ}^*)|\leq 2\frac{|\calX||\calY||\calZ|}{n}\log n+\frac{|\calX||\calY||\calZ|}{n}\log\frac{1}{t_{n,min}}\triangleq \overline{\delta}_n$, where $t_{n,min}$ is defined in
 (\ref{eq: W min dfn}). The last step follows by definition of $P_{\widetilde{X}|XZ}^{(i_0)}$ as the maximizing $\mu_{\widetilde{X}|XZ}^{(i)}$ of:
$
\max_{\mu_{XZ\widetilde{X}}:\; \mu_{XZ\widetilde{X}}=\mu_{\widetilde{X}ZX},\; \mu_{XZ}=P_{XZ}^{(i_0)}}F(\mu_{XZ\widetilde{X}}, W_{Y|XZ})
$.

Now, 
\begin{flalign}
&D(P_{Z|X}^*\|W_{Z|X}|P_X)+ \max_{\mu_{XZ\widetilde{X}}:\; \mu_{XZ\widetilde{X}}=\mu_{\widetilde{X}ZX},\; \mu_{XZ}=P_{XZ}^{(i_0)}}F(\mu_{XZ\widetilde{X}}, W_{Y|XZ})\nonumber\\
&\geq  D(P_{Z|X}^{(i_0)}\|W_{Z|X}|P_X)+ \max_{\mu_{XZ\widetilde{X}}:\; \mu_{XZ\widetilde{X}}=\mu_{\widetilde{X}ZX},\; \mu_{XZ}=P_{XZ}^{(i_0)}}F(\mu_{XZ\widetilde{X}}, W_{Y|XZ})-\epsilon_{1,n}\label{eq: aofdiv}\\
&\geq \min_{\substack{
\widehat{P}_{XZ}\in\calP_n(\calX\times \calZ):\;\widehat{P}_{X}=P_n,\\
I(\widehat{P}_{XZ})\leq R+\epsilon_{1,n}}}D(\widehat{P}_{Z|X}\|W_{Z|X}|\widehat{P}_X)+ \max_{\substack{\mu_{XZ\widetilde{X}}:\; \mu_{XZ\widetilde{X}}=\mu_{\widetilde{X}ZX},\\ \mu_{XZ}=\widehat{P}_{XZ}}}F(\mu_{XZ\widetilde{X}}, W_{Y|XZ})-\epsilon_{1,n},\label{eq: aifugviufdg}
\end{flalign}
where both (\ref{eq: aofdiv}) and (\ref{eq: aifugviufdg}) hold since $\|P_{XZ}^*-P_{XZ}^{(i_0)}\|\leq\frac{|\calX|\calZ|}{n}$, which implies $|I(P_{XZ}^*)-I(P_n\times
 P_{XZ}^{(i_0)})|\leq 2 \frac{|\calX||\calZ|}{n}\log n\triangleq \epsilon_{1,n}$. The last step follows since $P_{Z|X}^{(i_0)}$ is an empirical distribution of order $n$, and since $I(P_{XZ}^*)\leq R$.

%
%
%
\end{document}